\documentclass{article}
\usepackage[margin=1in]{geometry}
\usepackage[round]{natbib}

\usepackage[]{graphicx}
\usepackage{color, soul, threeparttable, multirow, multicol, setspace}
\usepackage{bbm, caption, tikz, booktabs, colortbl, dcolumn, float, booktabs, amsfonts, amsthm, amsmath, threeparttable, makecell, multicol, multirow, rotating, algorithmic, algorithm, listings}
\usepackage{enumitem}

\usepackage{xcolor}
\usepackage{xr}
\usepackage{xr-hyper}
\usepackage{hyperref}
\hypersetup{
    colorlinks=true,
    citecolor=blue,  
    linkcolor=red, 
    urlcolor=red     
}

\usepackage{listings}
\newcommand{\bd}{\boldsymbol}
\newcommand{\Ex}{\mathbb{E}}
\newcommand{\Vx}{\mathbb{V}}
\newcommand{\Px}{\mathbb{P}}

\newcommand{\mc}{\mathcal}
\newcommand{\mb}{\mathbf}

\newcommand{\simple}{\tauest_{\text{simple}}}
\newcommand{\tauest}{\widehat\tau}
\newcommand{\Gset}{\mc G_{\bd\beta}^g}
\newcommand{\bigCI}{\mathrel{\text{\scalebox{1}{$\perp\mkern-10mu\perp$}}}}

\newcommand{\simp}{\text{simple}}
\newcommand{\aipw}{\text{aipw}}

\theoremstyle{thmstyleone}%
\newtheorem{theorem}{Theorem}
\newtheorem{assumption}{Assumption}
\newtheorem{condition}{Condition}
\theoremstyle{thmstyletwo}%
\newtheorem{remark}{Remark}%
\theoremstyle{thmstylethree}%

\definecolor{purpleR}{rgb}{0.627, 0.125, 0.941}
\definecolor{greenR}{rgb}{0, 1, 0}
\definecolor{cyanR}{rgb}{0, 1, 1}

\begin{document}

\begin{center}
\doublespacing
    \bf 
    {\LARGE COADVISE: Covariate Adjustment with Variable Selection in Randomized Controlled Trials}
\end{center}

\bigskip

\begin{center}
    {\large Yi Liu$^{1,2}$, Ke Zhu$^{1,3}$, Larry Han$^{4}$, and Shu Yang$^{*1}$} \\
    \bigskip

    $^1$Department of Statistics, North Carolina State University, Raleigh, NC, USA
    \bigskip

    $^2$Duke Clinical Research Institute, Duke University, Durham, NC, USA
    \bigskip

    $^3$Department of Biostatistics and Bioinformatics, Duke University School of Medicine, Durham, NC, USA
    \bigskip

    $^4$Department of Public Health and Health Sciences, Northeastern University, Boston, MA, USA

    \bigskip

    $^*$Corresponding author: Shu Yang, \texttt{syang24@ncsu.edu}
\end{center}

\bigskip

\abstract{Adjusting for covariates in randomized controlled trials can enhance the credibility and efficiency of treatment effect estimation. However, handling numerous covariates and their complex (non-linear) transformations poses a challenge. Motivated by the case study of the Best Apnea Interventions for Research (BestAIR) trial data from the National Sleep Research Resource (NSRR), where the number of covariates ($p=114$) is comparable to the sample size ($N=196$), we propose a principled \underline{Co}variate \underline{Ad}justment with \underline{V}ar\underline{i}able \underline{Se}lection (COADVISE) framework. COADVISE enables variable selection for covariates most relevant to the outcome while accommodating both linear and nonlinear adjustments. This framework ensures consistent estimates with improved efficiency over unadjusted estimators and provides robust variance estimation, even under outcome model misspecification. We demonstrate efficiency gains through theoretical analysis, extensive simulations, and a re-analysis of the BestAIR trial data to compare alternative variable selection strategies, offering cautionary recommendations. A user-friendly R package, \texttt{Coadvise}, is available to facilitate practical implementation.}

\bigskip


\doublespacing

\section{Introduction}\label{sec:intro}

\subsection{Background}\label{subsec:background}

Randomized controlled trials (RCTs) are widely regarded as the gold standard for evaluating the effects of treatments or interventions \citep{neyman1923applications, rubin1974estimating}. Through randomization, covariates are balanced on average across treatment groups, enabling unbiased estimation of treatment effects. However, in practice, randomization does not guarantee perfect balance in finite samples, particularly when dealing with a large number of covariates. Such covariate imbalances can obscure the interpretation of trial outcomes and introduce accidental biases. Incorporating covariates into the analysis can enhance both the interpretability and efficiency of RCT results. While covariate balance can be pursued during both the design \citep{fisher1926,taves1974min, pocock1975seq, rosen2008, hu2012asy, hu2014adap,morgan2012rerandomization, li2018asymptotic,zhao2024no} and analysis phases of an RCT, this paper focuses on methods for covariate adjustment during the analysis of already-conducted trials. 

Regulatory agencies, including the U.S. Food and Drug Administration (FDA) and the European Medicines Agency (EMA), have emphasized the importance of covariate adjustment in RCTs. The FDA recently issued guidelines encouraging its use to improve the precision of treatment effect estimates \citep{fda2023adjusting}. The EMA's 2015 guideline highlights that ``balancing treatment groups with respect to one or more specific prognostic covariates can enhance the credibility of the trial results'' \citep{ema2015guideline}. The establishment of the American Statistical Association (ASA) Biopharmaceutical Section (BIOP) Covariate Adjustment Scientific Working Group (SWG) (\url{https://carswg.github.io/}) further underscores the growing focus on advancing statistical methods for covariate adjustment in clinical trials.

Despite these regulatory endorsements, significant challenges remain due to high-dimensional covariates and nonlinear relationships. These complexities necessitate robust and comprehensive analytical strategies. However, the comparative advantages and limitations of existing methods remain insufficiently explored, leaving practitioners with limited guidance and uncertainty in selecting the most appropriate approaches. For instance, in the case study presented in Section \ref{sec:data}, the Best Apnea Interventions for Research (BestAIR) trial includes data from $N=169$ participants and $p=114$ covariates, making $p$ comparable to $N$. When adjusting for covariates using data from a single treatment arm---where $N_1=83$ (treated) and $N_0=86$ (controls)---$p$ exceeds the sample sizes. This high-dimensional setting can lead to biased and imprecise covariate-adjusted average treatment effect (ATE) estimates, as the sample size may be insufficient for reliable parameter estimation. Nevertheless, prior studies, such as \cite{gao2024adjusting}, have identified baseline covariate imbalances in the BestAIR trial data, underscoring the need for efficient covariate-adjusted methods. This motivates our approach to improving estimation precision while addressing challenges aforementioned. 

This paper presents a unified framework for covariate adjustment in RCT analysis, with a primary focus on selecting important covariates for reducing data dimensionality for efficiency gains. We provide a comprehensive tutorial on commonly used covariate-adjusted approaches, clarifying their underlying assumptions, strengths, and limitations. Additionally, we offer cautionary recommendations to guide practitioners in navigating potential pitfalls and selecting appropriate methods for specific scenarios. By addressing the aforementioned challenges, this framework aims to improve the reliability and interpretability of RCT findings, aligning with the evolving needs of researchers and regulatory agencies. 

The rest of this section reviews related work on covariate adjustment and outlines our contributions. Section \ref{sec:setup} covers key statistical preliminaries, including notation, assumptions, and the covariate adjustment methods employed in our framework. Section \ref{sec:coadvise} introduces the {\underline{Co}}variate {\underline{Ad}}justment with {\underline{V}}ar{\underline{i}}able {\underline{Se}}lection (COADVISE) framework, highlighting its integration of variable selection and covariate-adjusted estimators. Numerical experiments assess finite-sample performance are presented in Section \ref{sec:simu}. Section \ref{sec:data} demonstrates the application of COADVISE through a re-analysis of BestAIR trial data. Finally, Section \ref{sec:conclu} concludes the paper, discussing implications, limitations, and future directions. Additional tutorial guidance and R code are provided in Appendix \ref{app:addtuto}. 

\subsection{Related work}\label{subsec:relatework}

Covariate adjustment in RCTs has a rich history in the statistical literature. One of the earliest and most widely used approaches is the analysis of covariance (ANCOVA), introduced by \cite{fisher1966design}. This method estimates the ATE by fitting a linear regression model where the binary treatment indicator and covariates serve as predictors of the outcome. The ordinary least squares (OLS) estimator is then applied to the treatment coefficient to obtain the ATE estimate. ANCOVA is appealing due to its intuitive structure: by projecting the outcome onto the covariates, it accounts for variability explained by those covariates, thereby improving precision \citep{reluga2024unified}. Notably, this remains true even when the actual relationship between covariates and outcomes is more complex than a linear model \citep{wang2019analysis, wang2023model, harrell2024comment, harrell2017regression}.

Despite its strengths, ANCOVA is not universally superior to the simple estimator. Its efficiency and credibility can diminish when the conditional outcome models for treatment and control groups differ significantly \citep{freedman2008regression}. To address these limitations, \cite{lin2013agnostic} proposed including interactions between the treatment and covariates in the model. This method, known as analysis of heterogeneous covariance (ANHECOVA) \citep{cassel1976some, yang2001efficiency, ye2023toward}, allows the regression model to capture heterogeneity in the treatment effect across covariates. Although ANHECOVA has historical roots, \cite{lin2013agnostic} was instrumental in establishing its advantages through a design-based lens. This estimator is not only unbiased but also asymptotically more efficient than the simple estimator under a wide range of conditions, without relying on restrictive model assumptions \citep{lin2013agnostic}. The adoption of ANHECOVA has expanded in recent years, particularly for its ability to enhance efficiency in RCTs employing covariate-adaptive randomization schemes \citep{bannick2025general}. The U.S. FDA has endorsed its use, especially in evaluating conditional treatment effects, further strengthening its role in modern controlled trials \citep{fda2023adjusting, van2023use}. Recent work by \cite{reluga2024unified} provided a unified perspective on the relationships between the simple estimator, ANCOVA, and ANHECOVA, providing conditions under which ANHECOVA uniformly dominates other methods in terms of efficiency and precision.

Beyond ANCOVA and ANHECOVA, a prominent method in the analysis of RCTs is augmented inverse probability weighting (AIPW) \citep{robins1994estimation, tsiatis2006semiparametric, glynn2010introduction, cao2009improving, kang2007demystifying, yuan2012variable, zhang2008improving, tsiatis2008covariate}. The AIPW estimator combines outcome modeling with inverse {treatment} probability weighting, effectively leveraging information from both observed outcomes and treatment assignment probabilities. 
{This dual-model approach allows AIPW to achieve semiparametric efficiency—that is, to attain the semiparametric variance lower bound—when both the treatment and outcome models are correctly specified. In the setting of RCTs considered in this paper, the treatment assignment mechanism is known by design, so only correct specification of the outcome model is required. As a result, AIPW can be asymptotically most efficient among regular estimators of the ATE in this context. Moreover, AIPW remains consistent even when the outcome model is misspecified, owing to the robustness conferred by randomization.} 
This robustness, combined with its efficiency, makes AIPW particularly attractive in practice. {Recent work has extended AIPW methodology to accommodate more complex randomization schemes. For example, \cite{bannick2025general} consider covariate-adaptive randomization, in which treatment assignment probabilities are stratified and potentially correlated with baseline covariates. In certain structured settings, such as when allocation probabilities are constant within strata, AIPW can still achieve semiparametric efficiency with only a correctly specified outcome model. However, under more general covariate-adaptive designs, correct modeling of the treatment assignment mechanism becomes necessary, mirroring challenges found in observational studies. Additionally, \cite{ye2023robust, ye2023toward} propose robust variance estimators for AIPW under multi-arm RCT settings, which we adopt in our implementation (see Section \ref{subsec:AIPW}). } 

Moreover, AIPW offers flexibility by accommodating a wide range of parametric and nonparametric models for outcome estimation, making it more adaptable than ANCOVA or ANHECOVA, which depend on fixed linear model structures. Interestingly, when linear regression is used for outcome modeling within the AIPW framework, the AIPW estimator becomes equivalent to ANHECOVA {in RCTs considered in this paper}. In this sense, ANHECOVA can be viewed as a special case of AIPW \citep{van2023use}. However, this equivalence also underscores a limitation of ANHECOVA: if the linear model is misspecified, the efficiency gains achieved by ANHECOVA are limited compared to those of AIPW when the latter employs correctly specified---or at least better specified---models. For this reason, AIPW is also regarded as a fully adjusted estimator in the literature, capable of optimizing efficiency without being overly reliant on strict parametric assumptions \citep{li2023estimating}. 

As mentioned in Section \ref{subsec:background}, all covariate adjustment methods encounter difficulties in ensuring parameter convergence when faced with high-dimensional covariate data. As a result, a key preliminary step is variable selection---ideally choosing a set of variables smaller than the sample size $N$---that are most predictive of the outcome. \cite{cho2024variable} and \cite{belloni2014inference} proposed strategies tailored to covariate adjustment in observational studies, and \cite{yang2020doubly} presented complementary methods in the context of data integration. Following these prior works, we aim to select covariates that strongly predict the outcome. 

Several methods have been studied for data-driven variable selection procedures in RCTs. For example, the least absolute shrinkage and selection operator (Lasso) \citep{tibshirani1996regression} has been used in conjunction with ANHECOVA to select relevant covariates \citep{bloniarz2016lasso,schochet2022lasso,liu2023lasso,zhu2025design}. Another approach involves selecting the top $k$ covariates with the highest marginal correlations with the outcome, where $k<p$ \citep{senn1994testing}.  Alternatively, covariates can be screened using preliminary significance tests, such as two-sample $t$-tests or $z$-tests, to identify those most strongly associated with the treatment assignment or outcome \citep{permutt1990testing, zhao2024randomization}. These methods have primarily focused on continuous outcomes and the application of ANCOVA and ANHECOVA estimators. As we noted, identifying which covariate sets are most beneficial for covariate adjustment is the objective. Cho and Yang, along with additional references, propose various strategies tailored to specific purposes of covariate adjustment in observational studies. Similarly, \cite{yang2020doubly} offered aligned strategies in the context of data integration. In the context of RCTs, where bias is controlled, covariate adjustment primarily aims to enhance efficiency. Therefore, consistent with previous research \citep{cho2024variable, yang2020doubly}, we recommend selecting covariates that are strong prognostic factors for the outcome.

For binary or categorical outcomes, research on covariate adjustment has also made significant progress \citep{ye2023robust, moore2009covariate, jiang2017covariate, guo2023generalized, cohen2024no}, though variable selection has not been a primary focus in these studies. Notably, \cite{van2024automated} introduced a unified framework for variable selection combined with the AIPW estimator. Their approach also leverages double/debiased machine learning algorithms \citep{chernozhukov2018double}, employing techniques such as sample splitting and cross-fitting for model fitting and outcome prediction during covariate adjustment. This integration of machine learning with semiparametric methods represents a promising direction for addressing high-dimensional covariates while maintaining the robustness and efficiency of AIPW-based estimators.

\subsection{Our contributions}\label{subsec:contribution}

The contributions of this paper center on advancing covariate adjustment methods for RCTs through the development of a unified framework, COADVISE. Accompanying this framework is the \texttt{Coadvise} R package (\url{https://github.com/yiliu1998/Coadvise}), designed to make these methods accessible and user-friendly for researchers. Our work builds on existing methodologies by addressing several limitations and enhancing flexibility in the analysis of RCT data.

First, under a complete randomization scheme, we extend traditional covariate adjustment methods, such as ANCOVA, ANHECOVA, and AIPW, by incorporating a variety of variable selection techniques, including Lasso, adaptive Lasso, marginal correlation, and preliminary testing into the analysis stage of RCTs. This integration addresses the challenges of high-dimensional covariates, enabling efficient and robust estimation of treatment effects. Furthermore, due to concerns regarding the validity of post-selection inference arising from using the same dataset for both variable selection and covariate adjustment \citep{van2024automated, belloni2017program}, in Section \ref{subsec:postsel}, we outline regularity conditions that variable selection procedures must satisfy to ensure valid post-selection inference, and we justify that the AIPW estimator remains valid when using Lasso and adaptive Lasso for variable selection.

Second, as detailed in Section \ref{subsec:extend}, we introduce two key extensions to the COADVISE framework beyond complete randomization. The first extension incorporates externally trained super-covariates---or outcome foundation models---into the variable selection process. The second extension adapts the framework to covariate-adaptive randomization, allowing treatment allocation to be stratified based on a factor-type variable to achieve improved balance across prognostic strata.

All of the above contributions are implemented in our \texttt{Coadvise} R package. Unlike existing tools such as \texttt{RobinCar}, \texttt{Coadvise} explicitly supports variable selection, enabling users to identify covariates most predictive of outcomes and improve estimation efficiency. To clarify the distinctions between the two packages, we provide a detailed overview of the COADVISE framework and a comparison with \texttt{RobinCar} within the context of covariate adjustment for RCTs (see Table~\ref{tab:summary}). For user convenience, the package includes several missing data imputation methods for handling missing covariates and outcomes, which are common in RCTs; details are provided in Appendix \ref{subapp:miss}. Our package also promotes interoperability by allowing users to export selected covariates for use in \texttt{RobinCar} or other tools and custom analyses. This flexibility ensures that \texttt{Coadvise} complements rather than replaces existing tools. Based on empirical evidence in Section \ref{sec:simu}, we recommend using Lasso or adaptive Lasso for variable selection and set Lasso as the default in our package. 


Finally, we apply the COADVISE framework to a real-world dataset from the BestAIR trial, using data from the National Sleep Research Resource (NSRR). This case study illustrates the practical benefits of our framework and highlights its relevance for both academic research and regulatory applications. The open-source \texttt{Coadvise} package includes detailed documentation and reproducible examples, ensuring accessibility for a broad audience.

\begin{table}
    \centering
    \small
    
    \begin{tabular}{rcccccccccccccccc}
    \toprule
    \textbf{Variable selection} & \multicolumn{4}{l}{\makecell[l]{Lasso \citep{tibshirani1996regression}, Adaptive Lasso \citep{zou2006adaptive}, \\ Marginal correlation \citep{permutt1990testing}, \\ Preliminary testing \citep{senn1994testing, zhao2024randomization}}} \\
    \midrule
    \textbf{\makecell[r]{Covariate-adjusted \\ estimator}} & \makecell[c]{Model-based \\ variance estimator \\ (see Section \ref{sec:setup})} & \makecell[c]{Robust to model \\ misspecification \\ (see Section \ref{sec:setup}) } & \makecell[c]{Nonlinear \\ adjustment \\ (see Section \ref{sec:coadvise})} & \makecell[c]{Efficiency gain \\ guarantee \\ (see Section \ref{sec:coadvise})} \\
    \midrule
    \makecell[r]{ANCOVA \\ \citep{fisher1966design}} & Sandwich & $\checkmark$ & $\times$ & 
    \makecell[c]{ $\checkmark$ if $\pi_1=\pi_0^\dagger$ or \\ under a constant \\ treatment effect}
     \\
    \addlinespace
   \makecell[r]{ANHECOVA \\ \citep{lin2013agnostic, ye2023toward}} & Sandwich & $\checkmark$ & $\times$ & $\checkmark$ \\
    \addlinespace
    \makecell[r]{AIPW \\ \citep{bannick2025general}} & Delta method & $\checkmark$ & $\checkmark$ & \makecell[c]{ $\checkmark$ if (i) or (ii) in \\ Theorem \ref{thm:effgaintheory} holds} \\
    \midrule
    \textbf{Software} & \makecell[c]{Multi-valued \\ treatments} & \makecell[c]{Missing data \\ imputation}  & \makecell[c]{Variable \\ selection} & \makecell[c]{Covariate-adaptive \\ randomization} \\
     \texttt{Coadvise} (ours) & $\checkmark$ & $\checkmark$ & $\checkmark$ &  $\checkmark$ \\
     \texttt{RobinCar} \citep{ye2023toward} & $\checkmark$ & $\times$ & $\times$ &  $\checkmark$ \\
    \bottomrule
    \end{tabular}
    \begin{tablenotes}
        \item $\dagger$: $\pi_a=P(A=a)$ is the probability of being assigned to the treatment group $a$, where $a=0,1$. 
    \end{tablenotes}
    \caption{ Overview of the COADVISE Framework. The top two sections summarize all available methods for variable selection and covariate-adjusted estimation, including key properties of the supported estimators. The bottom section compares COADVISE with the \texttt{RobinCar} package. Details on missing data imputation methods are provided in Table~\ref{tab:missmethod} in Appendix \ref{subapp:miss}. }
    \label{tab:summary}
\end{table}

\section{Preliminaries}\label{sec:setup}

\subsection{Notation and set-up}\label{subsec:notation}

The notation in this paper follows the ``super-population'' framework \citep{ding2017bridging}, which assumes that each observation is a random sample drawn from an infinite population under certain inclusion/exclusion criteria \citep{harrell2024comment}. For instance, when estimating the ATE of an active drug for patients with a certain disease, it may be reasonable to assume that individuals are drawn from such a super-population, and the findings from the RCT can inform whether the drug is effective to the target population. In contrast, the ``finite-population'' framework \citep{bloniarz2016lasso, greenland1999confounding, schochet2010regression, stuart2011use} is applicable when focusing on the ATE for a policy applied to a fixed number of units, such as the states in the U.S., where the number of units is finite and fixed. A key difference between these two frameworks lies in the source of randomness: in the finite-population framework, randomness arises from treatment assignment and outcomes, whereas in the super-population framework, additional randomness comes from the covariates. 

We consider a binary treatment, denoted by $A \in \{0,1\}$, where $1$ represents the treated group and $0$ represents the control group. Each participant in the sample is characterized by a baseline covariate vector $\mb X$, such as clinical characteristics. We adopt the Neyman-Rubin potential outcomes framework \citep{neyman1923applications, imbens2015causal}, in which each participant has two potential outcomes, $Y(0)$ and $Y(1)$, representing the potential outcomes under control and treatment, respectively.

\subsection{Causal parameter, identification assumptions, and the unadjusted estimator}\label{subsec:assmps}

We are interested in estimating the ATE from RCT data, which is defined as: 
\begin{align*}
    \tau = \Ex\{Y(1)-Y(0)\},
\end{align*}
where the expectation $\Ex\{\cdot\}$ is taken over the target population represented by the trial sample. However, we can only observe one outcome for each participant, denoted by $Y$. 
{Throughout, we assume that (i) the observed outcome is consistent with the potential outcome under the received treatment, i.e., $Y=AY(1)+(1-A)Y(0)$; and (ii) there is only one version of the treatment $A$, and the potential outcome of each participant neither depends on nor affects the treatment received by others (also known as the stable-unit treatment value assumption [SUTVA])}. In addition, we make the following standard identification assumptions for RCT data. 

\begin{assumption}[Randomization]\label{assm:random}
    $A \bigCI (Y(0), Y(1), \mb X)$, where ``$\bigCI$'' denotes independence. 
\end{assumption}
\begin{assumption}[Positivity]\label{assm:overlap}
    There exist a constant $\eta$ such that $0 < \eta \leq \pi_a = P(A=a) \leq 1-\eta < 1$ almost surely, for $a=0,1$. 
\end{assumption}
Assumption \ref{assm:random} is reasonable because, in RCTs, the treatment assignment is controlled by the experimenters. Assumption \ref{assm:overlap} is also commonly satisfied, as the probability of assigning each participant to either the treatment or control group is fixed and non-zero in RCTs, such as $\pi_1 = 0.5$ in the case of a 1:1 randomization.  

Under these assumptions, the ATE $\tau$ from an RCT sample can be identified using the following formula:
\begin{align*}
    \tau = \frac{\Ex(AY)}{\Ex(A)} - \frac{\Ex\{(1-A)Y\}}{\Ex(1-A)}. 
\end{align*}
Thus, given a set of independent and identically distributed (i.i.d.) data, denoted by $\mc O = \{(\mb X_i, A_i, Y_i), i=1,\dots, N\}$, we can estimate $\tau$ using the \textit{simple estimator}: 
\begin{align}\label{eq:simple}
\widehat\tau_{\text{simple}} = \frac 1{N_1}\sum_{i=1}^N A_iY_i - \frac 1{N_0}\sum_{i=1}^N (1-A_i)Y_i, 
\end{align}
where $N_1 = \sum_{i=1}^N A_i$ and $N_0 = \sum_{i=1}^N (1-A_i)$. The variance estimator for $\widehat\tau_{\text{simple}}$, often referred to as the \textit{Neyman-type variance estimator}, is given by:
\begin{align*}
    \widehat\sigma^2_\simp = \frac{\widehat S_0^2}{N_0} +\frac{\widehat S_1^2}{N_1},
\end{align*}
where $\widehat S_a^2 = (N_a - 1)^{-1}\sum_{i=1}^{N_a} I(A_i = a)(Y_i - \widehat{\bar{Y}}(a))^2$, with $\widehat{\bar{Y}}(a) = {N_a}^{-1} \sum_{i=1}^{N_a} I(A_i = a) Y_i$. The Neyman-type variance estimator is unbiased for the true variance of $\widehat\tau_{\text{simple}}$ under the super-population framework. Additionally, the estimator $\widehat\tau_{\text{simple}}$ when divided by the Neyman-type variance estimator, corresponds to the usual two-sample t-test statistic (or the F-statistic in one-way ANOVA). As a result, $\widehat\tau_{\text{simple}}$ is sometimes referred to as the ANOVA estimator in the literature \citep{st1989analysis, ye2023toward, reluga2024unified}.  

The estimator $\widehat\tau_{\text{simple}}$ is consistent for $\tau$ and asymptotically normal, and it is attractive in practice due to its simplicity and ease of interpretation. However, incorporating covariate information can improve both the efficiency and credibility of ATE estimation, as discussed in Section  \ref{subsec:relatework}. Therefore, we introduce several covariate-adjusted estimators in the following sections.

\subsection{Linear covariate adjustment}\label{subsec:linAdj}

Two historically popular methods for covariate adjustment are analysis of covariance (ANCOVA) and analysis of heterogeneous covariance (ANHECOVA) \citep{fisher1966design, cassel1976some}. Both methods adjust for covariates using linear regression models. The ANCOVA model \citep{fisher1966design} is specified as:
\begin{align}\label{eq:ANCOVA}
    \Ex\{Y\mid\mb X, A\} = \gamma_0 + \beta_1 A + \mb X'\bd\gamma_1,
\end{align}
where the ATE estimator from ANCOVA is the ordinary least squares (OLS) estimator for the coefficient $\beta_1$, i.e., $\widehat\tau_{\text{ANCOVA}} = \widehat\beta_1^{\text{ols}}$.

A refinement of the ANCOVA model, known as the ANHECOVA model, includes interaction terms between the covariates and the treatment, and is specified as:
\begin{align}\label{eq:ANHECOVA}
    \Ex\{Y\mid\mb X, A\} = \gamma_0 + \beta_2 A + \mb X'\bd\gamma_1 + A(\mb X-\bar{\mb X})'\bd\gamma_2,
\end{align}
where the ATE estimator from ANHECOVA is the OLS estimator for coefficient $\beta_2$, i.e., $\widehat\tau_{\text{ANHECOVA}} = \widehat\beta_2^{\text{ols}}$.

A robust variance estimator for both the ANCOVA and ANHECOVA estimators is the Huber-White (HW) sandwich variance estimator \citep{lin2013agnostic, zhao2021covariate}. For ANCOVA, the HW variance estimator is written as
\begin{align}\label{eq:HWvar}
\widehat\Vx(\widehat\tau_{\text{ANCOVA}}) = 
(\mb V'\mb V)^{-1}\{\mb V'\text{diag}(\widehat\varepsilon_1^2,\dots,\widehat\varepsilon_N^2)\mb V\}(\mb V'\mb V)^{-1},
\end{align}
where $\mb V = (A, \mb X)$, and $(\widehat\varepsilon_1, \dots, \widehat\varepsilon_N)$ is the vector of estimated regression residuals. For the ANHECOVA estimator, $\mb V$ is modified to include interaction terms as $\mb V = (A, \mb X, A(\mb X - \bar{\mb X}))$. In the finite-population framework, equation \eqref{eq:HWvar} provides robust and consistent variance estimates for both ANCOVA and ANHECOVA estimators. However, under the super-population framework, additional uncertainty arises from estimating $\bar{\mb X}$ as an approximation of $\Ex(\mb X)$. Therefore, the corrected HW variance estimator for the ANHECOVA estimator is given by \citep{zhao2021covariate}: 
\begin{align*}
\widehat\Vx(\widehat\tau_{\text{ANHECOVA}}) = 
(\mb V'\mb V)^{-1}\{\mb V'\text{diag}(\widehat\varepsilon_1^2,\dots,\widehat\varepsilon_N^2)\mb V\}(\mb V'\mb V)^{-1} + (\widehat\alpha_1-\widehat\alpha_0)'{S_X^2}(\widehat\alpha_1-\widehat\alpha_0)/N,
\end{align*}
where $S_X^2$ is the sample covariance of $\mb X$, and $\widehat\alpha_a$ is the OLS estimator from the regression model $\Ex(Y(a) \mid \mb X) = \xi_a + \alpha_a \mb X$, using data from the group with $A=a$, for $a = 0,1$.

It has been shown that the ANHECOVA estimator, $\widehat\tau_{\text{ANHECOVA}}$, is consistent, asymptotically normal, and asymptotically more efficient than both $\widehat\tau_{\text{simple}}$ and $\widehat\tau_{\text{ANCOVA}}$ \citep{lin2013agnostic, ye2023toward}. However, both $\widehat\tau_{\text{ANCOVA}}$ and $\widehat\tau_{\text{ANHECOVA}}$ have a limitation: they cannot accommodate non-linear adjustments. In some cases, non-linear adjustments may lead to greater efficiency gains, particularly when the true outcome model follows a known non-linear form given some prior knowledge. The augmented inverse probability weighted (AIPW) estimator, introduced in the next section, offers a more flexible approach to covariate adjustment by accommodating both linear and non-linear outcome regression models in the two treatment groups.

\subsection{Non-linear covariate adjustment}\label{subsec:AIPW}

The AIPW estimator for RCT data is given by \citep{bannick2025general, glynn2010introduction}: 
\begin{align}\label{eq:AIPW}
    \widehat\tau_{\text{AIPW}} = \widehat\theta_1 -\widehat\theta_0,
\end{align}
where for $a=0,1$,
$$
\widehat\theta_a = \frac{1}{N_a}\sum_{i=1}^NI(A_i=a)[Y_i-\widehat\mu_a(\mb X_i)] + \frac1N\sum_{i=1}^N\widehat\mu_a(\mb X_i), \text{ with }N_a = \sum_{i=1}^N I(A_i=a). 
$$
In \eqref{eq:AIPW}, $N_a$ is the number of subjects in group $A = a$, and $\widehat\mu_a(\mb X)$ is the predicted value of $\Ex\{Y(a) \mid \mb X\}$, obtained by fitting a working model $\mu_a(\cdot)$ under treatment $A = a$ (e.g., a generalized linear model (GLM) $\mu_a(\mb X) = g(\mb X'\bd\beta_a)$). 
This model may not be correctly specified, where correct specification means $\mu_a(\mb X) \to_p \Ex\{Y(a) \mid \mb X\}$, where $\to_p$ denotes convergence in probability. It can be shown that when $\mu_a(\mb X)$ is a linear model for $Y(a)$, the AIPW estimator coincides with the ANHECOVA estimator, provided that the same set of covariates are used as predictors \citep{ye2023toward, van2023use, reluga2024unified}. Thus, ANHECOVA can be viewed as a special case of the AIPW estimator.

\begin{remark}\label{rmk:MLmodel}
    In practice, the working model for $\mu_a(\mb X)$ in the AIPW estimator can be fit using black-box machine learning (ML) models and combined through ensemble methods such as \texttt{SuperLearner} \citep{chernozhukov2018double, van2007super}. However, our focus on parametric models is driven by the theoretical results on efficiency gain presented in Theorem \ref{thm:effgaintheory} (Section \ref{subsec:effgaintheory}). Specifically, we establish efficiency conditions based on the estimation approach for $\bd\beta_a$. While ML-based estimation methods are flexible, deriving unified theoretical results on efficiency gain in this setting remains an open question, which we leave for future research. Further discussion on our decision not to employ ML models in this work is provided in Remark \ref{rmk:MLmodel-2}, where we formally introduce the COADVISE framework and outline additional considerations.
\end{remark}

A robust variance estimator for the AIPW estimator can be derived using the Delta method (Taylor's expansion) \citep{ye2023robust}. Under certain regularity conditions \citep{bannick2025general} and denoting $\bd\theta = (\theta_0, \theta_1)'$, it can be shown that: 
\begin{align}\label{eq:asyVarAIPW}
    \sqrt{N}(\widehat{\bd\theta}-\bd\theta)\to_d\mc N(\bd 0, \mb V),
\end{align}
where $\to_d$ denotes convergence in distribution, and
\begin{align}\label{eq:AIPWvarmatrix}
    \mb V = \begin{pmatrix}
    v_{00} & v_{01} \\ v_{10} & v_{11}
\end{pmatrix},
\end{align}
with
\begin{align*}
    v_{aa} & = \pi_a^{-1}\Vx\{Y_a-\mu_a(\mb X)\} + 2\text{Cov}\{Y_a, \mu_a(\mb X)\} - \Vx\{\mu_a(\mb X)\}, \text{ where }\pi_a=P(A=a), \\
    v_{ab} & = \text{Cov}\{Y_a, \mu_b(\mb X)\} + \text{Cov}\{Y_b, \mu_a(\mb X)\} - \text{Cov}\{\mu_a(\mb X), \mu_b(\mb X)\}\text{ for }a\not=b. 
\end{align*}
The function $\mu_a$ is the assumed limit of $\widehat\mu_a$ used in \eqref{eq:AIPW}, and it satisfies $\Vert\widehat\mu_a - \mu_a\Vert_{2} \to_p 0$, where $\Vert\cdot\Vert_2$ represents a finite $L_2$ norm. However, $\mu_a$ is not necessarily equal to the true outcome model $\Ex\{Y(a) \mid \mb X\}$, unless the working model is correctly specified. 

The asymptotic variance for $\widehat\tau_{\text{AIPW}} = \widehat\theta_1 - \widehat\theta_0 = c' \widehat{\bd\theta} := f(\widehat{\bd\theta})$ is given by $\{\nabla f(\bd\theta)\}' \mb V\{\nabla f(\bd\theta)\} = v_{00} - 2v_{01} + v_{11}$, where $c = (1\ \ -1)'$. Therefore, a consistent variance estimator is: 
\begin{align*}
    N^{-1}\{\nabla f(\widehat{\bd\theta})\}'\widehat{\mb V}\{\nabla f(\widehat{\bd\theta})\} = N^{-1}\{\widehat v_{00}-2\widehat v_{01}+\widehat v_{11}\},
\end{align*} 
where $\widehat v_{00}$, $\widehat v_{01}$, and $\widehat v_{11}$ are sample-based estimates of $v_{00}$, $v_{01}$, and $v_{11}$, respectively, using the sample versions of the variances, covariances, and proportions.

\section{Methodology}\label{sec:coadvise}

All of the covariate-adjusted estimators, as discussed in Section \ref{sec:setup}, can suffer from high-dimensional issues. When the number of available covariates is large relative to the sample size, or when overfitting occurs \citep{bloniarz2016lasso}, the finite-sample performance of these estimators may deteriorate, potentially leading to efficiency losses compared to simpler estimators. In addition, the consistency and robustness of variance estimators can be compromised. For example, computing matrix inverses in sandwich variance estimators may become unstable or infeasible, resulting in unavailable or unreliable standard error estimates \citep{matsouaka2023variance, matsouaka2024overlap}. 

To address these concerns, we incorporate a variable selection step before performing covariate-adjusted ATE estimation. The goal is to retain only the most important variables—identified through data-driven criteria—while ensuring that the number of selected covariates remains small relative to the sample size (i.e., of lower order than $N$). This approach balances the inclusion of relevant covariates with the exclusion of less informative ones, improving finite-sample efficiency. It yields an estimator that lies between the simple estimator (using no covariates) and the fully adjusted estimator (using all covariates), achieving higher efficiency while maintaining consistency.

We also note that our COADVISE framework generates predicted mean potential outcome vectors for all participants in multi-valued treatment settings, allowing users to compute any pairwise contrasts between mean outcomes. However, for consistency with the notation in Section \ref{sec:setup}, the following Sections \ref{subsec:workflow} and \ref{subsec:effgaintheory} focus on the AIPW estimator under binary treatment to illustrate our workflow. This focus is motivated by the AIPW estimator's superior theoretical properties, intuitive interpretation, and straightforward generalization to multi-valued treatments. Additional practical guidelines for users are provided in Appendix \ref{app:addtuto}.

For simplicity, we assume here that the observed data contain no missing values, which allows us to concentrate on high-dimensional issues and variable selection. In practice, if missing data are present, we first impute them within the COADVISE framework to obtain a complete dataset. By separating the imputation and variable selection steps, the subsequent theoretical results remain valid without loss of generality.

\subsection{The COADVISE workflow}\label{subsec:workflow}

\begin{figure}
    \centering
    \includegraphics[width=0.7\textwidth]{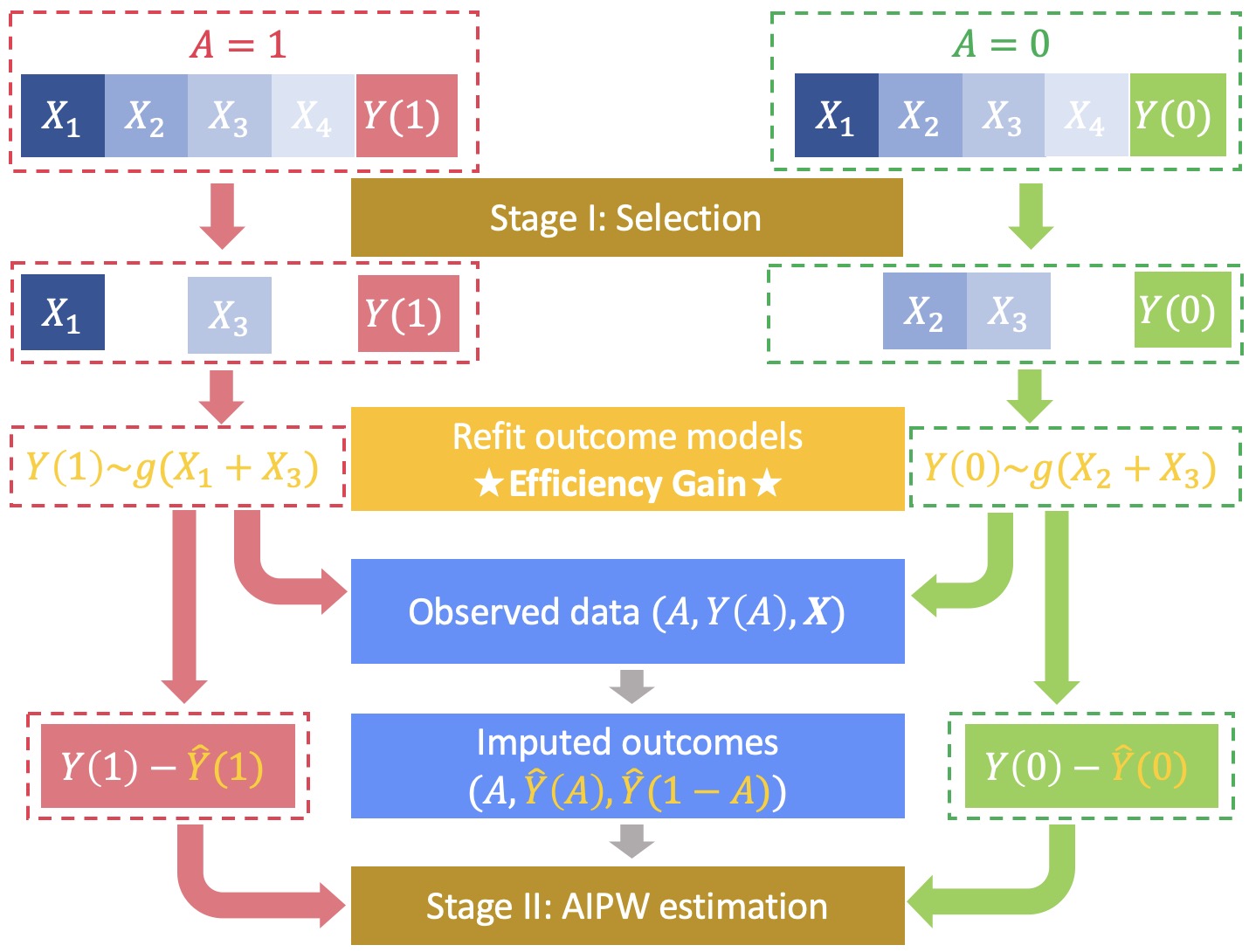}\caption{An illustration of the COADVISE framework using AIPW estimation for the covariate adjustment estimator consisting of two stages: \textbf{Stage I:} Perform variable selection for the conditional outcome models in both treatment groups. \textbf{Stage II:} Re-fit the conditional outcome models using the selected variables and compute the AIPW estimator. The model re-fitting step aligns with the theoretical results on efficiency gains, ensuring that the estimator benefits from covariate adjustment with variable selection. }\label{fig:flow}
\end{figure}

Figure \ref{fig:flow} illustrates our proposed workflow using the AIPW estimator as an example. The process begins with variable selection performed separately for each treatment group. This approach allows the outcome models $\mu_0(\mb X)$ and $\mu_1(\mb X)$ to include different sets of variables, thereby providing greater flexibility in capturing the most relevant features for each potential outcome based on the data. 

Next, we fit two parametric outcome regression models---one for $Y(1)$ and one for $Y(0)$---using a chosen link function $g(\cdot)$ from the GLM family (for example, logistic regression for a binary outcome). 

Once these models are estimated, each model is applied to the data from the opposite treatment group to predict the missing potential outcomes. In other words, a model $\widehat\mu_a(\mb X)$ fitted using data from group $A=a$ is used to predict the potential outcome $Y(a)$ for subjects in group $A=1-a$, for $a=0,1$. This procedure yields all the information necessary to compute the AIPW estimator in \eqref{eq:AIPW}. Finally, we quantify uncertainty by employing the robust variance estimator in \eqref{eq:asyVarAIPW} \citep{ye2023robust}. 

For a more technical illustration, Appendix \ref{subapp:lassoexample} provides an algorithm box detailing the Lasso variable selection combined with AIPW estimation.

\begin{remark}\label{rmk:MLmodel-2}
    Although variable selection methods like Lasso and adaptive Lasso can directly produce a fitted model by selecting variables with nonzero regression coefficients, we take an additional step by refitting a GLM using only the selected variables and estimating the coefficients via maximum likelihood estimation (MLE). This refitting step is essential for two reasons. First, it facilitates achieving the efficiency gains established in Theorem \ref{thm:effgaintheory} in the following section. Second, it enhances the interpretability of the covariate-adjusted estimator by relying on transparent parametric models rather than black-box approaches.
\end{remark}

\subsection{Theoretical results}\label{subsec:effgaintheory}

We establish theoretical results for efficiency gains via the AIPW estimator, with detailed proofs provided in Web Appendix A of \textit{Online Supplemental Material}. The key results are summarized in the following Theorem \ref{thm:effgaintheory}. 

\begin{theorem}\label{thm:effgaintheory}
    Suppose Assumptions \ref{assm:random}--\ref{assm:overlap} in Section \ref{subsec:assmps} hold, and denote the asymptotic variance of the AIPW and simple estimator as $\Vx_\aipw$ and $\Vx_\simp$, respectively. Then, regardless of the variable selection method used, $\Vx_\aipw\leq\Vx_\simp$ in any of the following cases: 
    \begin{enumerate}[label=(\roman*)]
        \item Both conditional outcome models $\mu_0(\mb X)$ and $\mu_1(\mb X)$, after variable selection, are specified by linear models with OLS used for estimating the regression coefficients. 
        \item If the outcome model for $Y(a)$ ($a=0,1$), after variable selection, is specified by a nonlinear GLM with link function $g$, then there exists a $\bd\beta_a$ as the probability limit of MLE $\widehat{\bd\beta}_a$ such that with probability 1: 
        \begin{enumerate}[label=(\alph*)]
            \item The sign of $g(\mb X'\bd\beta_a)$  matches that of $\Ex(Y(a)\mid \mb X)$; and 
            \item $\vert g(\mb X'\bd\beta_a)\vert \leq 2\vert \Ex(Y(a)\mid \mb X)\vert $. 
        \end{enumerate} 
    \end{enumerate}
\end{theorem}

For point (i) in Theorem \ref{thm:effgaintheory}, we highlight the connection with the ANHECOVA estimator. If both $\mu_0(\mb X)$ and  $\mu_1(\mb X)$ use the same set of covariates as ANHECOVA, then the AIPW estimator is identical to ANHECOVA, so in this case, the efficiency gain result is consistent with that established by \cite{lin2013agnostic} However, due to variable selection, $\mu_0(\mb X)$ and $\mu_1(\mb X)$ may use different covariates, making the AIPW estimator distinct but still guaranteeing efficiency gains when linear regressions are applied.

For point (ii), conditions (a) and (b) have the following interpretations: for (a), the GLM must correctly capture the sign of the true conditional outcome with probability 1; and for (b), the bias of the GLM is constrained to maintain efficiency. For example, with positive-valued outcomes, conditions (a) and (b) can be simplified to $0\leq g(\mb X'\bd\beta_a)\leq 2\Ex(Y(a)\mid \mb X)$, which is reasonable for some positive categorical outcomes, such as binary outcomes modeled by logistic regression. Further illustration and discussion of these conditions are provided in Web Appendix A of  \textit{Online Supplemental Material}. 

Comparing (i) and (ii), linear models with OLS ensure efficiency gains without additional bias constraints. However, nonlinear GLMs may offer higher efficiency gains in specific cases (e.g., binary outcomes), since the AIPW estimator's efficiency gain depends on the bias between the posited and true outcome models. When both outcome models are correctly specified, the AIPW estimator is the most efficient, achieving the asymptotic (semiparametric) variance lower bound \citep{hahn1998role, hirano2003efficient}. 

\begin{remark}[Linear calibration for nonlinear outcome models]\label{rmk:lincal}
When outcome models are specified via nonlinear GLMs, Condition (ii) for efficiency gains in Theorem~\ref{thm:effgaintheory} may fail, because it requires the model-induced bias to be sufficiently small---a condition that is typically untestable in practice. While Condition (i) avoids this bias requirement by using linear regressions with OLS coefficient estimates, such an approach is not always optimal; for example, nonlinear GLMs are often more natural for binary or categorical outcomes. To address this limitation, we adopt a linear calibration step motivated by recent literature \citep{bannick2025general, cohen2024no}. 

Specifically, after obtaining the fitted outcome predictions  
$\widehat\mu_1(\mb X)$ and $\widehat\mu_0(\mb X)$
from nonlinear GLMs, we perform an additional calibration regression within each treatment arm. For treated participants $(A=1)$, we fit  
$$
\texttt{lm}(Y \sim \widehat\mu_1(\mb X) + \widehat\mu_0(\mb X)),
$$ 
using the OLS estimates for regression coefficients, 
and we do the same for control participants $(A=0)$. Then, let $\widehat\mu_0^{\text{update}}(\mb X)$ and $\widehat\mu_1^{\text{update}}(\mb X)$ denote the predicted values for potential outcomes $Y(0)$ and $Y(1)$, respectively, from these regressions. We then replace the original $\widehat\mu_a(\mb X)$ by $\widehat\mu_a^{\text{update}}(\mb X)$ in the AIPW estimator $\widehat\tau_{\text{AIPW}} = \widehat\theta_1-\widehat\theta_0$, where for $a=0,1$, we update
\begin{align*}
    \widehat\theta_a = \frac{1}{N_a}\sum_{i=1}^NI(A_i=a)[Y_i-\widehat\mu_a^{\text{update}}(\mb X_i)] + \frac1N\sum_{i=1}^N\widehat\mu_a^{\text{update}}(\mb X_i), \text{ with }N_a = \sum_{i=1}^N I(A_i=a). 
\end{align*}
This correction aligns the predicted outcomes with the observed data within each treatment arm while retaining the flexibility of nonlinear GLMs, thereby ensuring that the AIPW estimator achieves efficiency gains without relying on the bias requirement in Theorem~\ref{thm:effgaintheory}. 

We allow users to decide whether to apply linear calibration in practice by including the option \texttt{lin.cal = TRUE} or \texttt{FALSE} in the \texttt{Coadvise()} function of our \texttt{Coadvise} R package. When users are uncertain about whether their postulated nonlinear GLM models may introduce substantial bias, enabling this additional step is recommended. 

\end{remark}

\subsection{Post-selection inference}\label{subsec:postsel}

Although Theorem \ref{thm:effgaintheory} guarantees efficiency gain under established mild conditions on outcome models, regardless of the choice of variable selection method, the uncertainty introduced by variable selection must be accounted for when conducting inference with the final AIPW estimator. This challenge is commonly referred to as the \textit{post-selection inference problem.} In this section, we outline the theoretical conditions under which a variable selection method enables valid inference using the variance estimator applied to the AIPW estimator constructed from the selected covariates. Design-based theoretical results related to Lasso-based variable selection, in conjunction with the ANHECOVA estimator, have been studied in \citet{bloniarz2016lasso}.

Suppose we have $p > 1$ covariates, denoted by $\mb X = (X_1, \dots, X_p)'$. Let $\mc S_0 \subset \{1, \dots, p\}$ and $\mc S_1 \subset \{1, \dots, p\}$ denote the sets of indices corresponding to the relevant covariates for the potential outcomes $Y(0)$ and $Y(1)$, respectively. Define $s_a = \vert\mc S_a\vert$ for $a = 0,1$. We draw upon the theoretical results of \citet{farrell2015robust} and \citet{van2024automated} to summarize the following technical conditions within our framework. The former reference considers the more general setting of observational studies, whereas our RCT framework represents a special case that avoids conditions related to the propensity score model. We assume that the estimated outcome regression coefficient $\widehat{\bd\beta}_a$ converges in probability to a limiting value $\bd\beta_a^*$, which may differ from the true coefficient vector $\bd\beta_a$ in the outcome model $\mu_a(\mb X) = \mu_a(\mb X;\bd\beta_a)$, as model misspecification is allowed in this setting.

\begin{condition}\label{cond:spar}
    $s_a\cdot\log\{\max(p, N)\}/\sqrt{N}= o(1)$ for $a=0,1.$
\end{condition}

\begin{condition}[Convergence and error rates of estimated coefficients]\label{cond:conver}
    $$G_N^a(\mb X) = \frac1N\sum_{i=1}^N\{\mu_a(\mb X_i;\widehat{\bd\beta}_a) - \mu_a(\mb X_i;\bd\beta_a^*)\}^2
    $$ is uniformly integrable with $\sqrt{G_N^a(\mb X)} = o_{P}(1)$, and $\Vert\widehat{\bd\beta}_a-\bd\beta_a^*\Vert_1 = O_{P}\left(s_a\sqrt{\log\{\max(p, N)\}/N}\right)$ for $a=0,1$. Here, $P$ is a data generating process that obeys the following regularity and boundedness conditions (below, let $U_a = Y(a)-\mu_a(\mb X)$ for $a=0,1$): 
    \begin{itemize}
        \item $\{(A_i,\mb X_i, Y_i)\}_{i=1}^N$ is an i.i.d. sample from population $\mc O = (A,\mb X, Y)$; 
        \item The covariates $\mb X$ have bounded support with $\max_{j\in\{1,\dots,p\}}\vert X_j\vert\leq M<\infty$;
        \item $\Ex\{\vert U_a\vert^4\mid\mb X\}\leq M<\infty$, $\min_{(j,a)\in\{1,\dots,p\}\times\{0,1\}}\Ex\{X_j^2U_a^2\}$ is bounded away from zero, and for some $r>0$, $\min\{\Ex\{\vert\mu_1(\mb X)\mu_0(\mb X)\vert^{1+r}\}, \Ex\{\vert U_0\vert^{4+r}\}, \Ex\{\vert U_1\vert^{4+r}\}\}\leq M<\infty$, for $a=0,1$.  
    \end{itemize}
\end{condition}


Condition \ref{cond:spar} concerns the relationship among the sparsity $s_a$, the number of covariates $p$, and the sample size $N$. This condition is common in the high-dimensional inference literature \citep{belloni2013least, belloni2014inference, belloni2017program, bloniarz2016lasso}. Condition \ref{cond:conver} requires that the fitted outcome models converge to a bounded limit, along with some regular and mild assumptions on the data-generating process. Both conditions may be strong in more general observational studies, but in our setting, where variable selection is employed and model misspecification is permitted, the dimensionality $p$ can be effectively limited and the existence of the limiting value $\bd\beta_a^*$ is reasonable. 

Outcome models by Lasso and adaptive Lasso variable selections with linear and logistic link functions have been shown to satisfy these conditions \citep{belloni2013least, belloni2017program}, supporting the validity of post-selection inference using the variance estimator adopted in our framework when either of these two variable selection methods is employed. 
Lasso also has a standardized implementation in R via the \texttt{glmnet} package, which can be fully pre-specified to prevent data dredging. For user convenience, we have incorporated this function for both Lasso and adaptive Lasso in our \texttt{Coadvise} R package. These methods also address the overfitting issues associated with OLS when the number of covariates is large relative to the sample size. 

Finally, we remark on the robustness of variance estimators for covariate-adjusted estimators. Thanks to the RCT design, the propensity score (treatment model) is known and correctly specified. In our framework, variable selection is applied only to the outcome models used in the covariate-adjusted estimators. For all estimators considered---ANCOVA, ANHECOVA, and AIPW with Lasso-based variable selection---Condition \ref{cond:spar} ensures that the residuals in their HW sandwich or model-based variance estimators capture the unexplained variation, including uncertainty due to variable selection \citep{liu2023lasso, bloniarz2016lasso, zhu2025design}. As a result, the robust variance estimators constructed from these residuals implicitly account for this uncertainty. 

\subsection{Extensions}\label{subsec:extend}

Thus far, we have discussed our framework under the setting of complete (simple) randomization. In this section, we outline two possible extensions of the COADVISE framework that address emerging challenges in RCT analysis and showcase how our framework may be combined with some advanced techniques. While these extensions are included to illustrate practical implementation possibilities, they are not central contributions of this paper and require more rigorous theoretical investigation in future work. We present them here to help motivate future research directions. 

\subsubsection{Pre-specified super-covariate}

The first extension considers the recently emerging methodology of leveraging super-covariates to enhance RCT analyses. A super-covariate is typically defined or termed as a transformation, summary score, model output, or projection of the existing baseline covariates (potentially high-dimensional) that captures key prognostic information about the outcome, often learned from external data (can be randomized or observational). As noted in several studies \citep{gagnon2023precise, de2025efficient}, the limited sample sizes in RCTs, together with the increasing availability of large external observational datasets containing the same covariates and outcomes, have motivated growing interest in constructing super-covariates from external data to improve the efficiency of ATE estimation. 

Prior research has investigated the super-covariate adjustment approaches. \cite{gagnon2023precise} proposed applying machine learning algorithms on multiple external data to obtain foundation models. \cite{de2025efficient} introduced a hybrid AIPW estimator that integrates available foundation models to improve the efficiency of the AIPW estimator. \cite{schuler2022increasing, hojbjerre2025powering, liao2025prognostic} and \cite{sun2024pretrained} considered leveraging historical data to learn prognostic regression models. \cite{poulet2025prediction} extended the idea of prediction powered inference to RCTs, providing externally trained prognostic scores as a function of baseline covariates for all RCT participants. 

However, a key concern is that the external data may differ from the RCT data due to distributional shifts, such as covariate shift or outcome shift \citep{liu2025targeted, lee2023improving, lee2022doubly}. As a result, it is generally uncertain whether the super-covariates constructed from external data are truly predictive of the outcomes in the RCT population. To address this issue, we propose a general, data-driven strategy that incorporates both the super-covariates trained from external data and the original covariates in the RCT data for covariate adjustment. 

Let $\mc O = (\mb X, A, Y)$ denote a copy of the RCT data, and in general, suppose we have $K\geq 1$ other mutually independent external data sources for super-covariates. The combination of the COADVISE framework and super-covariates is illustrated in the following Figure \ref{fig:supercov}. We first obtain or train $K$ super-covariates (or models) from each of the $K$ external data sources, denoted by $f_1(\mb X), \dots, f_K(\mb X)$. These super-covariates are then applied to the RCT participants to generate corresponding representations, e.g., by predictions. The resulting features are combined with the original covariates $\mb X$ to form an augmented dataset $\mc O^{\text{aug}}$, which is subsequently analyzed using the COADVISE framework. Both the original covariates and the super-covariates are treated as candidate variables for selection in the adjustment process. 

\begin{figure}[H]
    \centering
    \includegraphics[width=0.9\linewidth]{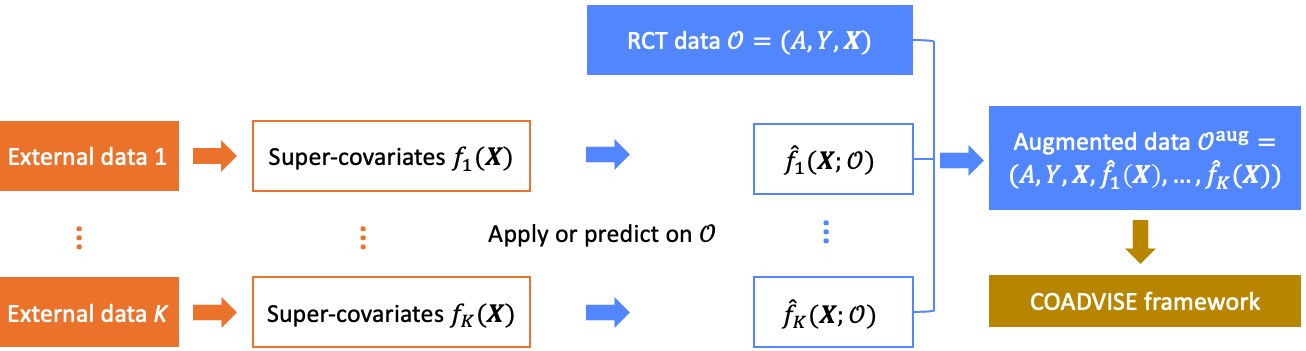}
    \caption{Illustration of incorporating super-covariates into the COADVISE framework in Figure \ref{fig:flow}.  }
    \label{fig:supercov}
\end{figure}

In practice, we recommend that users first obtain the $K$ super-covariates and apply them to the RCT data. They can then use our \texttt{Coadvise} R package to perform the analysis based on the augmented RCT dataset, as illustrated in Figure \ref{fig:supercov}. Our proposal offers several flexible features. First, we allow each external dataset to construct super-covariates of arbitrary dimensions, which are then aggregated and incorporated into the augmented dataset as additional covariate information. Second, since COADVISE employs a data-driven variable selection strategy---such as Lasso---even if some of the $K$ super-covariates are not predictive of the outcome in the RCT data, they may be excluded prior to the final adjustment step. This enables COADVISE to automatically guard against the inclusion of irrelevant covariates or super-covariates and to maintain an appropriate balance between sample size and model complexity.

\subsubsection{Covariate-adaptive randomization}

The preceding discussion in this paper focused on complete (simple) randomization. However, more complex randomization schemes are commonly employed in practice to ensure balance in treatment allocation across key subgroups. For instance, stratified permuted block randomization often assigns participants within predefined blocks (e.g., based on age or clinical site) to treatment arms in a way that maintains a predetermined allocation ratio, thereby achieving better balance within each block \citep{zelen1974randomization}.

Our second extension centers around the covariate-adaptive randomization \citep{bugni2018inference, ye2022inference, rafi2023efficient, bannick2025general} including stratified permuted block randomization \citep{zelen1974randomization} and Pocock and Simon's minimization \citep{pocock1975seq, taves1974min} as special cases. In general, we let $\mb Z$ be a discrete baseline covariate vector used in covariate-adaptive randomization, where $\mb Z$ has $L<\infty$ distinct joint levels $\mb z_1,\dots, \mb z_L$. Also, $\mb Z$ and $\mb X$ can share some common components. For example, the gender can be a factor for stratified randomization at the design stage, but it can also be incorporated into the covariate adjustment in the analysis stage. Following \cite{bannick2025general}, we impose the following mild assumption on treatment assignment in covariate-adaptive RCTs. We use $\pi_a(\mb z)=P(A=a\mid\mb Z=\mb z)$ to denote the treatment assignment probability for arm $a$ within stratum $\mb z$. 

\begin{assumption}\label{assp:covaradp}
(i) $A \bigCI (Y(0), Y(1), \mb X) \mid \mb Z$; (ii) For every level $\mb z$ of $\mb Z$, $\pi_a(\mb z) = \pi_a = P(A = a)$; and (iii) The sequence $\sqrt{N}\{N_a(\mb z)/N(\mb z) - \pi_a\}$ is bounded in probability as $N \to \infty$, where $N(\mb z)$ denotes the number of participants with $\mb Z = \mb z$, and $N_a(\mb z)$ the number assigned to treatment $A = a$ within stratum $\mb Z = \mb z$, for $a = 0, 1$.
\end{assumption}

Assumption \ref{assp:covaradp}(ii) ensures that each treatment is assigned with the same marginal probability across all strata defined by $\mb Z$, which is a primary goal of covariate-adaptive randomization. The estimand of interest remains the ATE, $\tau = \Ex\{Y(1) - Y(0)\}$. As shown in \cite{bannick2025general}, the AIPW estimator \eqref{eq:AIPW} remains consistent for $\tau$ even when the outcome model is misspecified. To incorporate covariate-adaptive randomization into the COADVISE framework, we propose using the same AIPW estimation procedure outlined in Figure \ref{fig:flow} following variable selection. The only modification lies in the final step, where a generalized variance expression derived by \cite{bannick2025general} is applied. 

We first review key aspects of the asymptotic variance expression. Recalling the notation in Equation~\eqref{eq:asyVarAIPW} from Section~\ref{subsec:AIPW}, we now present its modification under covariate-adaptive randomization, following \citet{bannick2025general}, with stratification by $\mb Z$:
\begin{align}\label{eq:asyVarAIPW-strata}
\sqrt{N}(\widehat{\bd\theta}-\bd\theta) \to_{d} \mathcal{N}(\mathbf{0}, \mb V^*),
\end{align}
where
\begin{align}\label{eq:AIPWvarmatrix-strata}
\mb V^* = \mb V - \Ex[\mb Q(\mb Z) \left\{\mb\Omega_{\text{SR}} - \mb\Omega(\mb Z) \right\}\mb Q(\mb Z)],
\end{align}
with
\begin{align*}
\mb Q(\mb z) = \text{diag}\left\{\pi_a^{-1}\Ex\{Y(a) -\theta_a\mid\mb Z=\mb z\}-\pi_a^{-1}\Ex[\mu_a(\mb X) - \Ex\{\mu_a(\mb X)\} \mid \mb Z=\mb z ]\right\}, \quad \text{for } a = 0, 1.
\end{align*}
Furthermore, $\mb\Omega(\mb z) = \text{diag}\left\{ \pi_0(\mb z), \pi_1(\mb z) \right\} - \left( \pi_0(\mb z) ~~ \pi_1(\mb z) \right)' \left( \pi_0(\mb z) ~~ \pi_1(\mb z) \right)$ is a $2 \times 2$ covariance matrix determined by the covariate-adaptive randomization design. In contrast, $\mb\Omega_{\text{SR}} = \text{diag}\{\pi_0, \pi_1\} - (\pi_0 ~~ \pi_1)'(\pi_0 ~~ \pi_1)$ denotes the corresponding matrix under simple randomization. 

Under Assumption~\ref{assp:covaradp}, the covariate-adaptive randomization mechanism assigns equal treatment probabilities within each stratum, implying $\pi_a(\mb z) = \pi_a$ for all $\mb z$ and $a$, and thus $\mb\Omega(\mb Z) = \mb\Omega_{\text{SR}}$. Consequently, $\Ex\left\{ \mb Q(\mb Z) (\mb\Omega_{\text{SR}} - \mb\Omega(\mb Z)) \mb Q(\mb Z) \right\} = \mb 0$ and the asymptotic variance still simplifies to $\mb V^* = \mb V$. Nevertheless, to accommodate potential deviations in finite samples, we estimate and retain this correction term in our variance estimator. Specifically, we estimate the treatment probabilities by $\widehat\pi_a(\mb z) = N_a(\mb z) / N(\mb z)$ and $\widehat\pi_a = N_a / N$ for $a = 0, 1$. Consistent with the estimator of $\mb V$ in Equation~\eqref{eq:AIPWvarmatrix}, we use sample-based estimates for all components, including proportions, variances, and covariances, to construct the finite-sample version of the variance estimator. Finally, by the Delta method, the variance estimator for the AIPW estimate of the ATE can be obtained accordingly, following the same derivation presented at the end of Section \ref{subsec:AIPW}.

Additionally, we include two basic estimators without covariate adjustment: the simple unadjusted estimator \eqref{eq:simple}, which remains consistent under Assumption~\ref{assp:covaradp}, and a stratified unadjusted estimator following \citet{bugni2018inference}, defined as
\begin{align}\label{eq:strata}
\widehat\tau_{\text{strata}} = \sum_{\mb z \in \mc Z} \frac{N(\mb z)}{N} \left\{ \bar{Y}(1; \mb z) - \bar{Y}(0; \mb z) \right\},
\end{align}
where $\mb z$ indexes the strata defined by $\mb Z$, $\mc Z$ denotes the support of $\mb Z$, $N(\mb z)$ is the number of participants in stratum $\mb z$, and $\bar{Y}(a; \mb z)$ is the average outcome in treatment arm $a$ within stratum $\mb z$. A natural plug-in variance estimator for $\widehat\tau_{\text{strata}}$ is given by:
\begin{align}\label{eq:strata-var}
\widehat{\Vx}\left(\widehat\tau_{\text{strata}}\right) = \sum_{\mb z \in \mc Z} \left( \frac{N(\mb z)}{N} \right)^2 \left\{ \frac{\widehat S_1^2(\mb z)}{N_1(\mb z)} + \frac{\widehat S_0^2(\mb z)}{N_0(\mb z)} \right\},
\end{align}
where $N_a(\mb z)$ is the number of participants receiving treatment $a$ in stratum $\mb z$, and $\widehat S_a^2(\mb z)$ is the sample variance of outcomes in treatment arm $a$ within stratum $\mb z$, for $a = 0,1$. We provide an R function, \texttt{CoadviseCAR()}, in our \texttt{Coadvise} R package to implement the proposed framework for covariate-adaptive randomized trials, where ``CAR'' stands for covariate-adaptive randomization. A toy example implementing the above three estimators using Lasso variable selection (for AIPW) is given in Appendix \ref{subapp:CAR}. 

\section{Numerical Experiments}\label{sec:simu}

In this section, we present a series of Monte Carlo simulation experiments to evaluate different configurations within our COADVISE framework. We use a comprehensive data-generating process (DGP), detailed in Section \ref{subsec:DGP}, and generate two distinct types of datasets for each DGP.

First, we generate 20 independent super-population full datasets, each containing $5 \times 10^7$ independent units, with both potential outcomes $(Y(0), Y(1))$ available for each unit. We then compute the mean of $Y(1) - Y(0)$ across these 20 large datasets to obtain the true ATE. Given the large sample size and 20 replications, the associated uncertainty is negligible.

Second, we generate $M = 500$ independent random sample datasets of size $N$, with $N$ varying {in \{40, 100, 200, 500\}} to represent different sample size scenarios. {The choice of 200 is is intended to mimic the sample size in the case study of Section \ref{sec:data}, where $N = 169$. } In each Monte Carlo replication, the dataset consists of $N_1$ treated units and $N_0$ control units, with an expected 1:1 ratio such that $N_0 + N_1 = N$. 

\subsection{Data generating process}\label{subsec:DGP}

In line with the nature of an RCT, we generate the binary treatment assignment as $A \sim \text{Bern}(0.5)$, independent of any covariates or outcomes, without imposing any restrictions on the sample size. We consider two sets of covariates: $\mb X = (X_1, \dots, X_5)'$ and $\mb V = (V_1, \dots, V_{50})'$. The set $\mb X$ is associated with the outcomes, while $\mb V$ consists of noise variables that are uncorrelated with both the outcomes and the treatment.

We then specify two models for the potential outcomes: 
\begin{align*}
    \text{Continuous outcome: }& Y(a) = 30 + 20\mb X'\bd\beta_{0} + a\delta(\mb X) + \epsilon, \\
    \text{Binary outcome: }& Y(a)\sim \text{Bern}(e(\mb X, a)), \text{ where } \\ & e(\mb X, a)=\{1+\exp(-20\mb X'\bd\beta_{0}-a\delta(\mb X))\}^{-1}, 
\end{align*}
for $a = 0, 1$, where $\epsilon \sim \mc N(0, 1)$ i.i.d., and $\delta(\mb X)$ characterizes the individual treatment effect. The details of $\bd\beta_0$, different cases of $\delta(\mb X)$ (linear and nonlinear), and the distributions used for $\mb X$ and $\mb V$ can be found in Web Appendix B of \textit{Online Supplemental Material} for reproducibility.

\subsection{Methods for comparison}\label{subsec:compmethods}

We consider the following competing methods for estimating ATE: Simple, ANCOVA, ANHECOVA, and AIPW estimators. For each of these methods, we apply the variable selection procedures described in Table \ref{tab:methods} to select covariates from $\{\mb X, \mb V\}$ defined in Section \ref{subsec:DGP}. All estimation and variable selection methods are available through our R package, \texttt{Coadvise}.

\begin{table}
\caption{Competing methods for variable selection in simulation}\label{tab:methods} 
    \centering
    \begin{tabular}{rl}
    \toprule 
    Method abbreviation & Description \\ 
    \midrule All & \makecell[l]{Include all $\mb X$ and $\mb V$, i.e., no covariate selection} \\ 
    \addlinespace
    Lasso & \makecell[l]{Select covariates with non-zero regression \\ coefficient estimates in Lasso regression \citep{bloniarz2016lasso}} \\
    \addlinespace
    Adaptive Lasso & \makecell[l]{Select covariates with non-zero regression \\ coefficient estimates in adaptive Lasso regression \citep{zou2006adaptive}} \\
    \addlinespace
    Corr ($k$) & \makecell[l]{Select the $k$ covariates with the highest \\ marginal correlations with $Y$, where $k = 1, 3, 10$ \citep{permutt1990testing}} \\
    \addlinespace
    Corr ($\xi$) & \makecell[l]{Select covariates whose marginal correlations \\ with $Y$ are greater than $\xi$, where $\xi = 0.10, 0.25$ \citep{permutt1990testing}} \\
    \addlinespace
    Pre-test ($\alpha$) & \makecell[l]{Perform a univariate two-sample $t$-test on each covariate to test \\ for differences between two groups at level $\alpha$. Select covariates \\ where the test is rejected, with $\alpha = 0.05, 0.10$ \\ \citep{senn1994testing, zhao2024randomization}} \\
    \bottomrule
    \end{tabular}
\end{table}

\subsection{Performance criteria}\label{subsec:criteria}

The simulation results are presented using the following criteria: bias (via boxplots), coverage probability (CP\%), and empirical percent power (Power\%). Denote $\tau$ as the true ATE value and $\tauest$ as the estimated value from each Monte Carlo replication. Across $M = 500$ replications, these measurements are defined as follows: 
\begin{itemize}
    \item Bias: $\tauest-\tau$, represented in boxplots. From the width of a box, we can evaluate the estimation efficiency. 
    \item CP\%: The proportion of replications where $\tau$ falls within the 95\% confidence interval constructed from $\tauest$ and its estimated standard error, using a normal approximation.
    \item Power\%: The proportion of replications where the p-value is $\leq 0.05$ for testing $H_0: \tau = 0$ versus $H_a: \tau \neq 0$.
\end{itemize}
With $M = 500$ simulations, a CP\% is considered significantly different from the nominal 95\% coverage level if it falls outside the range $95 \pm 1.96 \times \sqrt{95 \times 5/500} = [93, 97]$.

\subsection{Results}\label{subsec:results}

For brevity, we present the simulation results for the nonlinear continuous and nonlinear outcomes under sample sizes of 40 and 200 in this section. Complete simulation results, including figures, are available in Web Appendix B of \textit{Online Supplemental Material}.

\begin{figure}
    \centering
    \textbf{Nonlinear continuous outcome}
    \includegraphics[width=\textwidth]{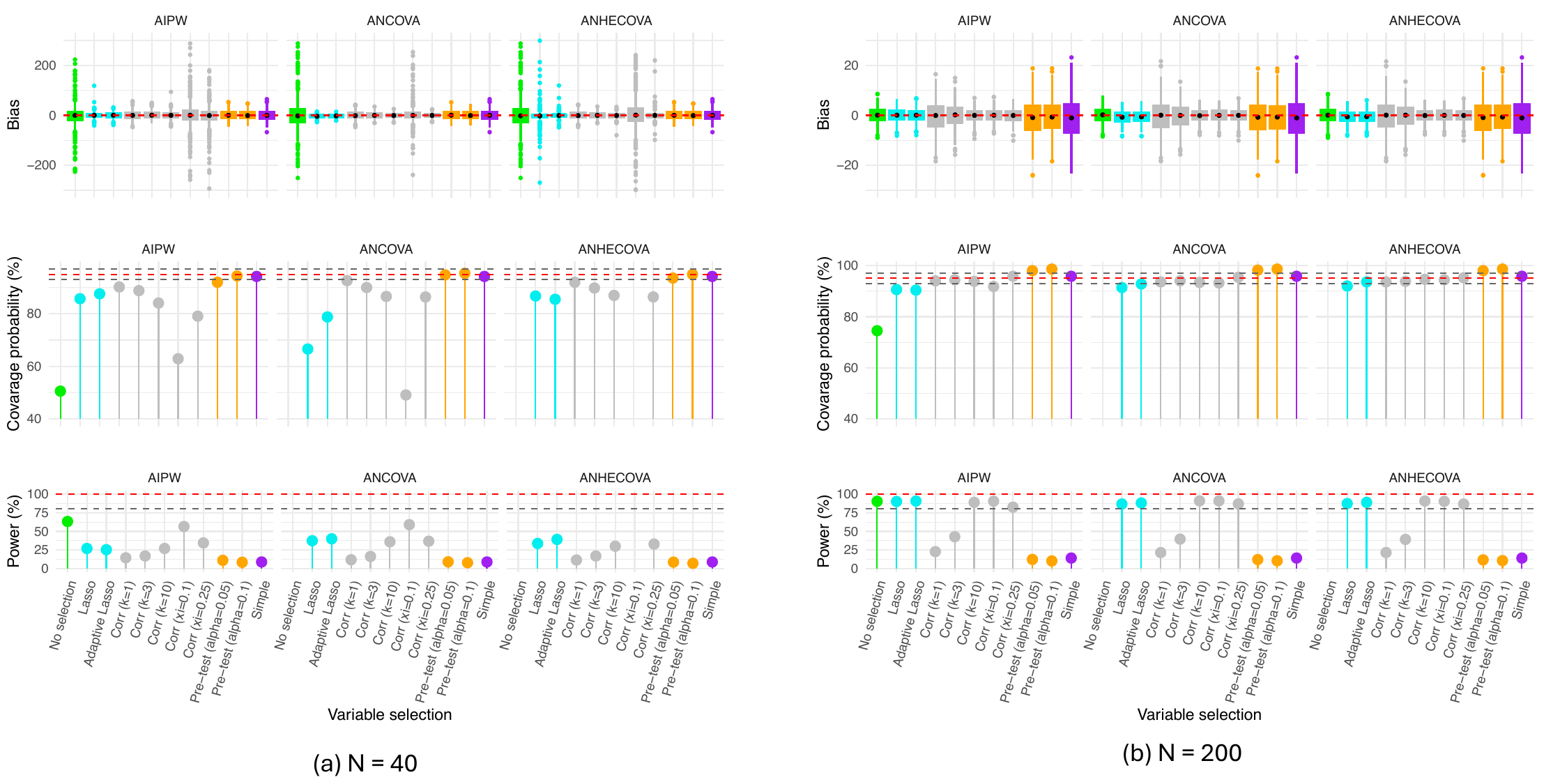}

    \textbf{Nonlinear binary outcome}
    \includegraphics[width=\textwidth]{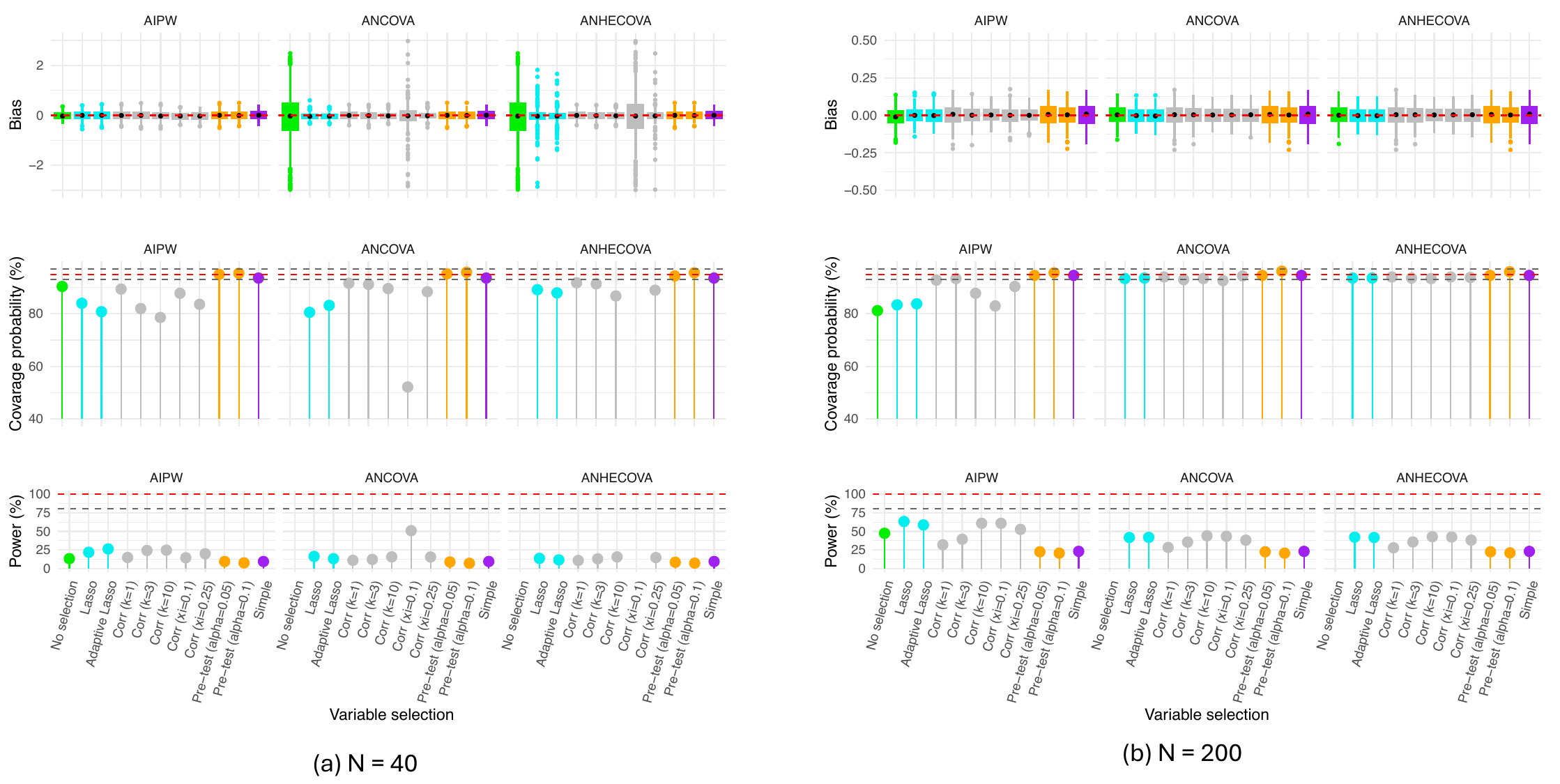}
    \caption{Simulation results for nonlinear and binary outcomes with sample sizes (a) $N=40$ and (b) $N=200$. In the Bias boxplots, black dots indicate medians. In the CP\% plots, the \textcolor{red}{red dashed lines} mark the 95\% coverage level, while the \textcolor[gray]{0.5}{gray dashed lines} mark the 93\% and 97\% levels. In the Power\% plots, the \textcolor{red}{red dashed lines} indicate 100\% power, and the \textcolor[gray]{0.5}{gray dashed lines} indicate 80\% power. Missing values in CP\% and Power\% for ANHECOVA and ANCOVA arise from the non-obtainable sandwich variance estimators in high-dimensional settings.}
    \label{fig:results-main}
\end{figure}

\begin{figure}
    \centering
    \includegraphics[width=\linewidth]{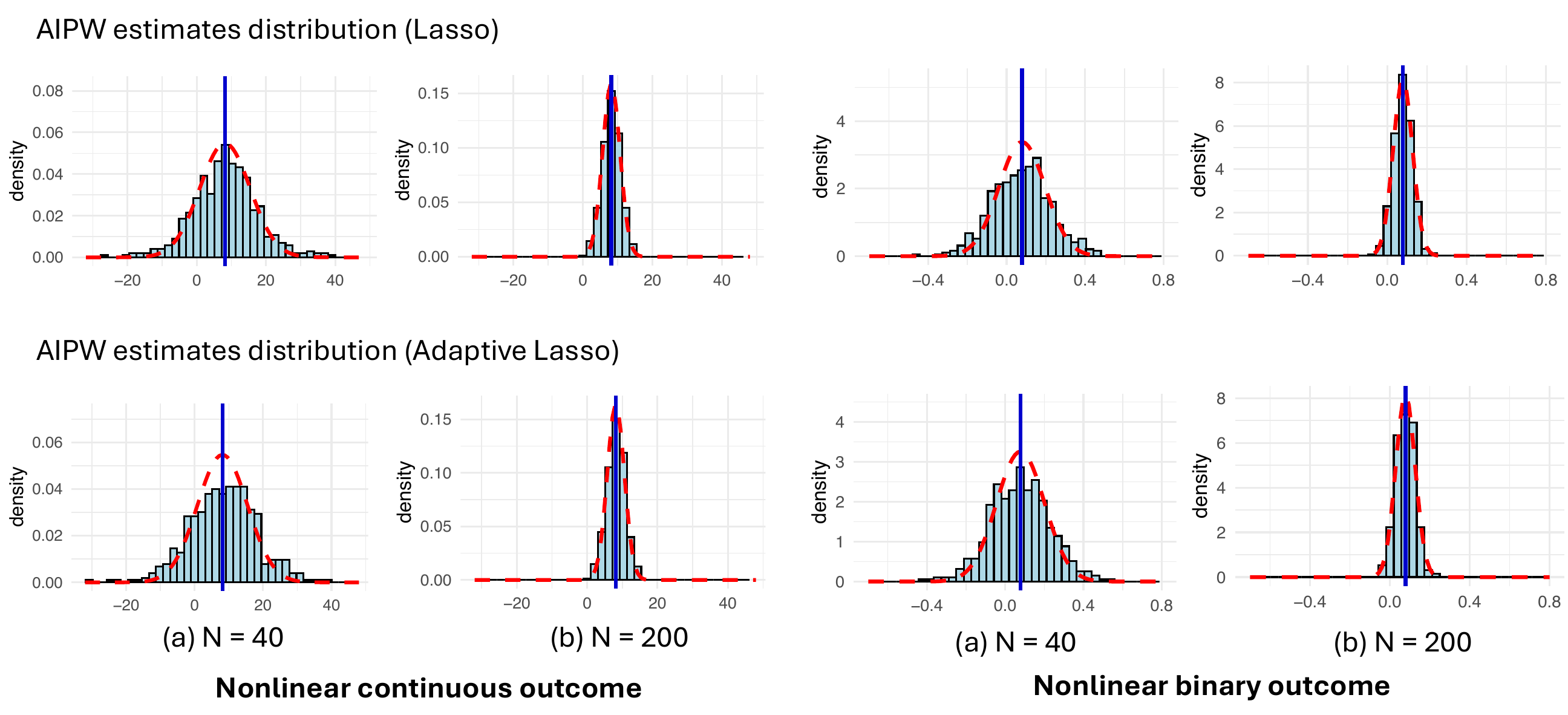}
    \caption{Distributions of the AIPW estimates across 500 Monte Carlo replicates under nonlinear continuous and binary outcomes, with sample sizes $N = 40$ and $N = 200$. The \textcolor{red}{red dashed lines} represent the theoretical normal distributions based on the ATE truth and standard deviations of the 500 point estimates. The \textcolor{blue}{blue lines} indicate the true ATE. }
    \label{fig:results-estdist}
\end{figure}

Figure \ref{fig:results-main} presents the results of biases via boxplots in the upper panels, and the results of CP\% and Power\% in the lower four panels {under each outcome}. We used different colors to distinguish between variable selection methods: \textcolor{greenR}{green} represents the estimator without variable selection (fully adjusted including all covariates), \textcolor{cyanR}{cyan} represents Lasso and adaptive Lasso, \textcolor[gray]{0.5}{gray} indicates selection based on marginal correlation (as described in Table \ref{tab:methods} for pre-specified $k$ or $\xi$), \textcolor{orange}{orange} represents preliminary testing, and \textcolor{purpleR}{purple} represents the simple estimator (which does not include any covariates). 

From Figure \ref{fig:results-main}, we observe that when the sample size is {very small ($N=40$)}, there are many outliers in the boxplots for the ``No selection'' case for all covariate-adjusted estimators. This is expected, since they include a relatively large number of covariates compared to the sample size. Across both sample sizes, the Simple, Pre-test (both $\alpha = 0.05$ and $0.1$) estimators exhibit wider bias box widths, with the Simple estimator widest. The Corr method does not exhibit consistent performance across different parameter choices of $k$ and $\xi$. For example, Corr with $k = 1$ or $k = 3$ is notably less efficient when the sample size is $N = 200$, yet performs relatively well under $N = 40$. In contrast, Corr with $\xi = 0.10$ shows poor performance under $N = 40$ but improves substantially with a larger sample size of $N = 200$. Lasso and adaptive Lasso exhibit similar performance patterns under the same estimator and are overall the most efficient and consistently performing methods across simulation settings. An exception is observed with the ANHECOVA estimator under $N = 40$ for both outcomes, where some outliers appear. Nevertheless, in all other scenarios---particularly when paired with the AIPW estimator---these methods consistently demonstrate stable gains in efficiency and statistical power. These findings suggest that Simple, Pre-test, and Corr methods are generally less efficient in our simulation setting, particularly when compared to Lasso and adaptive Lasso. 

In terms of CP\% and power, all methods exhibit greater stability around the 95\% nominal level at the larger sample size ($N=200$), reflecting improved consistency of the variance estimators with increasing sample size. However, even when $N=200$, we observe below nominal or missing CP\% for all three covariate-adjusted estimators (due to the inability to obtain HW sandwich variance estimators in high-dimensional settings). Interestingly, across both sample sizes, the Simple and Pre-test (both $\alpha=0.05$ and $0.1$) estimators result in the lowest power. This finding for the Pre-test estimator is consistent with \cite{zhao2024randomization}, where they advise against its use in practice based on theory and simulations. {Lasso and adaptive Lasso, when combined with all estimators, yield satisfactory CP\% under the continuous outcome with $N=200$. However, under the binary outcome setting with $N = 200$, they exhibit under-coverage when used with the AIPW estimator, though this performance improves when the sample size increases to $N=500$, as shown in Web Appendix B of \textit{Online Supplemental Material}. This empirical finding suggests that, for binary outcomes using logistic regression as the outcome model in AIPW, it is valuable to investigate more robust variance estimation methods for smaller sample sizes in future research. Overall, Lasso and adaptive Lasso selection methods combined with AIPW generally perform well (in terms of both point and variance estimations) across both sample sizes and both outcomes. } 

{Futhermore, although} we do not include the results of other DGPs, we briefly summarize them here. The results for {other two outcomes (linear continuous and linear binary)} are similar to those shown in Figure \ref{fig:results-main} and our comments above. 
{At the largest sample size ($N=500$), the variance estimators following Lasso and adaptive Lasso selections perform more reliably with similar efficiency gains, yielding CP\% values that are closer to the 95\% nominal level across all outcomes. However, the issue of non-obtainable sandwich variance estimators for ANHECOVA and ANCOVA without variable selection persists even at $N=500$. This indicates that, despite a seemingly adequate sample size relative to the data dimension, the sandwich variance estimator may remain computationally unstable or fail to converge when the number of covariates is large. This finding further highlights the advantage of applying variable selection to control the effective dimensionality, thereby facilitating the computation of model-based variance estimators. }

{Finally, to empirically assess the asymptotic normality of post-selection estimators following Lasso and adaptive Lasso, we present histograms of AIPW estimates after variable selection in Figure \ref{fig:results-estdist}. Additional results for ANCOVA, ANHECOVA, and other simulation scenarios are provided in Web Appendix B of \textit{Online Supplemental Material}. As shown in Figure \ref{fig:results-estdist}, the empirical distributions closely align with the corresponding theoretical normal densities (indicated by red dashed lines). As sample size increases, the estimates become more concentrated around the mean, and the agreement with the theoretical curves improves under both continuous and binary outcomes. Supplemental figures further support these findings, demonstrating strong normality even at moderate sample sizes (e.g., $N=100$ and $N=500$). Taken together, these results provide empirical evidence for the asymptotic normality of covariate-adjusted AIPW estimators following variable selection via Lasso or adaptive Lasso. }

\section{Case Study}\label{sec:data}

We apply the COADVISE framework to analyze the Best Apnea Interventions for Research (BestAIR) trial data, available through the National Sleep Research Resources (NSRR) (\url{https://sleepdata.org/datasets/bestair/files/datasets}). The dataset includes 169 participants, with 83 assigned to the active treatment group (continuous positive airway pressure [CPAP]) and 86 to a combined control arm. The primary objective of the study is to assess the impact of CPAP on 24-hour systolic blood pressure (SBP). The data includes 114 covariates, making it high-dimensional given the sample size ($N=169$). We consider two outcomes: (i) continuous SBP values (in mmHg); and (ii) binary SBP values, classifying SBP as high ($\geq 130$ mmHg) or low ($<130$ mmHg) based on a 130 mmHg threshold.

Both the outcome and covariates contain missing values. Since only 25 observations are complete, we use the random forest method \citep{stekhoven2012missforest} (\texttt{missForest} option in our package) to impute all missing values. {To justify the missingness pattern, we note that \texttt{missForest} assumes an MAR mechanism, which, while untestable in practice, is less restrictive than the MCAR assumption. For missing outcomes, we include all baseline covariates (e.g., age, gender, race, study site) in the imputation model, making the MAR assumption more plausible. Since treatment is randomized and all covariates are collected prior to treatment assignment, the outcome missingness mechanism is unlikely to depend on unobserved outcomes. Furthermore, missingness in baseline covariates is less concerning in randomized trials, as these covariates are independent of both treatment and outcome by design. In this application, which primarily aims to demonstrate the COADVISE framework, we impute all missing covariates under the MAR assumption and include them in the variable selection step to take full advantage of the high-dimensional structure of the data. }

Table \ref{tab:bestAIR} presents the results of the analysis using three approaches to variable selection: (i) no covariate selection (all covariates included), (ii) Lasso, and (iii) adaptive Lasso. When no variable selection is applied, only the simple estimator provides complete results for both outcomes. The standard errors (SEs) for the other methods are unavailable (NA), except for the AIPW estimator in the binary SBP case. This is due to the instability of the HW sandwich variance estimator for ANCOVA and ANHECOVA, which, given the sample size relative to the number of covariates, becomes unstable when inverting high-dimensional matrices, causing it to fail to converge. Additionally, in the case of continuous SBP, the variance estimated for AIPW using the Delta method was negative, leading to an unavailable (NA) SE. Therefore, without variable selection, the results of all covariate-adjusted estimators are not reliable. 

However, when variable selection is applied, all covariate-adjusted methods return complete results, including SEs, CIs, and p-values. Interestingly, the results from the covariate-adjusted methods often differ notably from those of the simple estimator. For the continuous SBP outcome, the simple estimator indicates a significantly non-zero ATE with p-values less than 0.05, whereas none of the variable selection methods identify a significant ATE at the 0.05 level. Additionally, with variable selection, the SEs from the covariate-adjusted methods are smaller than those from the simple estimator for both outcomes. This suggests that proper covariate adjustment can enhance efficiency.

\begin{table}
\centering 
\small
\caption{Estimated average treatment effects by CPAP on blood SBP by data from the BestAIR trial}\label{tab:bestAIR}
    \begin{tabular}{rlccccccc}
      \toprule
      Variable selection & Estimator & \multicolumn{3}{c}{\bf Continuous SBP} & \multicolumn{3}{c}{\bf Binary SBP} \\
      \cmidrule(lr){3-5}\cmidrule(lr){6-8}
        & & Estimate (CI) & SE & p-value & Estimate (CI) & SE & p-value \\ 
        
        \cmidrule(lr){1-5}\cmidrule(lr){6-8}
        \addlinespace
                    & Simple & -3.82 (-7.03, -0.62) & 1.64 & 0.02 & -0.12 (-0.25, 0.00) & 0.06 & 0.06 \\ 
        No selection &  ANCOVA & -0.19 (NA, NA) & NA & NA & 0.08 (NA, NA) & NA & NA \\ 
                    &  ANHECOVA  & 48.68 (NA, NA) & NA & NA & 0.48 (NA, NA) & NA & NA \\ 
                    &  AIPW  & 29.84 (NA, NA) & NA & NA & -0.05 (-0.12, 0.02) & 0.04 & 0.18 \\ 
        \cmidrule(lr){1-5}\cmidrule(lr){6-8}
          \addlinespace
                & Simple & -3.82 (-7.03, -0.62) & 1.64 & 0.02 & -0.12 (-0.25, 0.00) & 0.06 & 0.06 \\ 
        Lasso &  ANCOVA & -0.01 (-0.32, 0.30) & 0.16 & 0.94 & -0.00 (-0.08, 0.07) & 0.04 & 0.93 \\ 
                &  ANHECOVA  & -0.08 (-0.42, 0.27) & 0.18 & 0.67 & 0.00 (-0.08, 0.09) & 0.04 & 0.92 \\ 
                &  AIPW  & -0.39 (-0.90, 0.12) & 0.26 & 0.13 & -0.04 (-0.13, 0.04) & 0.04 & 0.30 \\ 
        \cmidrule(lr){1-5}\cmidrule(lr){6-8}
          \addlinespace
            & Simple & -3.82 (-7.03, -0.62) & 1.64 & 0.02 & -0.12 (-0.25, 0.00) & 0.06 & 0.06 \\ 
Adaptive Lasso & ANCOVA & -0.09 (-0.44, 0.27) & 0.18 & 0.63 & -0.03 (-0.11, 0.05) & 0.04 & 0.45 \\ 
            & ANHECOVA  & -0.29 (-1.05, 0.48) & 0.39 & 0.46 & -0.03 (-0.11, 0.05) & 0.04 & 0.41 \\ 
            &  AIPW  & -0.32 (-0.79, 0.15) & 0.24 & 0.19 & -0.09 (-0.16, -0.01) & 0.04 & 0.03 \\ 
        \bottomrule
    \end{tabular}
    \begin{tablenotes}
        \item CI: confidence interval; SE: standard error; NA: not available.  
    \end{tablenotes}
\end{table}

\section{Concluding Remarks}\label{sec:conclu}

\subsection{Summary}

COADVISE is a comprehensive and flexible framework for covariate adjustment with variable selection, designed to enhance the reliability and efficiency of treatment effect estimation in RCTs. The framework accommodates several variable selection procedures, including no selection, Lasso, adaptive Lasso, marginal correlation methods, and preliminary testing, prior to covariate adjustment. We also implement three widely used covariate-adjustment methods---ANCOVA, ANHECOVA, and AIPW---in the estimation stage. For AIPW, we provide a broad range of parametric models for the conditional outcome models used in the estimator. Notably, we do not implement machine learning (ML) models in the final model fitting step after variable selection due to their black-box nature, which complicates theoretical guarantees for efficiency gains and interpretability. Instead, we focus on transparent parametric models to ensure clarity and alignment with the theoretical conditions established in our work. 

Our framework stands out from existing methods for several reasons. To our knowledge, this tutorial is the first to offer comprehensive guidance on variable selection preceding covariate adjustment, while systematically comparing various variable selection and covariate-adjusted methods from a practical statistical perspective. Moreover, the data-driven variable selection techniques identify key predictors in a balanced manner, optimizing both efficiency and robustness and ensuring that the number of selected variables remains manageable given the sample size. Although some relevant variables may be omitted, this approach strikes a careful balance between the extremes of unadjusted and fully adjusted methods, thereby enhancing efficiency without compromising consistency.  

Furthermore, our empirical studies suggest that Lasso and adaptive Lasso should be considered as priority in practice, since they generally outperform other methods for variable selection in our framework. Their key advantages include (i) flexibility in model specification aligning with the AIPW estimator's outcome model(s), and (ii) the absence of some tuning parameters that must be selected by users, such as thresholds $k$ and $\xi$ in marginal correlation methods, or preliminary testing level $\alpha$, making them more practical for application. Although Lasso and adaptive Lasso do require a tuning parameter $\lambda$ in the penalty term (see Algorithm \ref{algo:Coadvise} in Appendix \ref{subapp:lassoexample}), our package implements \texttt{lambda.min}, which selects the $\lambda$ value that minimizes the cross-validated error, making this process straightforward for users. Adaptive Lasso further has adaptive weights in the penalty term, which are inversely proportional to an initial estimate of the coefficients, obtained from an unpenalized OLS regression in our package. 

\subsection{Discussion}

We acknowledge several limitations of the COADVISE framework. First, we need to further investigate the post-selection inference arising from variable selection methods beyond Lasso and adaptive Lasso. Although our simulations suggest that all variable selection methods included in our package yield valid inferences under larger sample sizes, additional theoretical work is warranted to support these findings. In support of our method, we note that our model-based variance estimators (such as the sandwich or Delta methods) are robust to model misspecification, ensuring valid uncertainty quantification regardless of the variables selected for the outcome model(s). Furthermore, fitting the post-selection models using MLE or OLS improves efficiency, as demonstrated by the theory established in Section \ref{subsec:effgaintheory}. 

A related concern was raised in a recent commentary \citep{harrell2024comment}, which argued that complex methods like ANHECOVA and AIPW, unlike ANCOVA, rely on the assumption that RCT participants are randomly sampled from the clinical population. This assumption ignores inclusion/exclusion criteria and may limit external validity. We respectfully contend that this issue is not unique to these complex estimators; even simpler estimators, including ANCOVA, target the ATE for the population defined by the trial’s criteria, since the RCT sample represents only those who met the criteria. Our primary contribution is to address the high-dimensional covariate challenges that affect internal validity, thereby ensuring robust ATE estimates for the trial population across varying sample sizes. Future extensions might explore data integration and external information borrowing to further enhance the generalizability of RCT findings \citep{lee2022doubly, lee2023improving, lee2024transporting, lee2024genrct}. 

{Additionally, while our \texttt{Coadvise} R package includes several commonly used missing data imputation methods for user convenience, we have not undertaken a comprehensive investigation into their theoretical robustness or optimal integration within our framework, particularly in combination with variable selection. Future work could explore more robust imputation strategies that align with different variable selection mechanisms. We also plan to incorporate more advanced and recently developed imputation methods with theoretical guarantees that are compatible with our framework in future versions of the package. }

There are also several promising avenues for extending our framework. For instance, alternative adjustment methods could be incorporated to further improve efficiency. Existing literature suggests that integrating propensity score modeling, $e(\mb X)=P(A=1\mid \mb X)$, into the AIPW estimator \citep{gao2024adjusting}, or applying propensity score weighting methods without outcome modeling (such as inverse probability weights \citep{shen2014inverse}, overlap weights \citep{li2018balancing, yang2021covariate, zeng2021propensity}, or other schemes \citep{li2013weighting, liu2024average}) can enhance efficiency, even when the propensity score is known (e.g., $e(\mb X)=0.5$ for 1:1 randomization across all $\mb X$) \citep{williamson2014variance, shen2014inverse}. However, extending our framework to include propensity score modeling introduces an additional variable selection challenge, effectively leading to a double selection problem for the AIPW estimator \citep{koch2018covariate, belloni2014inference}. Moreover, although logistic regression is commonly used for propensity score estimation \citep{thomas2020overlap, hirano2001estimation, connors1996effectiveness} it remains uncertain whether it consistently enhances efficiency or if alternative methods with clearer regularity conditions might be preferable. Another potential extension involves prognostic covariate adjustment (PROCOVA), which employs machine learning techniques to generate prognostic scores based on baseline characteristics \citep{harrell1984regression, harrell1996multivariable, vanderbeek2022prognostic, rubin2000combining, li2024prognostic}. Future work could extend our variable selection methods to PROCOVA or explore the joint use of propensity and prognostic scores for covariate adjustment \citep{yang2023multiply, zhang2022practical}. 

Furthermore, our framework can be extended to a wide range of more general and complex settings. These include more general covariate-adaptive randomization \citep{bannick2025general, kernan1999stratified, broglio2018randomization, zhu2025design, yu2024sharp},  rerandomization \citep{zhu2023pair, morgan2012rerandomization}, time-to-event outcomes \citep{hua2024inference, yi2020cox, benkeser2021improving}, restricted mean survival times \citep{han2023breaking, zhong2022adjusting, karrison2018restricted, hanada2024comparison}, cluster-randomized trials \citep{murray1998design, donner2000design, stephens2012augmented, zhou2022constrained, balzer2023two}, and ordinal or composite outcomes with novel estimands such as the win ratio or generalized odds ratio \citep{agresti1980generalized, pocock2012win, barnhart2025trial, matsouaka2022robust, oakes2016win, zhang2024sequential}, as well as the Mann-Whitney rank-based statistic \citep{follmann2020analysis}. Other promising directions include conformal inference on individual or heterogeneous treatment effects \citep{yang2024doubly, wang2024conformal, gao2024role}, the integration of multi-source RCT and observational data \citep{zhuang2025assessment, lee2024genrct, lee2023improving, yang2023elastic, han2025federated, liu2024multi, yang2022rwd, liu2025targeted}, synthetic controls \citep{abadie2015comparative, abadie2010synthetic, ben2021augmented}, and external controls \citep{li2023estimating, lee2024doubly, gao2025improving, wang2025integrating}. Additionally, alternative strategies for estimating regression coefficients, such as debiased machine learning algorithms \citep{chernozhukov2018double, wang2025rate, carlsson2014comparison}, debiased regressions \citep{lu2023debiased, chang2024exact, chiang2023regression}, and linear calibration \citep{bannick2025general, guo2023generalized, cohen2024no} are of interest for future investigation.

\section*{Acknowledgment}

The authors are grateful to Marlena Bannick for insightful discussions during the early stages of this research endeavor. 

\section*{Funding}

Yi Liu was supported by the National Heart, Lung, and Blood Institute (NHLBI) of the National Institutes of Health (NIH) under Award Number T32HL079896. This project is supported by the Food and Drug Administration (FDA) of the U.S. Department of Health and Human Services (HHS) as part of a financial assistance award U01FD007934 totaling \$1,674,013 over two years funded by FDA/HHS. It is also supported by the National Science Foundation under Award Number SES 2242776, totaling \$225,000 over three years. The contents are those of the authors and do not necessarily represent the official views of, nor an endorsement by, FDA/HHS, the NIH, or the U.S. Government.

\section*{Data and Code Availability Statement}

The BestAIR trial data used in Section \ref{sec:data} is available upon reasonable request at \url{https://sleepdata.org}. The R package \texttt{Coadvise} implementing the methodology of this paper is available at \url{https://github.com/yiliu1998/Coadvise}. 

\bibliographystyle{plainnat}
\bibliography{arXiv}

\clearpage

\appendix
\renewcommand{\thesection}{S}
\newcounter{Appendix}[section]
\numberwithin{equation}{subsection}
\renewcommand\theequation{\Alph{section}.\arabic{equation}}
\numberwithin{table}{section}
\numberwithin{figure}{section}

\section{Appendix: R Code and Tutorial to COADVISE}\label{app:addtuto}

\subsection{Summary of missing data imputation methods in COADVISE}\label{subapp:miss}

\begin{table}[H]
    \centering
    \small
    \singlespacing
    \begin{tabular}{rccccccccc}
    \toprule
         \textbf{\makecell[r]{Imputation method}} & \makecell[c]{Handles missing \\ covariates } & \makecell[c]{Handles missing \\ outcome } & Assumption \\
    \midrule
        Complete-case analysis (\texttt{cc}) & $\checkmark$ & $\checkmark$ & MCAR \\
        \addlinespace
       \makecell[r]{Multiple imputation by chained equations \\ (\texttt{mice}; \cite{van2011mice})}  & $\checkmark$ & $\checkmark$ & MAR \\
        \addlinespace
        \makecell[r]{Random forest \\ (\texttt{missForest}; \cite{stekhoven2012missforest})} & $\checkmark$ & $\checkmark$ & MAR \\
        \addlinespace
        \makecell[r]{Inverse probability weighting \\ (\texttt{ipw}; \cite{robins1994estimation}) } & $\times$ & $\checkmark$ & MAR  \\
        \addlinespace
        \makecell[r]{Missingness indicator \\ (\texttt{missInd}; \cite{zhao2024covariate}; \\ \cite{zhao2024adjust, song2021missing})} & $\checkmark$ & $\times$ & MAR \\
    \bottomrule
    \end{tabular}
    \caption{Summary of missing data imputation methods available in the current \texttt{Coadvise} R package.}
    \label{tab:missmethod}
\end{table}

\subsection{R code for BestAIR trial study}\label{subapp:bestair}

We show R code of the analysis for our case study in Section \ref{sec:data}, as a tutorial for the use of our R package \texttt{Coadvise}. For the demonstration purpose, we only present the code by Lasso selection. For other methods, readers can visit the package website (\url{https://github.com/yiliu1998/Coadvise}) or type code \texttt{?Coadvise} in R (assuming the package is correctly installed) to see more details. 

In the code below, \texttt{A} is the binary treatment (CPAP) indicator vector, \texttt{Y} is the continuous SPB value, and \texttt{X} is the covariate matrix (169 rows and 114 columns). We illustrate the additional arguments used in the code in the following Table \ref{tab:argumentsR}. 

\singlespacing
\begin{lstlisting}
#### Install the Coadvise package and load the dataset
if(!require("devtools")) install.packages("devtools") 
devtools::install_github("yiliu1998/Coadvise")
library(Coadvise)     # load the R package
load("BestAIR.Rdata") # the original dataset

#### Analysis using Coadvise() function
result.cont <- Coadvise(Y=Y, A=A, X=X, trt.name=1, ctrl.name=0, 
                        conti.out=TRUE, var.sel.method="Lasso",
                        lasso.family="gaussian", out1.model.aipw="linear", 
                        out0.model.aipw="linear", MI.method="missForest", 
                        seed=4399)
                        
#### Print the ATE results                        
print(result.cont$df.fit)
    method         tau        se     ci.lwr     ci.upr          p
1   Simple -3.82127278 1.6350819 -7.0259744 -0.6165711 0.01943644
2   ANCOVA -0.01156769 0.1567089 -0.3187115  0.2955761 0.94115646
3 ANHECOVA -0.07524842 0.1776052 -0.4233482  0.2728513 0.67179650
4     AIPW -0.38878980 0.2600589 -0.8984959  0.1209163 0.13491273

#### Print the mean outcome vector and covariance matrix by AIPW
print(result.cont$AIPW.out.means)
$mu
      tau0     tau1
1 125.0551 124.6664
$sigma
         [,1]      [,2]
[1,] 176.1196 106.39670
[2,] 106.3967  48.10336
\end{lstlisting}

\begin{table}[H]
    \centering
    \singlespacing
    \scriptsize
    \rotatebox{0}{
    \begin{tabular}{rlllllllllll}
    \toprule
        Arguments & Description & Required & Default & Available options  \\
    \midrule
        \texttt{A} & The treatment vector & $\checkmark$ \\
        \addlinespace
        \texttt{Y} & The outcome vector & $\checkmark$ \\
         \addlinespace
        \texttt{X} & The covariate vector/matrix/data frame & $\checkmark$ & \\
         \addlinespace
         \texttt{trt.name} & The value of treatment in \texttt{A} & $\checkmark$ \\
        \addlinespace
        \texttt{ctrl.name} & The value of control group in \texttt{A} & $\checkmark$ \\
         \addlinespace
        \texttt{conti.out} & Whether the outcome \texttt{Y} is continuous & $\checkmark$ & & \texttt{TRUE}, \texttt{FALSE} \\
         \addlinespace
        \texttt{var.sel.method} & Variable selection method & $\times$ & \texttt{"Lasso"} & \makecell[l]{\texttt{"No"},  \texttt{"Lasso"}, \texttt{"A.Lasso"}, \\ \texttt{"Corr.k"}, \texttt{"Corr.xi"}, \\ \texttt{"Pre.test"}}  \\
         \addlinespace
        \texttt{lasso.family} & Family used in Lasso regression & $\times$ & \texttt{"gaussian"} & \makecell[l]{\texttt{"gaussian"},  \texttt{"binomial"},\\ \texttt{"poisson"}, \texttt{"multinomial"} \\ (Same as the \texttt{glmnet} package)}   \\
         \addlinespace
        \texttt{A.lasso.family} & Family used in adaptive Lasso regression   & $\times$ & \texttt{"gaussian"} & \makecell[l]{\texttt{"gaussian"},  \texttt{"binomial"},\\ \texttt{"poisson"}, \texttt{"multinomial"}}  \\
         \addlinespace
        \texttt{out1.model.aipw} & $\Ex\{Y(1)\mid\mb X\}$ model in AIPW & $\times$ & \texttt{"linear"} & \makecell[l]{\texttt{"linear"},  \texttt{"logit"}, \texttt{"probit"} \\ \texttt{"log"}, \texttt{"cloglog"}, \texttt{"identity"}} \\
         \addlinespace 
        \texttt{out0.model.aipw} & $\Ex\{Y(0)\mid\mb X\}$ model in AIPW & $\times$ & \texttt{"linear"} &  \makecell[l]{\texttt{"linear"},  \texttt{"logit"}, \texttt{"probit"} \\ \texttt{"log"}, \texttt{"cloglog"}, \texttt{"identity"}} \\
        \addlinespace 
        \texttt{lin.cal} & Whether to use linear calibration in AIPW & $\times$ & \texttt{FALSE} & \texttt{TRUE}, \texttt{FALSE}\\
        \addlinespace 
        \texttt{k} & \makecell[l]{ Number of covariates with highest \\ correlations with outcome } & $\times$ & 1 & \makecell[l]{Any positive integers less \\ than the number of covariates } \\
        \addlinespace 
        \texttt{xi} & Lowest correlation with outcome & $\times$ & 0.25 & Numbers between 0 and 1 \\
         \addlinespace
        \texttt{pre.alpha} & Level of preliminary testing & $\times$ & 0.05 & Numbers between 0 and 1 \\
         \addlinespace
        \texttt{conf.level} & Level of returned confidence interval for ATE & $\times$ & 0.05 & Numbers between 0 and 1 \\
         \addlinespace 
        \texttt{MI.method} & Missing data imputation method & $\times$ & \texttt{"cc"} & \makecell[l]{\texttt{"cc"},  \texttt{"mice"}, \texttt{"missForest"}, \\ \texttt{"ipw"}, \texttt{"missInd"}} \\
         \addlinespace
        \texttt{seed} & seed used in \texttt{set.seed()} function & $\times$ & 4399 & Any real numbers \\
    \bottomrule
    \end{tabular}
    }
    \caption{Arguments in \texttt{Coadvise()} R function.}
    \label{tab:argumentsR}
\end{table}

\doublespacing
Both \texttt{trt.name} and \texttt{ctrl.name} need to be specified without default, which are the value of treatment and control groups in \texttt{A}, respectively. Note that \texttt{A} can have more than two levels, but we require users to input the treated (\texttt{trt.name}) and control (\texttt{ctrl.name}) each time for a pair-wise comparison. The \texttt{conti.out} must be specified from \texttt{TRUE} and \texttt{FALSE} as well, which allows the program to identify whether it needs to be processed as a categorical outcome for valid missing data imputation purpose. The \texttt{var.sel.method} argument is for the variable selection method, and we specify \texttt{"Lasso"} for its value. We also need to specify the model family via \texttt{lasso.family} if \texttt{"Lasso"} is used. The default is \texttt{"gaussian"} for using linear model with Gaussian random error assumption in Lasso regression, but users can choose other families, such as \texttt{"binomial"} for binary outcome if using logistic regression, and any families provided in the \texttt{glmnet} package \citep{friedman2021package, hastie2021introduction}. The arguments \texttt{out0.model.aipw} and \texttt{out1.model.aipw} are for specifying the link functions used in outcome models $\mu_0(\mb X)$ and $\mu_1(\mb X)$, respectively, in the AIPW estimator. For example, in the binary outcome case, we use \texttt{"logit"} to specify logistic regression models for both potential outcomes. 

The function \texttt{Coadvise()} returns a list comprising three components: (i) a data frame containing results for the Simple, ANCOVA, ANHECOVA, and AIPW estimators, including point estimates (\texttt{tau}), standard errors (\texttt{se}), confidence intervals (\texttt{ci.lwr} and \texttt{ci.upr}), and p-values (\texttt{p}); (ii) the covariate data after variable selection, used in all ANCOVA, ANHECOVA, and AIPW estimators. For the AIPW estimator, two sets of covariate data are returned, as variable selection is performed separately for the two treatment groups; and (iii) the estimated mean outcomes vector and covariance matrix from the AIPW estimator. In this example, we display components (i) and (iii) from the returned list. 

\subsection{Algorithm example}\label{subapp:lassoexample}

In this section, we present the proposed COADVISE framework, providing technical details in Algorithm \ref{algo:Coadvise}. For illustration, we use Lasso for variable selection and AIPW for estimation.

Let $\mc {R}^p$ denote the $p$-dimensional Euclidean space $(p > 0)$. Define $N_a = \sum_{i=1}^N I(A_i = a)$ as the number of units with treatment assignment $A_i = a$, $s_N^2(\mc W)$ as the sample variance of the i.i.d. data $\mc W = \{W_1, \dots, W_N\}$, and $\text{cov}_N(\mc W, \mc V)$ as the sample covariance of the data $\mc W = \{W_1, \dots, W_N\}$ and $\mc V = \{V_1, \dots, V_N\}$. Similarly, $s_{N_a}^2(\mc W)$ represents the sample variance for all $W_i \in \mc W$ with $A_i = a$, and $\text{cov}_{N_a}(\mc W, \mc V)$ denotes the sample covariance for all $W_i \in \mc W$ and $V_i \in \mc V$ with $A_i = a$, where $a = 0, 1$. Denote $\Px_{N}[f(W)] = N^{-1}\sum_{i=1}^N f(W_i)$ the empirical mean of any function of the i.i.d. observed data $\mc W$. 

\begin{algorithm}
  \caption{COADVISE by using Lasso for the variable selection and AIPW for ATE estimation}\label{algo:Coadvise}
  \begin{algorithmic}
  \STATE {\bf Input:} I.i.d. observed data $\mc O = \{(A_i, Y_i, \mb X_i), i=1,\dots,N\}$, where $\mb X_i\in\mc R^p$, $p>1$; a chosen generalized linear model (GLM) $g(\mb X'\bd\beta)$ with given link $g$, where $\bd\beta\in\mc R^p$ is the regression coefficient vector; a deviance function $D_g(\bd\beta)$ related to the given link $g$; tuning parameters $\lambda_0$ and $\lambda_1$ in Lasso. \\
  \STATE {\bf Output:} The estimate $\widehat\tau$, the variance estimation $\widehat\Vx(\tau)$, and the 95\% confidence interval $\widehat{\mc C}_{0.95}(\tau)$ for ATE $\tau = \Ex\{Y(1)-Y(0)\}$. \\
  \FOR {$a=0,1$} 
  \STATE $$
  \widehat{\bd\beta}_a^{\text{lasso}} = \text{arg}\min_{\bd\beta_a}\left[\frac1{N_a}D_g(\bd\beta_a) + \lambda_a\sum_{j=1}^p|\beta_a^{(j)}|\right],
  $$
  and let $\widehat S_a = \{j: \widehat{\beta}^{\text{lasso},(j)}_a\not=0, j=1,\dots,p\}$. 
  \ENDFOR 
  \FOR {$a=0,1$}
  \STATE $\widetilde{\bd\beta}_a = (\beta_a^{(1)}I(1\in\widehat S_a),\dots, \beta_a^{(p)}I(p\in\widehat S_a))'$; \\
  \STATE Find $\widehat{\bd\beta}_a$ that solves the MLE equation of $\bd\beta_a$, e.g., $ \Px_{N}\left[I(A=a)\{Y-g(\mb X'\widetilde{\bd\beta}_a)\}\mb X'\right]=\bd 0$ for linear, logistic or Poisson regressions; \\
  \STATE Calculate $$
  \widehat\theta_a = \frac1N\sum_{i=1}^N\left[\frac{I(A_i=a)}{\widehat P(A=a)}\{Y_i-g(\mb X_i'\widehat{\bd\beta}_a)\} + g(\mb X_i'\widehat{\bd\beta}_a)\right], 
  $$ 
  and let $\widehat\mu_a(\mb X) = g(\mb X'\widehat{\bd\beta}_a)$. 
  \ENDFOR
  \STATE Calculate
  \begin{align*}
      \widehat v_{11} & = (N/N_1)s_{N_1}^2(Y-\widehat\mu_1(\mb X)) + 2\text{cov}_{N_1}(Y, \widehat\mu_1(\mb X)) - s_N^2(\widehat\mu_1(\mb X)), \\
      \widehat v_{00} & = (N/N_0)s_{N_0}^2(Y-\widehat\mu_0(\mb X)) + 2\text{cov}_{N_0}(Y, \widehat\mu_0(\mb X)) - s_N^2(\widehat\mu_1(\mb X)), \\
    \widehat v_{10} & = \text{cov}_{N_0}(Y, \widehat\mu_0(\mb X)) + \text{cov}_{N_1}(Y, \widehat\mu_1(\mb X)) - \text{cov}_N\{\widehat\mu_1(\mb X), \widehat\mu_0(\mb X)\}. 
  \end{align*} \\
  \STATE {\bfseries Return: } $\widehat\tau = \widehat\theta_1-\widehat\theta_0, \widehat\Vx(\tau) = N^{-1}(\widehat v_{11}-2\widehat v_{10}+\widehat v_{00}),  \widehat{\mc C}_{0.95}(\tau) = \widehat\tau\pm1.96\cdot\widehat\Vx(\tau). $
  \end{algorithmic}
\end{algorithm} 

\clearpage

\subsection{Comparison of \texttt{Coadvise} and \texttt{RobinCar}}\label{subapp:compare}

In the code below, we use a toy example to compare and illustrate our \texttt{Coadvise} package and the well-known \texttt{RobinCar} package. 

\singlespacing
\begin{lstlisting}
#### We generate a small data with continuous outcome
n <- 300
X <- matrix(rnorm(10*n), ncol=10)                   # related covariates
V <- matrix(rt(200*n, df=10), ncol=200)             # noisy variables
A <- rbinom(n, size=1, prob=0.5)                    # treatment
Y <- X%*%(1:10)/10 + A*(4+X%*%(1:10)/50) + rnorm(n) # outcome

#### load R packages
if(!require("RobinCar")) { install.packages("RobinCar") } 
library(RobinCar)
library(Coadvise)
?robincar_linear # the function in RobinCar we will use later

#### We first use Coadvise to perform variable selection, using the Corr.k method with k=5, and results are saved in the "step1" object
step1 <- Coadvise(Y=Y, X=X, A=A, trt.name=1, ctrl.name=0, 
                  conti.out=TRUE, var.sel.method="Corr.k", k=5, 
                  out1.model.aipw="linear", out0.model.aipw="linear", 
                  seed=1335)

#### We then use covariates selected in step1 for ANCOVA
X.selected <- step1$stage1.covars$ANCOVA
covar.names <- colnames(X.selected)
df.RobinCar <- data.frame(X.selected, A=as.factor(A), Y)

#### The simple estimator by RobinCar
rc.simple <- robincar_linear(df=df.RobinCar, treat_col="A",
                             response_col="Y", 
                             contrast_h="diff", adj_method="ANOVA")
print(rc.simple)
  contrast    estimate    se `pval (2-sided)`
  <chr>          <dbl> <dbl>            <dbl>
1 treat 1 - 0     4.06 0.264         2.59e-53

#### The ANCOVA estimator by RobinCar
rc.ancova <- robincar_linear(df=df.RobinCar, treat_col="A", 
                             response_col="Y", covariate_cols=covar.names, 
                             contrast_h="diff", adj_method="ANCOVA")
print(rc.ancova)
  contrast    estimate    se `pval (2-sided)`
  <chr>          <dbl> <dbl>            <dbl>
1 treat 1 - 0     3.95 0.151        1.27e-151

#### The ANHECOVA estimator by RobinCar
rc.anhecova <- robincar_linear(df=df.RobinCar, treat_col="A", 
                               response_col="Y", contrast_h="diff",
                               covariate_cols=covar.names, 
                               adj_method="ANHECOVA")
print(rc.anhecova)
  contrast    estimate    se `pval (2-sided)`
  <chr>          <dbl> <dbl>            <dbl>
1 treat 1 - 0     3.95 0.151        1.54e-151

#### The ATE results by Coadvise using the selected covariates from step1 with no variable selection (var.sel.method="No")
step2 <- Coadvise(Y=Y, X=X.selected, A=A, trt.name=1, ctrl.name=0, 
                  conti.out=TRUE, var.sel.method="No", 
                  out1.model.aipw="linear", out0.model.aipw="linear", 
                  seed=1335)
print(step2$df.fit)
    method      tau        se   ci.lwr   ci.upr p
1   Simple 4.058923 0.2640761 3.541344 4.576503 0
2   ANCOVA 3.952416 0.1498078 3.658798 4.246034 0
3 ANHECOVA 3.951386 0.1497789 3.657825 4.244948 0
4     AIPW 3.951386 0.1494168 3.658535 4.244238 0

#### We finally check if using var.sel.method="No" in step2 is the same as the output of step1 (for Simple, ANCOVA, and ANHECOVA)
print(step1$df.fit)
    method      tau        se   ci.lwr   ci.upr p
1   Simple 4.058923 0.2640761 3.541344 4.576503 0
2   ANCOVA 3.952416 0.1498078 3.658798 4.246034 0
3 ANHECOVA 3.951386 0.1497789 3.657825 4.244948 0
4     AIPW 4.021081 0.1557283 3.715859 4.326303 0
\end{lstlisting}

\doublespacing
From the results above, we observe that \texttt{Coadvise} and \texttt{RobinCar} yield the same point estimates and standard errors when using identical estimation methods. Additionally, our \texttt{Coadvise} package allows for variable selection and outputs the selected covariates. Re-running the \texttt{Coadvise()} function with the selected covariates, but without further variable selection, produces results consistent with the initial variable selection process. 

Note that in the last two results (\texttt{step1\$df.fit} and \texttt{step2\$df.fit}), the AIPW estimators produce different point estimates. We would like to clarify that this occurs because for \texttt{step2\$df.fit}, we used the variables selected by ANCOVA for AIPW, whereas in \texttt{step1\$df.fit}, the variables selected in AIPW are different than ANCOVA, and the two outcome models in AIPW may use different sets of variables as predictors. 

\subsection{A toy example for COADVISE with covariate-adaptive randomization}\label{subapp:CAR}

In this section, we present a toy example to illustrate the use of the COADVISE framework for covariate-adaptive randomization via the \texttt{CoadviseCAR()} function. As detailed in Section~\ref{subsec:extend}, our \texttt{Coadvise} package includes three estimators for ATE estimation: the simple unadjusted estimator (Simple), the stratified unadjusted estimator (Strata), and the AIPW estimator. For AIPW, all variable selection methods supported in the \texttt{Coadvise()} function are available.

Below, we provide the R code used to generate the toy example. We simulate data with $N = 200$ observations and 100 covariates. Stratified randomization is performed across three strata---Low, Medium, and High---defined by a categorical variable. The potential outcomes are generated from a linear model involving the first three covariates, and the true ATE is set to $\tau = 1$. 

\singlespacing
\begin{lstlisting}
#### Data generation
n <- 200 # sample size
p <- 100 # number of covariates
strata_labels <- c("Low", "Medium", "High") # levels of strata
X <- matrix(rnorm(n*p), nrow=n, ncol=p) # covariates
colnames(X) <- paste0("X", 1:p)
A <- numeric(n)
# Generate treatment assignment with equal probability within each stratum
for (z in strata_labels) {
  idx <- which(strata==z)
  A[idx] <- rbinom(length(idx), 1, 0.5)
}

tau <- 1 # treatment effect (true ATE)
base <- X[,1] - 0.5*X[,2] + 0.3*X[,3]
Y0 <- base + rnorm(n) # potential outcome Y(0)
Y1 <- base + tau + rnorm(n) # potential outcome Y(1)
Y <- ifelse(A==1, Y1, Y0) # observed outcome

#### Analysis using CoadviseCAR() function and results
# Without variable selection for AIPW estimator
CoadviseCAR(Y=Y, A=A, 
            trt.name=1, ctrl.name=0, strata=strata, 
            X=X, conti.out=TRUE, var.sel.method="No")$df.fit
  method      tau        se    ci.lwr   ci.upr            p
1 Simple 1.078399 0.2258623 0.6357165 1.521081 1.800795e-06
2 Strata 1.063202 0.2298618 0.6126817 1.513723 3.738767e-06
3   AIPW 3.796362       NaN       NaN      NaN          NaN

# Lasso variable selection for AIPW estimator
CoadviseCAR(Y=Y, A=A, 
            trt.name=1, ctrl.name=0, strata=strata, 
            X=X, conti.out=TRUE, var.sel.method="Lasso")$df.fit
  method      tau        se    ci.lwr   ci.upr            p
1 Simple 1.078399 0.2258623 0.6357165 1.521081 1.800795e-06
2 Strata 1.063202 0.2298618 0.6126817 1.513723 3.738767e-06
3   AIPW 1.085008 0.1412289 0.8082046 1.361812 1.554312e-14
\end{lstlisting}

\doublespacing
From the analysis results, we observe that the Simple and Strata estimators yield similar point estimates and variance estimates. In contrast, the AIPW estimator without variable selection is highly biased and produces an unstable or unavailable variance estimate. However, after applying Lasso variable selection, the AIPW estimator provides an accurate point estimate with a smaller variance than both the Simple and Strata estimators, indicating a clear efficiency gain.

\clearpage

\begin{center} 
    {\bf\huge Online Supplemental Material}
\end{center}

\appendix
\numberwithin{equation}{subsection}
\renewcommand\theequation{\Alph{section}.\arabic{subsection}.\arabic{equation}}
\numberwithin{assumption}{subsection}
\numberwithin{condition}{subsection}
\numberwithin{table}{subsection}
\numberwithin{figure}{subsection}
\section{Technical Proofs}\label{app:tech}

In this section, we present our theoretical investigations into the efficiency gain achieved by the AIPW estimator for the ATE in RCT data. We adhere to the notation, setups, and assumptions established in the main text of the paper. 

\subsection{Minimizing the variance over a class of augmented estimators}\label{subapp:minVarGen}

First, consider a general class of augmented estimators for $\tau$ defined as follows:
\begin{align}\label{eq:class}
    \mc G = \{\tauest^{\mc G}(g_0, g_1): \Ex\{g_a(\mb X)^2\}<\infty, a=0,1\},
\end{align}
where
$$
\tauest^{\mc G} = \tauest^{\mc G}(g_0, g_1) = \frac{1}{N}\sum_{i=1}^N\left\{\frac{A_iY_i}{\widehat\pi_1}-\frac{A_i-\widehat\pi_1}{\widehat\pi_1} g_1(\mb X_i)\right\}-\frac1N\sum_{i=1}^N\left\{\frac{(1-A_i)Y_i}{1-\widehat\pi_1}-\frac{\widehat\pi_1-A_i}{1-\widehat\pi_1} g_0(\mb X_i)\right\} + o_p(N^{-1/2}), 
$$
with $\widehat\pi_1 = N_1/N$, the proportion of individuals with $A=1$ in the sample.

We note that when $g_1(\mb X) = g_0(\mb X) = 0$, the estimator $\tauest^{\mc G}$ reduces to $\tauest_{\text{simple}}$. Furthermore, it is straightforward to show that for any choice of $g_1$ and $g_0$, $\tauest^{\mc G} \to_p \tau$, using standard conditioning arguments and large sample theory.

The estimator $\tauest^{\mc G}$ provides a general form of the AIPW estimator, as it consists of $\tauest_{\text{simple}}$ augmented by functions $g_0$ and $g_1$. The term $o_p(N^{-1/2})$ indicates that two estimators in the class $\mc G$ differ by a negligible term that converges to 0 in probability with zero asymptotic variance. As can be observed, $\mc G$ is a large (uncountable) set, containing functions from an infinite-dimensional space. 

When $g_a(\mb X) = \Ex(Y(a) \mid \mb X)$, representing the fully specified model for $Y(a)$, $a=0,1$, the estimator $\tauest^{\mc G}$ achieves the semiparametric efficiency bound \citep{hahn1998role, hirano2003efficient}. However, since the true model of $Y(a)$ as a function of $\mb X$ is unknown, achieving this bound is generally not expected. Nevertheless, we can still improve efficiency through ``appropriate augmentation,'' even without precise model specification, compared to $\tauest_{\text{simple}}$. This is our objective in leveraging covariate information to gain efficiency \citep{bannick2025general}.

In practice, we focus on a subclass of $\mc G$, denoted as $\Gset \subset \mc G$, where $g_a(\mb X) = g(\mb X'\bd\beta_a)$ follows a generalized linear model (GLM) for $Y(a)$, for $a=0,1$. This subclass is characterized by the given function $g$, for all vectors $\bd\beta \in \mc R^p$ such that $\Ex\{g(\mb X' \bd\beta)^2\}<\infty$. If the true model for $Y(a)$ belongs to $\Gset$, then there exists a fixed $\bd\beta_a^0 \in \mc R^p$ such that $g(\mb X' \bd\beta_a^0) = \Ex(Y(a) \mid \mb X)$. This subclass $\Gset$ has practical appeal because, in many applications, we often fit both $Y(0)$ and $Y(1)$ using the same GLM with an appropriate link function $g$.

Next, we denote an AIPW estimator belonging to $\Gset$ by $\tauest^g$,  highlighting the dependence on the link function $g$. Using argument $\Vx(\cdot) = \Ex\{\Vx(\cdot\mid\mb X, Y(0), Y(1))\} + \Vx\{\Ex(\cdot\mid\mb X, Y(0), Y(1))\}$, we have the asymptotic variance of $\tauest^g$ as follows:
\begin{align}\label{eq:asyVar}
    \Vx(\tauest^{g}) & = \Vx\left\{\frac{AY}{\pi_1}-\frac{A-\pi_1}{\pi_1} g(\mb X'\bd\beta_1^*)-\frac{(1-A)Y}{1-\pi_1}+\frac{\pi_1-A}{1-\pi_1} g(\mb X'\bd\beta_0^*)\right\}\nonumber\\
    & = \Vx\left\{\frac{AY(1)}{\pi_1}-\frac{A-\pi_1}{\pi_1} g(\mb X'\bd\beta_1^*)-\frac{(1-A)Y(0)}{1-\pi_1}+\frac{\pi_1-A}{1-\pi_1} g(\mb X'\bd\beta_0^*)\right\}\nonumber\\
    & = \Ex\left\{\frac{1-\pi_1}{\pi_1}\{Y(1)-g(\mb X'\bd\beta_1^*)\}^2 + \frac{\pi_1}{1-\pi_1}\{Y(0)-g(\mb X'\bd\beta_0^*)\}^2\right\} + \Vx\{Y(1)-Y(0)\},
\end{align}
where $\bd\beta_a^*$ is the probability limit of $\widehat{\bd\beta}_a$, and $\widehat{\bd\beta}_a$ is an estimator (e.g., the ordinary least square [OLS] estimator) used to estimate $\bd\beta_a$ ($a=0,1$). 

Therefore, to minimize \eqref{eq:asyVar} over choices of $({\bd\beta_0^*}', {\bd\beta_1^*}')'$, it is equivalent to minimize
\begin{align}\label{eq:miniVarEq}
    \Ex\left\{\frac{1-\pi_1}{\pi_1}\{Y(1)-g(\mb X'\bd\beta_1^*)\}^2 + \frac{\pi_1}{1-\pi_1}\{Y(0)-g(\mb X'\bd\beta_0^*)\}^2\right\}.
\end{align}
Then, it is natural to consider setting the derivative with respect to $({\bd\beta_0^*}', {\bd\beta_1^*}')'$ to 0, i.e., to solve equations
\begin{align}\label{eq:zeroVarEq}
    0 & = \frac{1-\pi_1}{\pi_1}\Ex\left\{g_{\bd\beta_1}(\mb X'\bd\beta_1^*)[Y(1)-g(\mb X'\bd\beta_1^*)]\right\},\nonumber \\
    0 & = \frac{\pi_1}{1-\pi_1}\Ex\left\{g_{\bd\beta_0}(\mb X'\bd\beta_0^*)[Y(0)-g(\mb X'\bd\beta_0^*)]\right\},
\end{align}
where $g_{\bd\beta}(\mb X'\bd\beta^*)$ denotes $\dfrac{\partial}{\partial\bd\beta}g(\mb X'\bd\beta)\bigg|_{\bd\beta=\bd\beta^*}$. 

Observe that the solution of the following equation, say, $\widehat{\bd\beta}_a$,
\begin{align}\label{eq:olsEstEq}
\frac1N\sum_{i=1}^N I(A_i=a)g_{\bd\beta_a}(\mb X'\bd\beta_a)[Y_i-g(\mb X_i'\bd\beta_a)] = 0, 
\end{align}
converges to the solution of \eqref{eq:zeroVarEq}, for $a=0,1$. This is because when the outcome model is possibly misspecified, i.e, when $g(\mb X'\bd\beta_a)\not=\Ex(Y(a)\mid\mb X)$, the left-hand side of the above equation converges in probability to 
\begin{align}\label{eq:olsLimit}
\pi_a\Ex\left\{g_{\bd\beta_a}(\mb X'\bd\beta_a)[\Ex(Y(a)\mid\mb X)-g(\mb X'\bd\beta_a)]\right\} = \pi_a\Ex\left\{g_{\bd\beta_a}(\mb X'\bd\beta_a)[Y(a)-g(\mb X'\bd\beta_a)]\right\}, 
\end{align}
for $a=0,1$, which differs from \eqref{eq:zeroVarEq} only by a constant. Therefore, setting \eqref{eq:olsLimit} to be 0 results in the same solution to \eqref{eq:zeroVarEq}. We refer to this $\widehat{\bd\beta}_a$ as the OLS estimator in the following text. 

Moreover, to verify whether solving \eqref{eq:zeroVarEq} always uniquely minimizes \eqref{eq:miniVarEq}, we check if the second-order derivative of \eqref{eq:miniVarEq} with respect to $({\bd\beta_0^*}', {\bd\beta_1^*}')'$ is semi-positive definite (a sufficient but not necessary condition). The second-order derivative can be expressed as $\text{diag}(\bd\Sigma_0, \bd\Sigma_1)$ (diagonal matrix of two blocks), where
\begin{align}\label{eq:secDer}
    \bd\Sigma_a & = 2\frac{1-\pi_a}{\pi_a}\Ex\left\{-\{Y(a)-g(\mb X'\bd\beta_a^*)\}g_{\bd\beta_a\bd\beta_a}(\mb X'\bd\beta_a^*) + g_{\bd\beta_a}(\mb X'\bd\beta_a^*)g_{\bd\beta_a}(\mb X'\bd\beta_a^*)'\right\}\nonumber\\
    & = 2\frac{1-\pi_a}{\pi_a}\Ex\left\{g_{\bd\beta_a}(\mb X'\bd\beta_a^*)g_{\bd\beta_a}(\mb X'\bd\beta_a^*)'\right\} - 2\frac{1-\pi_a}{\pi_a}\Ex\left\{\{\Ex(Y(a)\mid \mb X)-g(\mb X'\bd\beta_a^*)\}g_{\bd\beta_a\bd\beta_a}(\mb X'\bd\beta_a^*)\right\},
\end{align}
for $a=0,1$, where $g_{\bd\beta\bd\beta}(\mb X'\bd\beta^*)$ denotes $\dfrac{\partial^2}{\partial\bd\beta\partial\bd\beta'}g(\mb X'\bd\beta)\bigg|_{\bd\beta=\bd\beta^*}$. We note that the first term of \eqref{eq:secDer} is always positive definite, while for the second term, further discussion is provided in Sections \ref{subapp:guaEffLin} and \ref{subapp:guaEffLog}. 

\subsection{Guaranteed efficiency gain using linear models}\label{subapp:guaEffLin}

Consider the special case where $g(\mb X'\bd\beta_a) = \mb X'\bd\beta_a$, the identical link function. In this case, the second term in \eqref{eq:secDer} is always zero because $g_{\bd\beta_a\bd\beta_a}(\mb X'\bd\beta_a^*)=\bd 0$. Thus, the minimum of \eqref{eq:miniVarEq} is always achieved by the OLS estimator $\widehat{\bd\beta}_a$ for $\bd\beta_a$. The efficiency gain is guaranteed, as $\simple\in\Gset$ by setting $\bd\beta_0=\bd\beta_1=\bd 0$. Consequently, the estimator in the class $\Gset$, which minimizes the variance, must be at least as efficient as $\simple$.

Furthermore, this result can easily be extended situations where a variable selection step is performed before AIPW estimation.  Suppose we select $s$ variables from $\mb X$ for $Y(a)$ model, where $0<s<p$. When fitting the OLS estimator, the only difference is that $\bd\beta\in\mc R^s$ instead of $\mc R^p$ for estimators in $\mc G^{\bd\beta}_g$. Here, $\simple$ remains a special case of estimators in $\mc G^{\bd\beta}_g$. Therefore, the proof remains straightforward.

Moreover, this illustrates that with more covariates (i.e., higher dimensions of $\bd\beta_a$), there is a greater opportunity to achieve higher efficiency. Without loss of generality, assume the first $s$ variables of $\mb X$ are selected. We can view the lower dimension $s$ as $\bd\beta_a\in\mc R^p$ where $\beta_a^{(s+1)}=\dots=\beta_a^{(p)}=0$ are fixed, with $\beta_a^{(j)}$ denoting the $j$-th element of $\bd\beta_a$. Therefore, the class of estimators $\mc G^{\bd\beta}_g$ when $\bd\beta_a\in\mc R^s$ can be viewed as a subset of $\mc G^{\bd\beta}_g$ when $\bd\beta_a\in\mc R^p$. Naturally, the variance by a minimizer found from a larger class must be at least as low as that by a minimizer found from a smaller subclass. However, when the sample size is not sufficiently large, including many covariates can negatively impact the finite-sample performance of the estimator, as all of the aforementioned results are based on large-sample asymptotics. 

\subsection{Conditions on efficiency gain under a nonlinear GLM}\label{subapp:guaEffLog}

nvestigating efficiency gains with a general link function $g$ is more challenging due to several factors: (i) the usual maximum likelihood estimator (MLE) for regression coefficients in a general GLM is not obtained by solving \eqref{eq:olsEstEq} to minimize the squared loss in \eqref{eq:miniVarEq}; (ii) for a general link function $g$, the second-order derivative \eqref{eq:secDer} is not always positive-definite; (iii) $\simple$ may not even belong to $\Gset$ for certain link functions. 

For (iii), consider logistic regression. The logistic link function is given by $g(u)=\{1+e^{-u}\}^{-1}>0$, which implies that the class $\Gset$, defined by the logistic function $g$, does not include $\simple$. This is because the influence function of $\simple$ cannot be approximated by any influence functions of estimators in $\Gset$ (in the sense that the difference between the two influence functions can be $O_p(N^{-r})$ with an $r\geq 1/2$). Achieving such an approximation would require $g(\mb X'\bd\beta_a^*)\to_p 0$ for all $\mb X\in\mc X$, which in trun would imply that there exists some fixed $\bd\beta_a^*$ such that $\mb X'\bd\beta_a^*\to_p-\infty$ for all $\mb X\in\mc X$. Consequently, $\simple$ cannot even lie on the boundary (or closure) of $\Gset$. 

To further explore the conditions for efficiency gain, we examine the difference between the asymptotic variances of $\simple$ and  $\tauest^g$, given by: 
\begin{align}\label{eq:varDiff}
    \Vx(\simple) - \Vx(\tauest^g) & = \Ex\left\{\frac{1-\pi_1}{\pi_1}Y(1)^2 + \frac{\pi_1}{1-\pi_1}Y(0)^2\right\}\nonumber \\
    & \qquad - \Ex\left\{\frac{1-\pi_1}{\pi_1}\{Y(1)-g(\mb X'\bd\beta_1^*)\}^2 + \frac{\pi_1}{1-\pi_1}\{Y(0)-g(\mb X'\bd\beta_0^*)\}^2\right\}\nonumber \\
    & = \Ex\left\{\frac{1-\pi_1}{\pi_1}g(\mb X'\bd\beta_1)\{2Y(1)-g(\mb X'\bd\beta_1)\} + \frac{\pi_1}{1-\pi_1}g(\mb X'\bd\beta_0)\{2Y(0)-g(\mb X'\bd\beta_0)\}\right\}\nonumber  \\
    & = \frac{1-\pi_1}{\pi_1}\underbrace{\Ex\left\{g(\mb X'\bd\beta_1)[2Y(1)-g(\mb X'\bd\beta_1)]\right\}}_{I_1} + \frac{\pi_1}{1-\pi_1}\underbrace{\Ex\left\{g(\mb X'\bd\beta_0)[2Y(0)-g(\mb X'\bd\beta_0)]\right\}}_{I_0}.
\end{align}
For $a=0,1$, we note that $I_a = \Ex\left\{g(\mb X'\bd\beta_a)[2Y(a)-g(\mb X'\bd\beta_a)]\right\} = \Ex\left\{g(\mb X'\bd\beta_a)[2\Ex(Y(a)\mid \mb X)-g(\mb X'\bd\beta_a)]\right\} = \Ex\{I_a(\mb X)\}$, where $I_a(\mb X) = g(\mb X'\bd\beta_a)[2\Ex(Y(a)\mid \mb X)-g(\mb X'\bd\beta_a)]$. Therefore, $I_a(\mb X)$ achieves its maximum when $g(\mb X'\bd\beta_a) = \Ex(Y(a)\mid \mb X)$. This implies that when $g(\mb X'\bd\beta_a)$ is the fully specified model for $\Ex(Y(a)\mid \mb X)$, $\Vx(\simple) - \Vx(\tauest^g)$ reaches its maximal value $\Ex\{\Ex\{Y(a)^2\mid\mb X\}\} = \Ex\{Y(a)^2\}$, corresponding to the highest efficiency gain. Additionally, in this case, the second term in the \eqref{eq:secDer} of the squared loss is always zero, which ensures $\bd\beta_a^*$ is the variance minimizer. However, this is typically unrealistic in practice, and if the model is not well-specified, it is possible that \eqref{eq:varDiff} can be negative. In the following, we outline some conditions for achieving efficiency gains.

\begin{condition}\label{cond:genGLM}
    For a nonlinear link function $g$, with probability 1, and there exists some $\bd\beta_a^*$ ($a=0,1$), we have: (a) the sign of $g(\mb X'\bd\beta_a^*)$ must match that of $\Ex(Y(a)\mid \mb X)$; and (b) $\vert g(\mb X'\bd\beta_a^*)\vert\leq 2\vert\Ex(Y(a)\mid \mb X)\vert$. 
\end{condition}

If Condition \ref{cond:genGLM} holds, then it is clear that \eqref{eq:varDiff} is always non-negative. This condition may be mild for some positive categorical outcomes. In the following illustration, we consider logistic regression as a more specific example, as it is commonly used in binary outcome modelling. For both $a=0,1$, we further write
\begin{align}\label{eq:derLogistic}
    g_{\bd\beta_a}(\mb X'\bd\beta_a^*) & = \frac{\exp(-\mb X'\bd\beta_a^*)}{\{1+\exp(-\mb X'\bd\beta_a^*)\}^2}\mb X' = g(\mb X'\bd\beta_a^*)\{1-g(\mb X'\bd\beta_a^*)\}\mb X',\nonumber \\
    g_{\bd\beta_a\bd\beta_a}(\mb X'\bd\beta_a^*) & = \frac{\exp(-\mb X'\bd\beta_a^*)\{\exp(-\mb X'\bd\beta_a^*)-1\}}{\{1+\exp(-\mb X'\bd\beta_a^*)\}^3}\mb X\mb X' = g(\mb X'\bd\beta_a^*)\{1-g(\mb X'\bd\beta_a^*)\}\{1-2g(\mb X'\bd\beta_a^*)\}\mb X\mb X'.
\end{align}
Therefore, for any general $\mb X\not=\bd 0$, we have $g_{\bd\beta_a}(\mb X'\bd\beta_a^*)\not=\bd 0$, since the logistic function satisfies $0<g(\mb X'\bd\beta_a^*)<1$. Additionally, $g_{\bd\beta_a\bd\beta_a}(\mb X'\bd\beta_a^*)=\bd 0$ only when $\bd\beta_a^*=\bd 0$ (so $g(\mb X'\bd\beta_a^*)=1/2$); for other $\bd\beta_a^*$, the sign of $g_{\bd\beta_a\bd\beta_a}(\mb X'\bd\beta_a^*)$ can be either positive or negative. We now introduce Condition \ref{cond:logit} for using logistic regression to achieve efficiency gain.  

\begin{condition}[Logistic outcome models]\label{cond:logit}
    (i) There exists a constant $\delta>0$ such that $\Ex(Y(1)\mid\mb X) = 1-\Ex(Y(0)\mid\mb X) \geq \delta$ with probability 1 over $\mb X\in\mc X$; (ii) For logistic regression model $g(\mb X'\bd\beta_a^*)$ we assigned to $Y(a)$, $|f_a(\mb X)-g(\mb X'\bd\beta_a^*)|< g(\mb X'\bd\beta_a^*)\{1-g(\mb X'\bd\beta_a^*)\}|1-2g(\mb X'\bd\beta_a^*)|^{-1}$ with probability 1. 
\end{condition}

If Condition \ref{cond:logit} is satisfied, it is straightforward to verify that the OLS estimator $\widehat{\bd\beta}_a$ minimizes the variance. We also have additional intuitions about Condition \ref{cond:logit} as follows. Condition \ref{cond:logit}(i) requires the true proportion of positive responses under both treatment and control is non-zero, which is a mild assumption for many binary outcomes in practice. In Condition \ref{cond:logit}(ii), the left-hand side of the inequality in (ii) can be interpreted as the (absolute local) bias of the model $g(\mb X'\bd\beta_a)$ at $\mb X$. This means that if the bias can be controlled by the right-hand side almost surely, along with (i), the proof in Section \ref{subapp:guaEffLin} regarding the minimizer still applies to logistic regression. 

\begin{figure}
    \centering
    \includegraphics[width=0.77\textwidth]{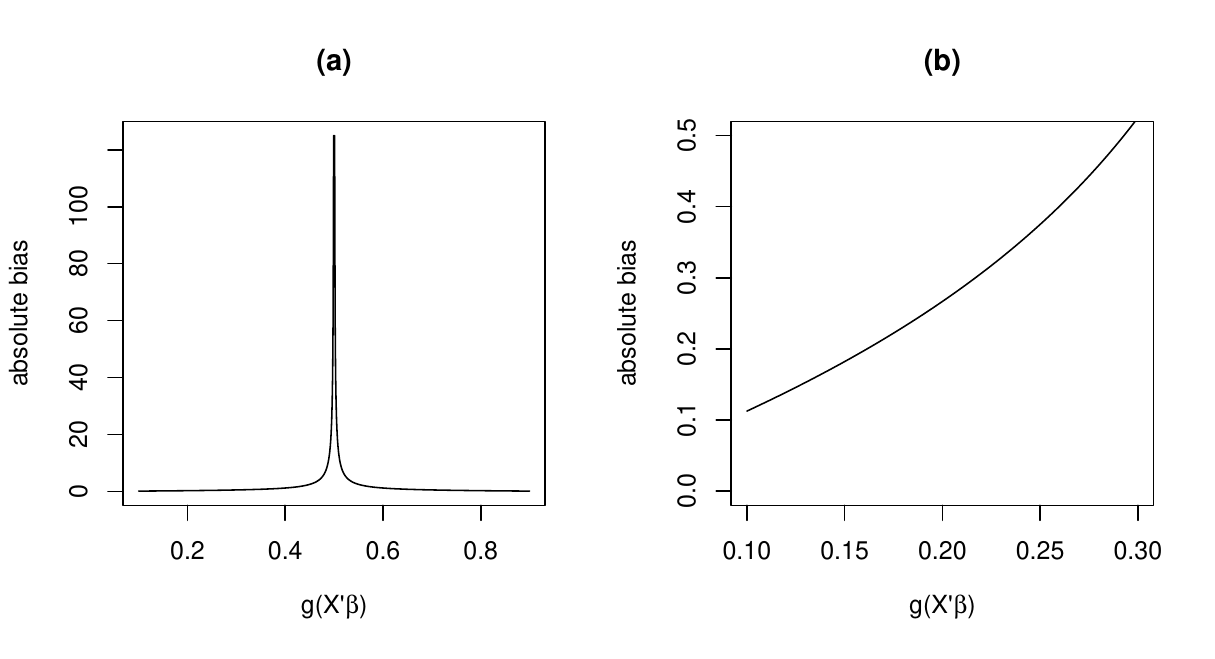}
    \caption{An illustration for the bound of absolute bias in Condition \ref{cond:logit}(ii) (right-hand side of the inequality). Panel (a) provides the absolute biases over $g(\mb X'\bd\beta)\in[0.1,0.9]$, which ranges from $[0.11, \infty)$. When $g(\mb X'\bd\beta)=0.5$, the absolute bias is actually $\infty$; Panel (b) provides a further look for $g(\mb X'\bd\beta)\in[0.1,0.3]$. }
    \label{fig:abias_logis}
\end{figure}

Figure \ref{fig:abias_logis} illustrates the right-hand side of inequality in (ii). From the figure, if $\bd\beta_a^*$ is such that $g(\mb X'\bd\beta^*)\approx 1/2$, the (absolute value of) bias is allowed to be relatively large. However, when $g(\mb X'\bd\beta^*)$ is either smaller or larger, the bias must be smaller, approaching zero when $g(\mb X'\bd\beta^*)\approx 0$ or $1$. In other words, when the true proportion of $Y(a)$ is very small or vary large, the model must better approximate the true value. But when the true proportion is around $50\%$, more error in the posited model is acceptable. The intuition is when the model $g(\mb X'\bd\beta^*)$  approaches a medium value, the uncertainty regarding whether the true $Y(a)$ is $0$ or $1$ increases. 

Moreover, it may be natural to consider  an alternative coefficient estimation strategy to gain efficiency without requiring the conditions mentioned above for a general GLM. Revisiting \eqref{eq:varDiff}, for $a=0,1$, write
\begin{align*}
    I_a & = \Ex\{g(\mb X'\bd\beta_a)[2Y(a)-2g(\mb X'\bd\beta_a) + g(\mb X'\bd\beta_a)]\} = 2\Ex\{g(\mb X'\bd\beta_a)[Y(a)-g(\mb X'\bd\beta_a)]\} + \Ex\{g(\mb X'\bd\beta_a)^2\}. 
\end{align*}
Therefore, to ensure $I_a\geq 0$, consider $\widetilde{\bd\beta}_a$ as the solution to the following joint equations
\begin{align}\label{eq:anotherEq}
    \frac1N\sum_{i=1}^N I(A_i=a)g(\mb X_i'\bd\beta_a)[Y_i-g(\mb X_i'\bd\beta_a)]\mb X' =\bd 0. 
\end{align}
Here $\mb X$ should include an intercept term so that setting \eqref{eq:anotherEq} to zero implies $\dfrac1N\displaystyle\sum_{i=1}^N I(A_i=a)g(\mb X_i'\bd\beta_a)[Y_i-g(\mb X_i'\bd\beta_a)]=0$. Therefore, the probability limit of $\widetilde{\bd\beta}_a$, denoted by ${\bd\beta}_a^{**}$, can satisfy $\Ex\{g(\mb X'\bd\beta_a^{**})[Y(a)-g(\mb X'\bd\beta_a^{**})]\}=0$, and thus, for $a=0,1$, we have
\begin{align*}
    I_a = 2\Ex\{g(\mb X'\bd\beta_a^{**})[Y(a)-g(\mb X'\bd\beta_a^{**})]\} + \Ex\{g(\mb X'\bd\beta_a^{**})^2\} = \Ex\{g(\mb X'\bd\beta_a^{**})^2\}\geq 0. 
\end{align*}
However, if we plug this ${\bd\beta}_a^{**}$ into \eqref{eq:zeroVarEq}, in general, $\Ex\{g_{\bd\beta_a}(\mb X'\bd\beta_a^{**})[Y(a)-g(\mb X'\bd\beta_a^{**})]\}\not=0$, meaning that $\widetilde{\bd\beta}_a$ is not the global (optimal) variance minimizer among all $\tauest^g\in\Gset$. Another issue with this approach is that $\widetilde{\bd\beta}_a$ is not the MLE for the regression coefficients of a nonlinear GLM, so it does not guarantee unbiased predictions of the outcomes \citep{van2024automated}. Finally, in practice, solving \eqref{eq:anotherEq} can be unstable. For example, in logistic regression, a $\widetilde{\bd\beta}_a$ with some large elements can make $g(\mb X'\widetilde{\bd\beta}_a)\approx 0$, but this $\widetilde{\bd\beta}_a$ might deviate significantly from the true value, leading to poor statistical properties. For these reasons, we did not incorporate this approach into our framework.

\section{Additional Simulation Details}

\subsection{Additional details of the data generating process}\label{subapp:DGP} 

Recall that  we generate the binary treatment assignment by $A\sim\text{Bern}(0.5)$, and we consider two sets of covariates $\mb X=(X_1,\dots,X_5)'$ and $\mb V=(V_1,\dots,V_{50})'$, where $(X_1,X_2)\sim\mc N_2(\bd\mu_X,\bd\Sigma_X)$ is a bivariate normal distribution with $\bd\mu_X=(0,0)'$, $\bd\Sigma_X=\begin{pmatrix}1 & 0.8 \\ 0.8 & 1\end{pmatrix}$, $X_3\sim\mc N(0,1)$, $X_4\sim t_{df=10}$, and $X_5\sim\text{Bin}(10,0.2)-2$. Additionally, $\mb V\sim\mc N_{50}(\bd\mu_V,\bd\Sigma_V)$, where $\bd\mu_V=(\underbrace{1,\dots,1}_{50\text{ elements}})'$, and $\bd\Sigma_V$ is the correlation matrix of matrix $\bd B$, with 
$$
\bd B = \underbrace{\begin{pmatrix}
    0.10 & 0.10 & \dots & 0.10\\
    0.11 & 0.11 & \dots & 0.11\\
    \vdots & \vdots & \ddots & \vdots \\
    0.59 & 0.59 & \dots & 0.59
\end{pmatrix} }_{50\text{ columns}} + 2\bd I_{50},
$$
where $\bd I_{50}$ is the $50\times 50$ identity matrix. 

We specify the following two models for potential outcomes: 
\begin{align*}
    \text{Continuous outcome: }& Y(a) = 30 + 20\mb X'\bd\beta_{0} + a\delta(\mb X) + \epsilon, \\
    \text{Binary outcome: }& Y(a)\sim \text{Bern}(e(\mb X, a)),\text{ with }e(\mb X, a)=\{1+\exp(-20\mb X'\bd\beta_{0}-a\delta(\mb X))\}^{-1}, 
\end{align*}
and we consider two specifications of $\delta(\mb X)$:
\begin{itemize}
    \item linear: $\delta(\mb X) = d_1 + c_1\mb X\bd\beta_1$, and
    \item nonlinear: $\delta(\mb X) = d_2 + c_2\mb W\bd\beta_1$, with $\mb W = (X_1^2, -X_2^2, \vert X_3\vert, X_4X_5, X_5)$, 
\end{itemize}
where $\bd\beta_0=(1,1,1,1,1)'$, and $\bd\beta_1=(2,4,6,2,4)'$. For continuous outcomes, $(c_1,d_1,c_2,d_2)=(1,8.15,2.92,0)$; for binary outcomes, $(c_1,c_2)=(10,2,2,2)$. The true ATEs, evaluating by the super-population data ($N=10^8$), are approximately $\tau = 8.15$ for both continuous outcomes and $\tau=0.08$ for both binary outcomes. 

\subsection{Additional simulation results of methods performance}

\begin{figure}[H]
    \centering
    \includegraphics[width=0.8\linewidth]{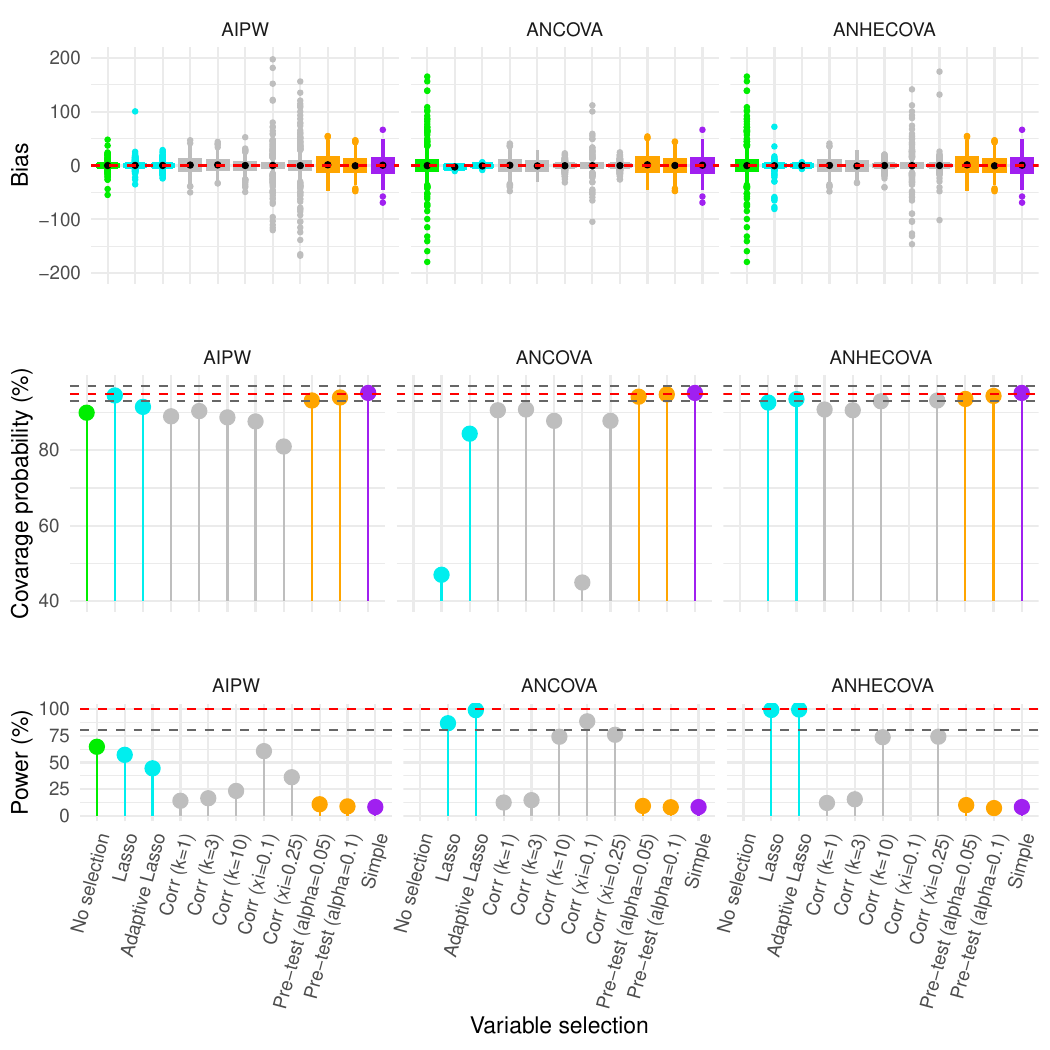}
    \caption{Simulation results under continuous outcome, linear $\delta(\mb X)$ and $N=40$. In the CP\% plots, the \textcolor{red}{red dashed} line indicates 95\% coverage level, and the two \textcolor{gray}{gray dashed} lines indicate 93\% and 97\% coverage levels. In the power plots, the \textcolor{red}{red dashed} line indicates 100\% power, and the \textcolor{gray}{gray dashed} line indicates 80\% power. }
    \label{fig:lin40_cont}
\end{figure}

\begin{figure}[H]
    \centering
    \includegraphics[width=0.8\linewidth]{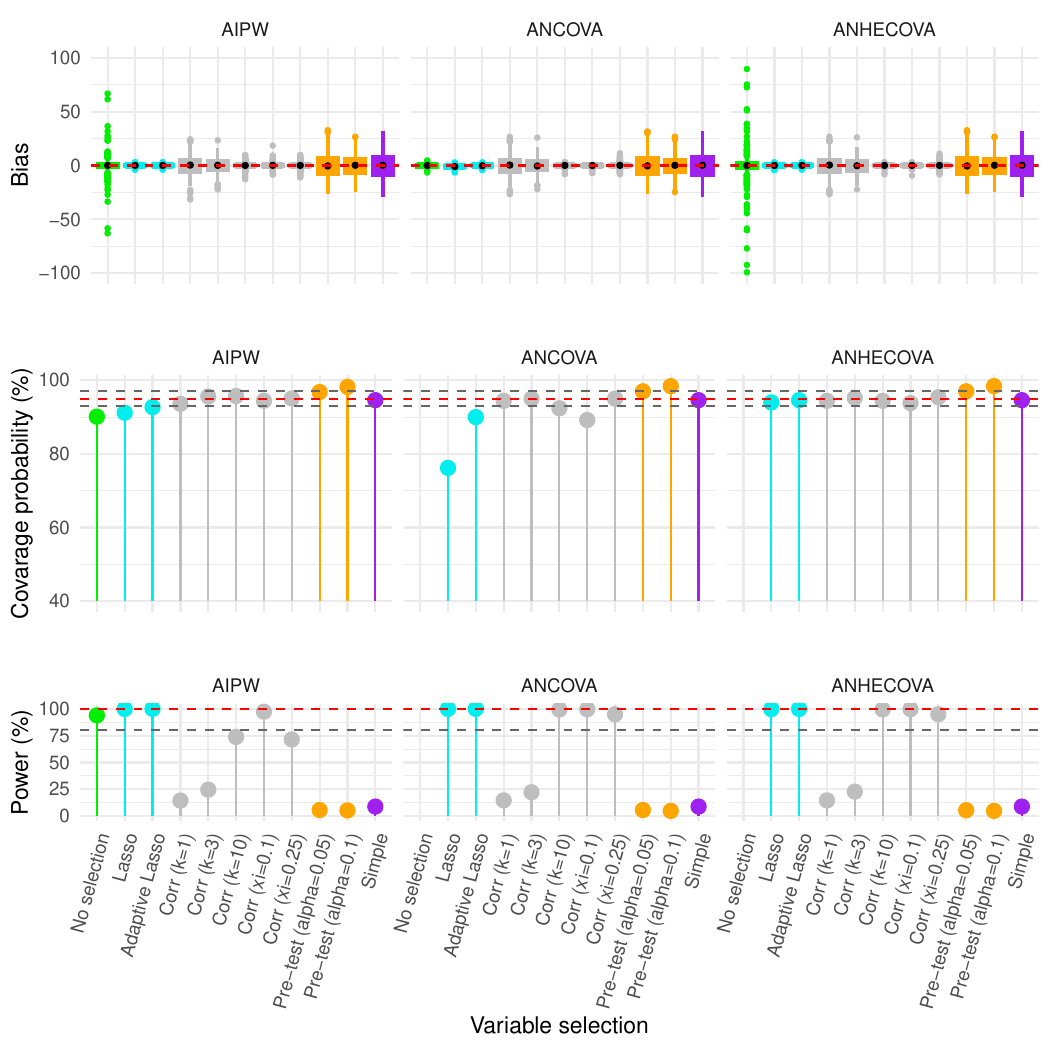}
    \caption{Simulation results under continuous outcome, linear $\delta(\mb X)$ and $N=100$. In the CP\% plots, the \textcolor{red}{red dashed} line indicates 95\% coverage level, and the two \textcolor{gray}{gray dashed} lines indicate 93\% and 97\% coverage levels. In the power plots, the \textcolor{red}{red dashed} line indicates 100\% power, and the \textcolor{gray}{gray dashed} line indicates 80\% power. }
    \label{fig:lin100_cont}
\end{figure}

\begin{figure}[H]
    \centering
    \includegraphics[width=0.8\linewidth]{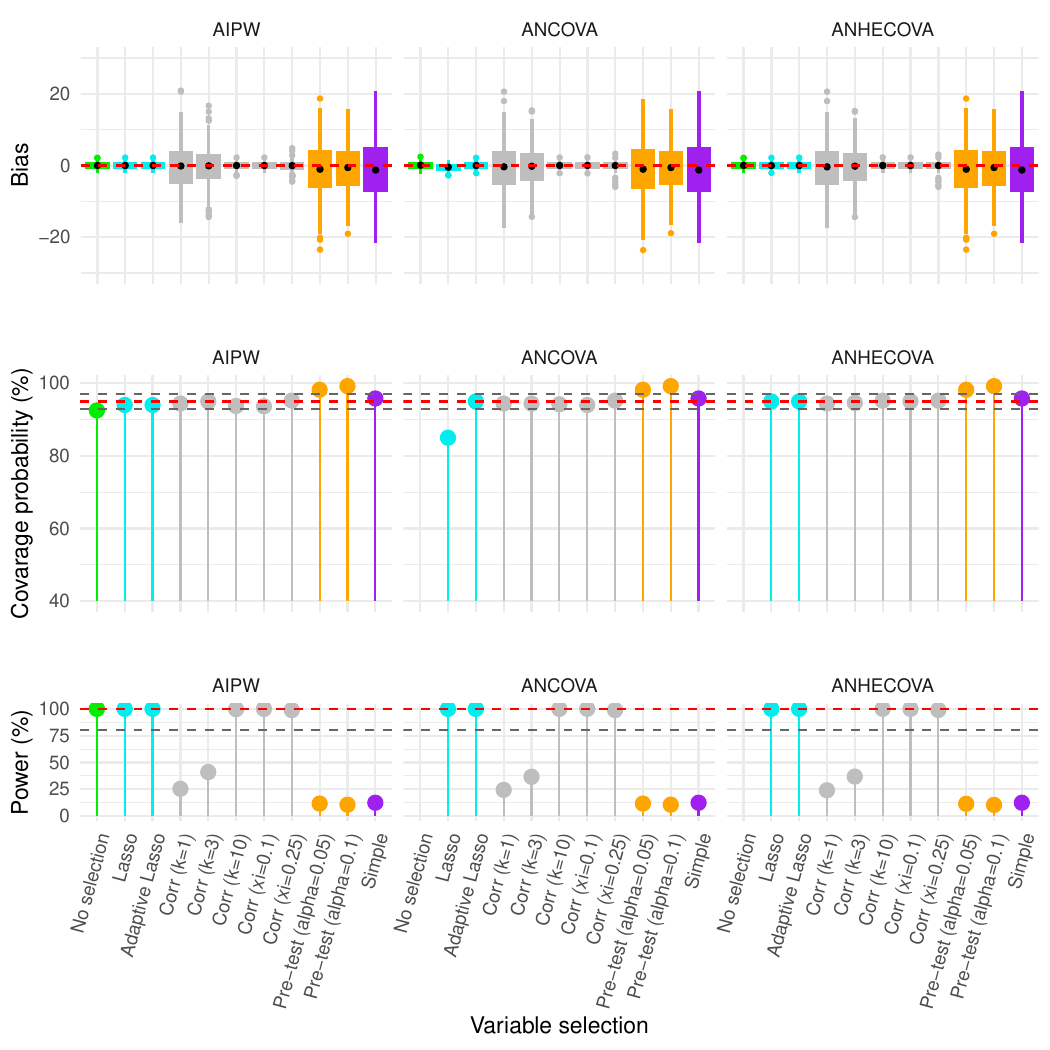}
    \caption{Simulation results under continuous outcome, linear $\delta(\mb X)$ and $N=200$. In the CP\% plots, the \textcolor{red}{red dashed} line indicates 95\% coverage level, and the two \textcolor{gray}{gray dashed} lines indicate 93\% and 97\% coverage levels. In the power plots, the \textcolor{red}{red dashed} line indicates 100\% power, and the \textcolor{gray}{gray dashed} line indicates 80\% power. }
    \label{fig:lin200_cont}
\end{figure}

\begin{figure}
    \centering
    \includegraphics[width=0.8\linewidth]{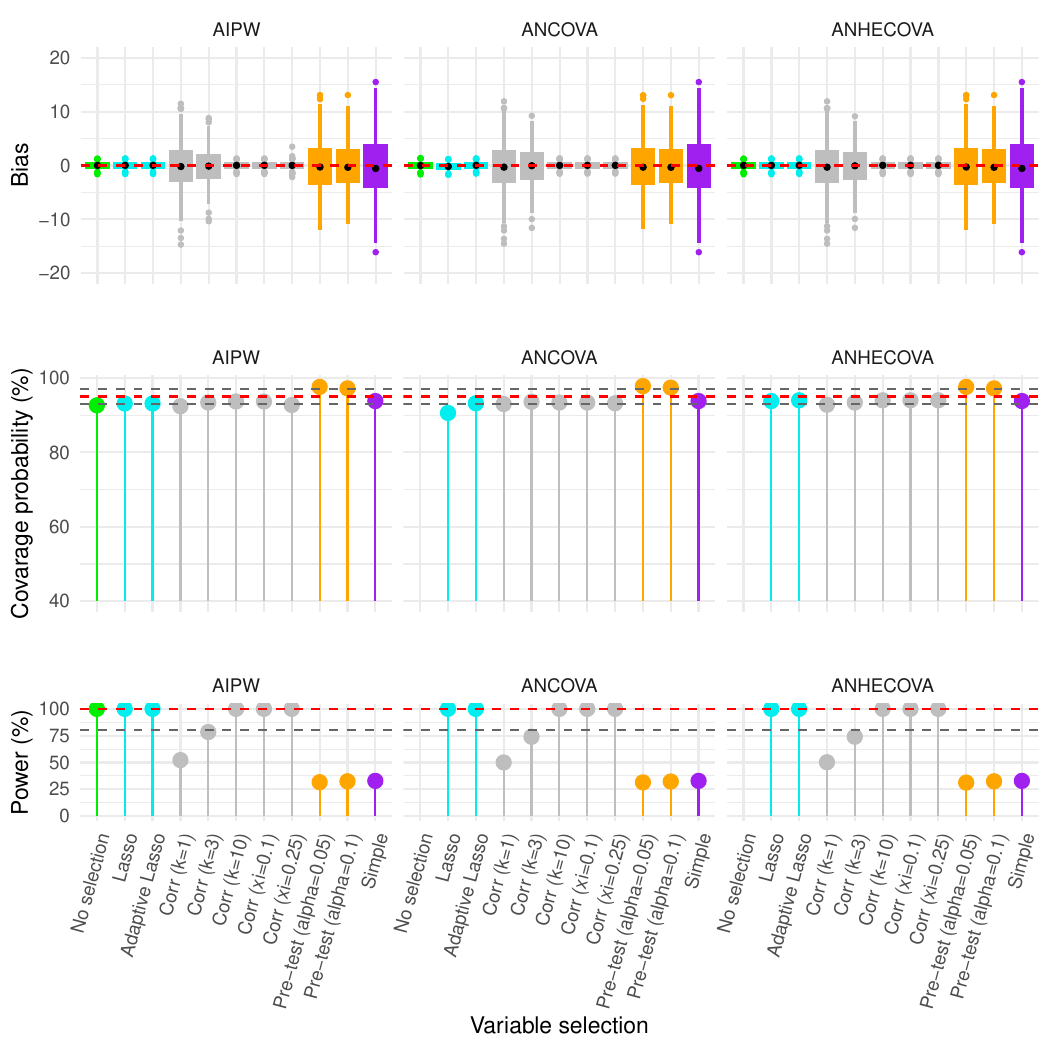}
    \caption{Simulation results under continuous outcome, linear $\delta(\mb X)$ and $N=500$. In the CP\% plots, the \textcolor{red}{red dashed} line indicates 95\% coverage level, and the two \textcolor{gray}{gray dashed} lines indicate 93\% and 97\% coverage levels. In the power plots, the \textcolor{red}{red dashed} line indicates 100\% power, and the \textcolor{gray}{gray dashed} line indicates 80\% power.  }
    \label{fig:lin500_cont}
\end{figure}

\begin{figure}
    \centering
    \includegraphics[width=0.8\linewidth]{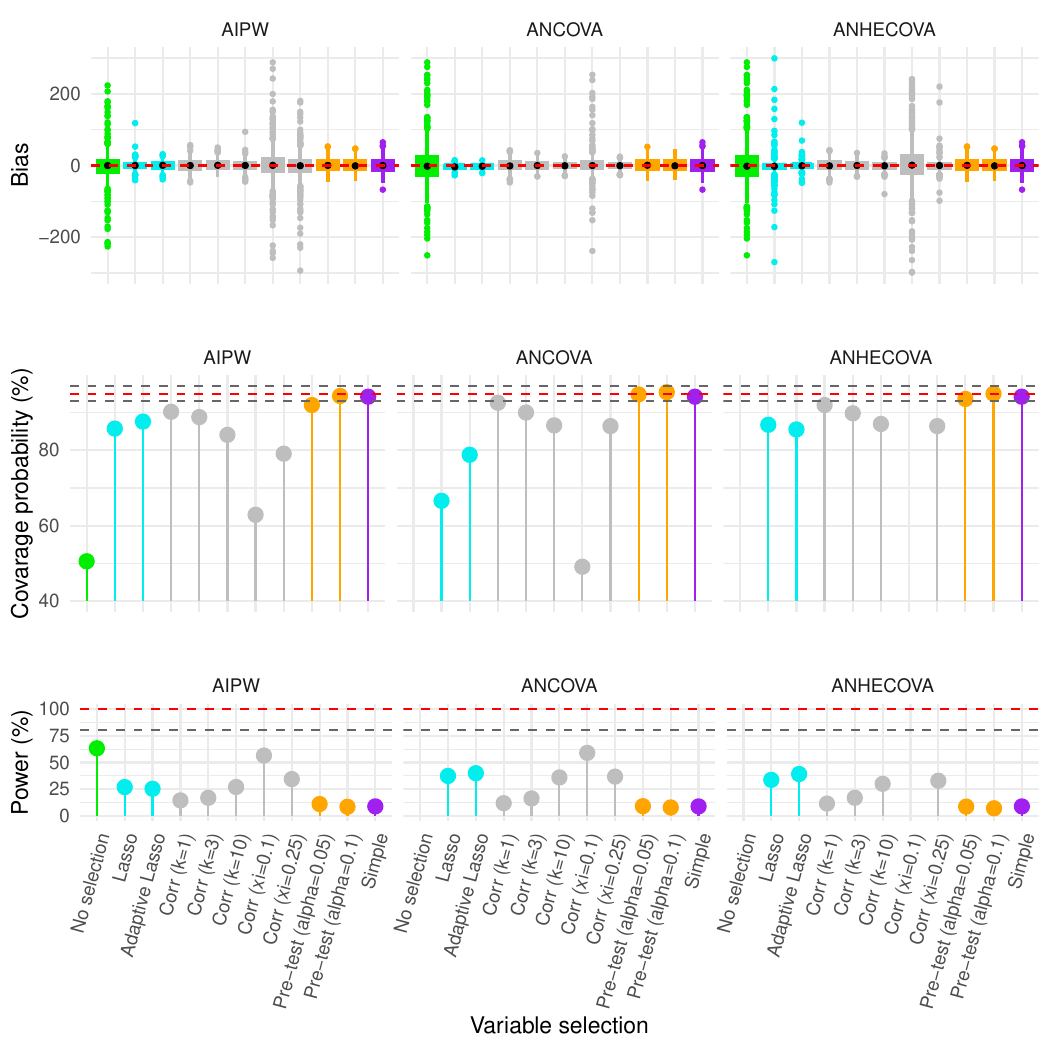}
    \caption{Simulation results under continuous outcome, nonlinear $\delta(\mb X)$ and $N=40$. In the CP\% plots, the \textcolor{red}{red dashed} line indicates 95\% coverage level, and the two \textcolor{gray}{gray dashed} lines indicate 93\% and 97\% coverage levels. In the power plots, the \textcolor{red}{red dashed} line indicates 100\% power, and the \textcolor{gray}{gray dashed} line indicates 80\% power. }
    \label{fig:nonlin40_cont}
\end{figure}

\begin{figure}
    \centering
    \includegraphics[width=0.8\linewidth]{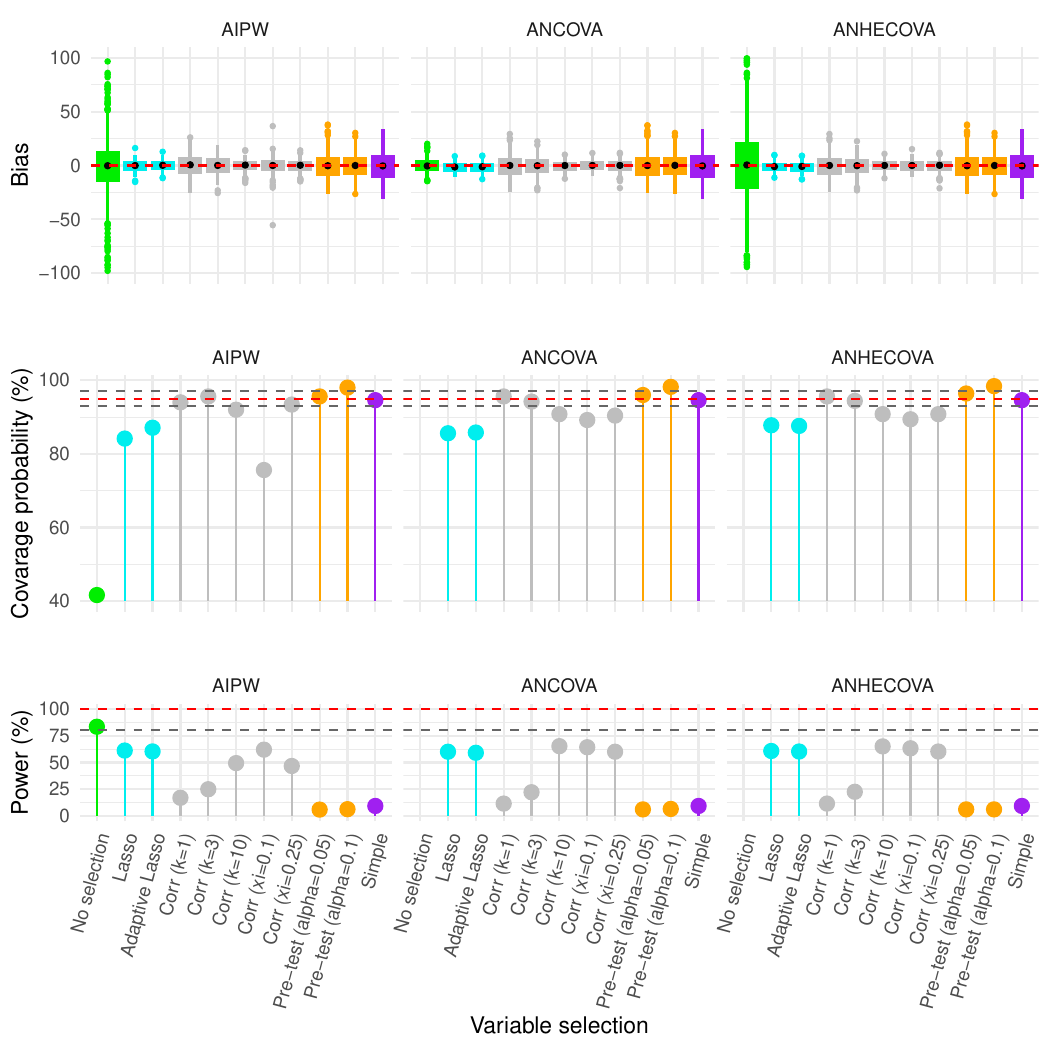}
    \caption{Simulation results under continuous outcome, nonlinear $\delta(\mb X)$ and $N=100$. In the CP\% plots, the \textcolor{red}{red dashed} line indicates 95\% coverage level, and the two \textcolor{gray}{gray dashed} lines indicate 93\% and 97\% coverage levels. In the power plots, the \textcolor{red}{red dashed} line indicates 100\% power, and the \textcolor{gray}{gray dashed} line indicates 80\% power. }
    \label{fig:nonlin100_cont}
\end{figure}

\begin{figure}
    \centering
    \includegraphics[width=0.8\linewidth]{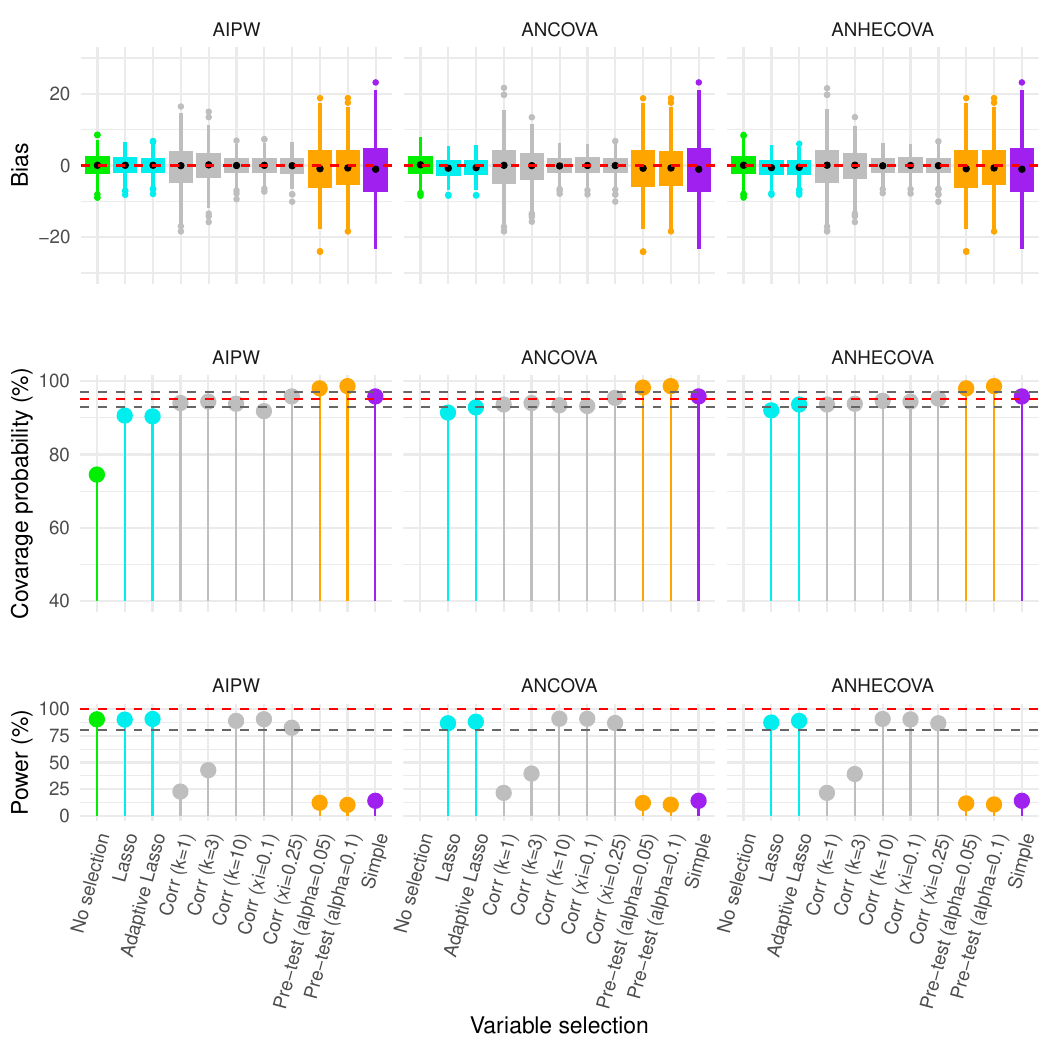}
    \caption{Simulation results under continuous outcome, nonlinear $\delta(\mb X)$ and $N=200$. In the CP\% plots, the \textcolor{red}{red dashed} line indicates 95\% coverage level, and the two \textcolor{gray}{gray dashed} lines indicate 93\% and 97\% coverage levels. In the power plots, the \textcolor{red}{red dashed} line indicates 100\% power, and the \textcolor{gray}{gray dashed} line indicates 80\% power. }
    \label{fig:nonlin200_cont}
\end{figure}

\begin{figure}
    \centering
    \includegraphics[width=0.8\linewidth]{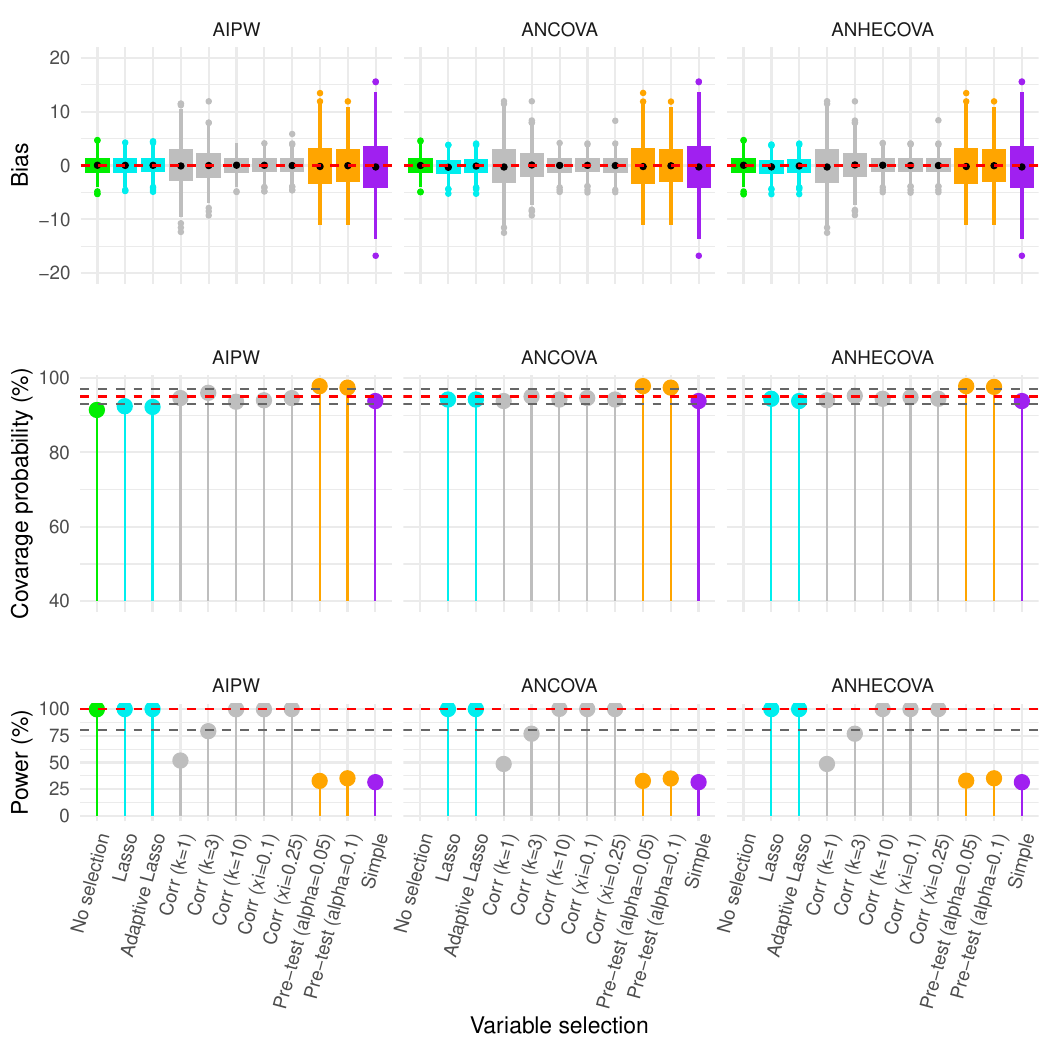}
    \caption{Simulation results under continuous outcome, nonlinear $\delta(\mb X)$ and $N=500$. In the CP\% plots, the \textcolor{red}{red dashed} line indicates 95\% coverage level, and the two \textcolor{gray}{gray dashed} lines indicate 93\% and 97\% coverage levels. In the power plots, the \textcolor{red}{red dashed} line indicates 100\% power, and the \textcolor{gray}{gray dashed} line indicates 80\% power.  }
    \label{fig:nonlin500_cont}
\end{figure}

\begin{figure}
    \centering
    \includegraphics[width=0.8\linewidth]{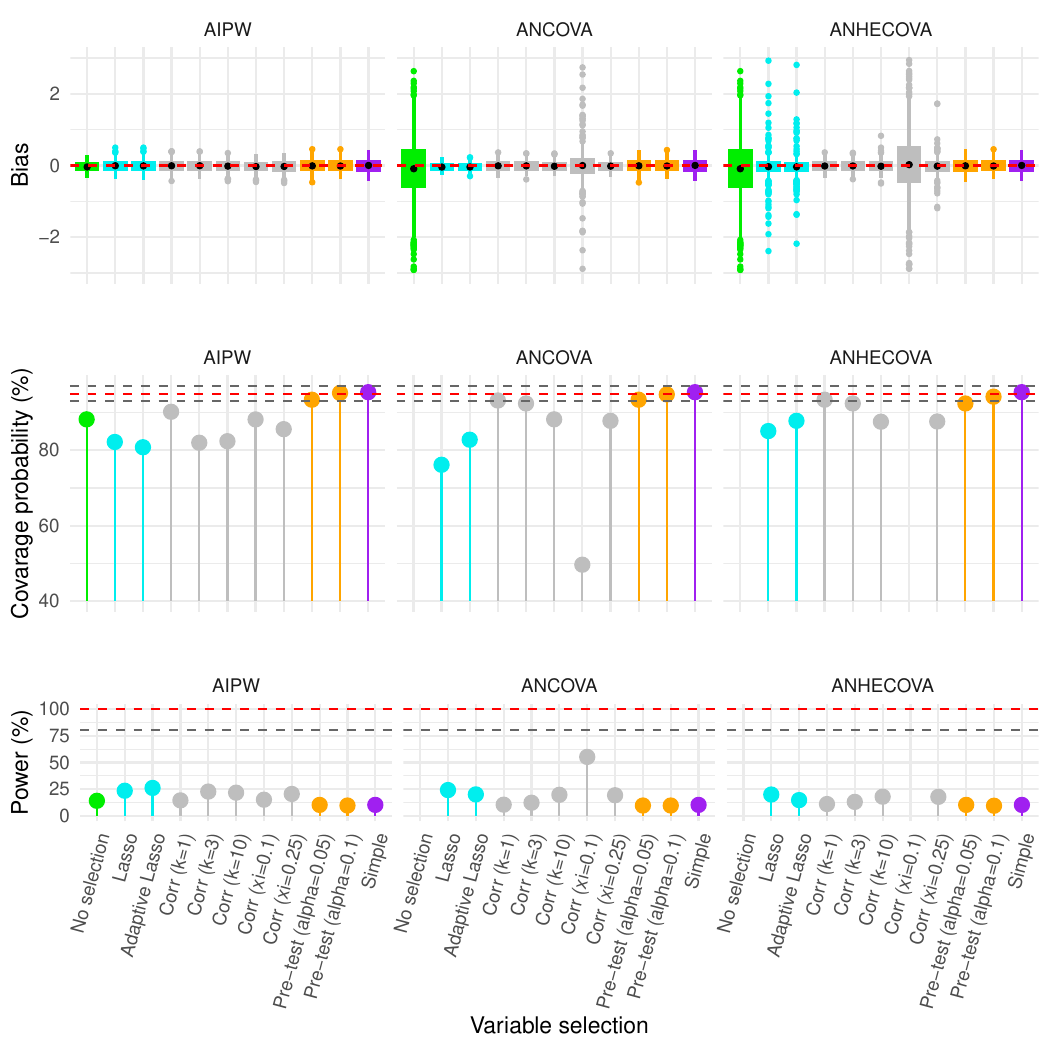}
    \caption{Simulation results under binary outcome, linear $\delta(\mb X)$ and $N=40$. In the CP\% plots, the \textcolor{red}{red dashed} line indicates 95\% coverage level, and the two \textcolor{gray}{gray dashed} lines indicate 93\% and 97\% coverage levels. In the power plots, the \textcolor{red}{red dashed} line indicates 100\% power, and the \textcolor{gray}{gray dashed} line indicates 80\% power. }
    \label{fig:lin40_bin}
\end{figure}

\begin{figure}
    \centering
    \includegraphics[width=0.8\linewidth]{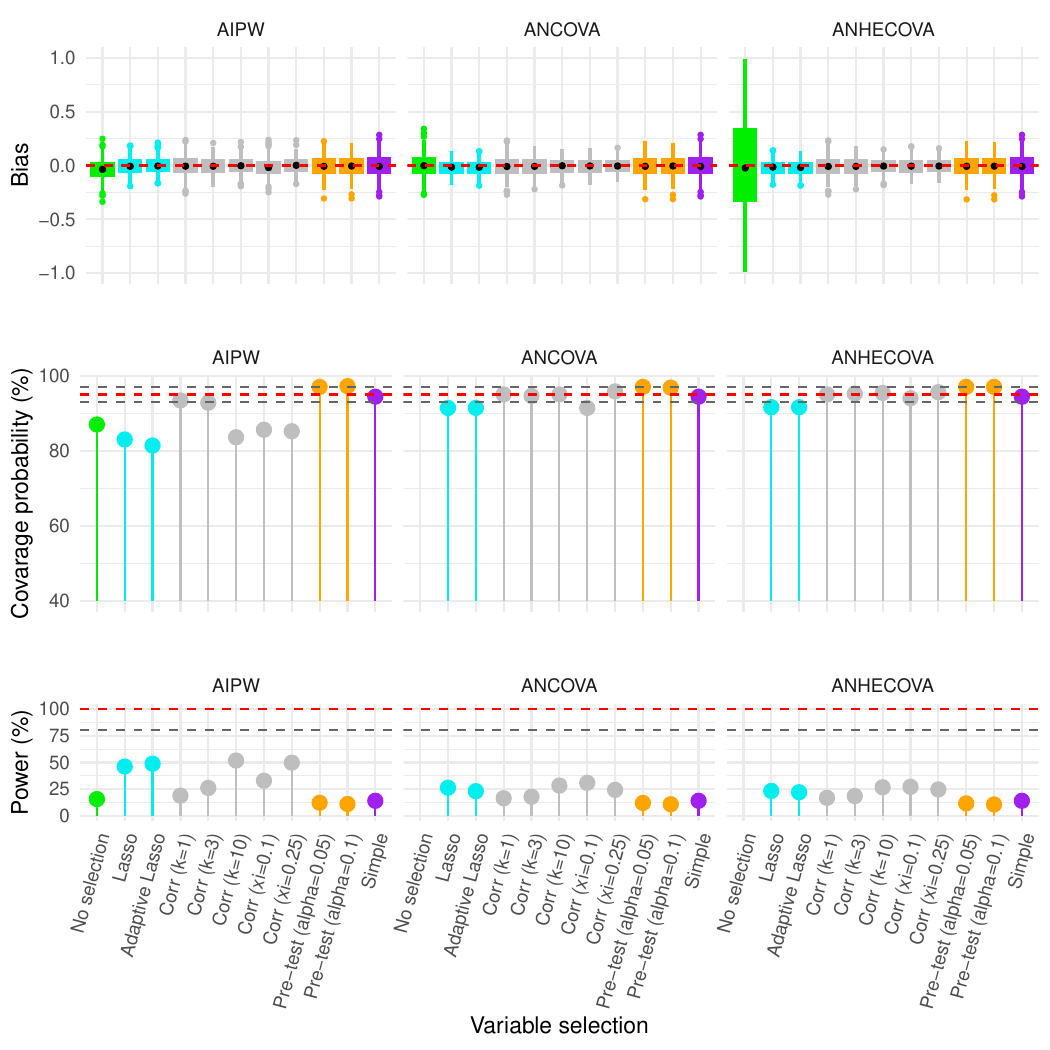}
    \caption{Simulation results under binary outcome, linear $\delta(\mb X)$ and $N=100$. In the CP\% plots, the \textcolor{red}{red dashed} line indicates 95\% coverage level, and the two \textcolor{gray}{gray dashed} lines indicate 93\% and 97\% coverage levels. In the power plots, the \textcolor{red}{red dashed} line indicates 100\% power, and the \textcolor{gray}{gray dashed} line indicates 80\% power. }
    \label{fig:lin100_bin}
\end{figure}

\begin{figure}
    \centering
    \includegraphics[width=0.8\linewidth]{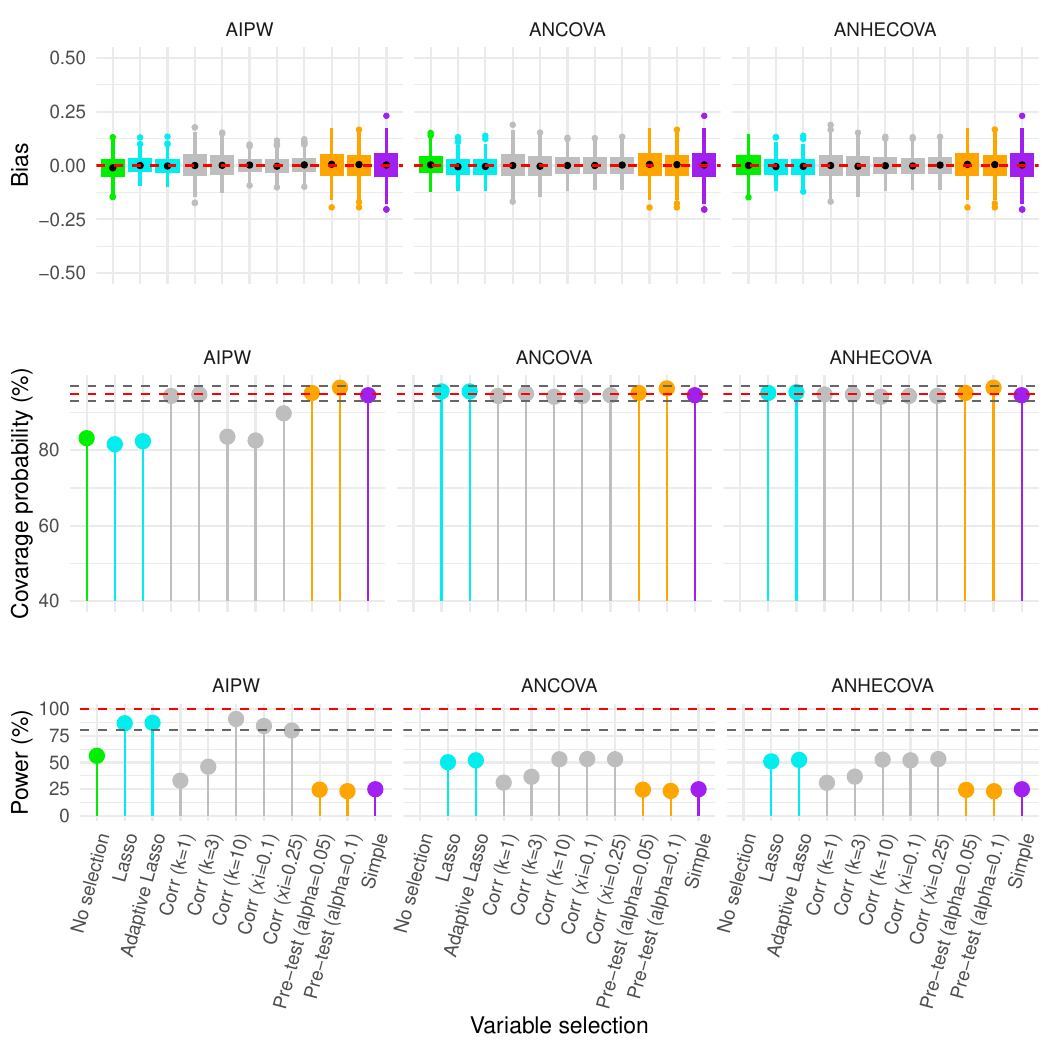}
    \caption{Simulation results under binary outcome, linear $\delta(\mb X)$ and $N=200$. In the CP\% plots, the \textcolor{red}{red dashed} line indicates 95\% coverage level, and the two \textcolor{gray}{gray dashed} lines indicate 93\% and 97\% coverage levels. In the power plots, the \textcolor{red}{red dashed} line indicates 100\% power, and the \textcolor{gray}{gray dashed} line indicates 80\% power. }
    \label{fig:lin200_bin}
\end{figure}

\begin{figure}
    \centering
    \includegraphics[width=0.8\linewidth]{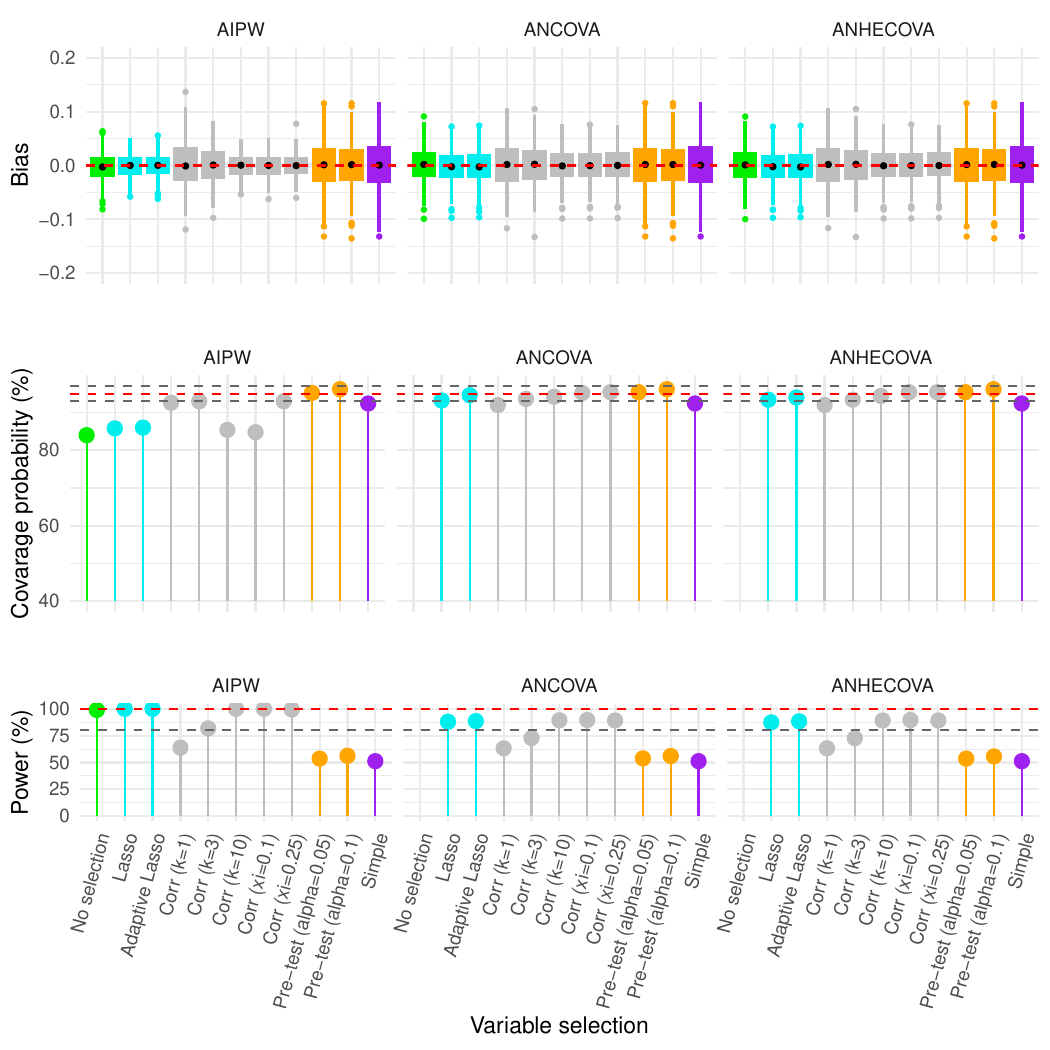}
    \caption{Simulation results under binary outcome, linear $\delta(\mb X)$ and $N=500$. In the CP\% plots, the \textcolor{red}{red dashed} line indicates 95\% coverage level, and the two \textcolor{gray}{gray dashed} lines indicate 93\% and 97\% coverage levels. In the power plots, the \textcolor{red}{red dashed} line indicates 100\% power, and the \textcolor{gray}{gray dashed} line indicates 80\% power.  }
    \label{fig:lin500_bin}
\end{figure}

\begin{figure}
    \centering
    \includegraphics[width=0.8\linewidth]{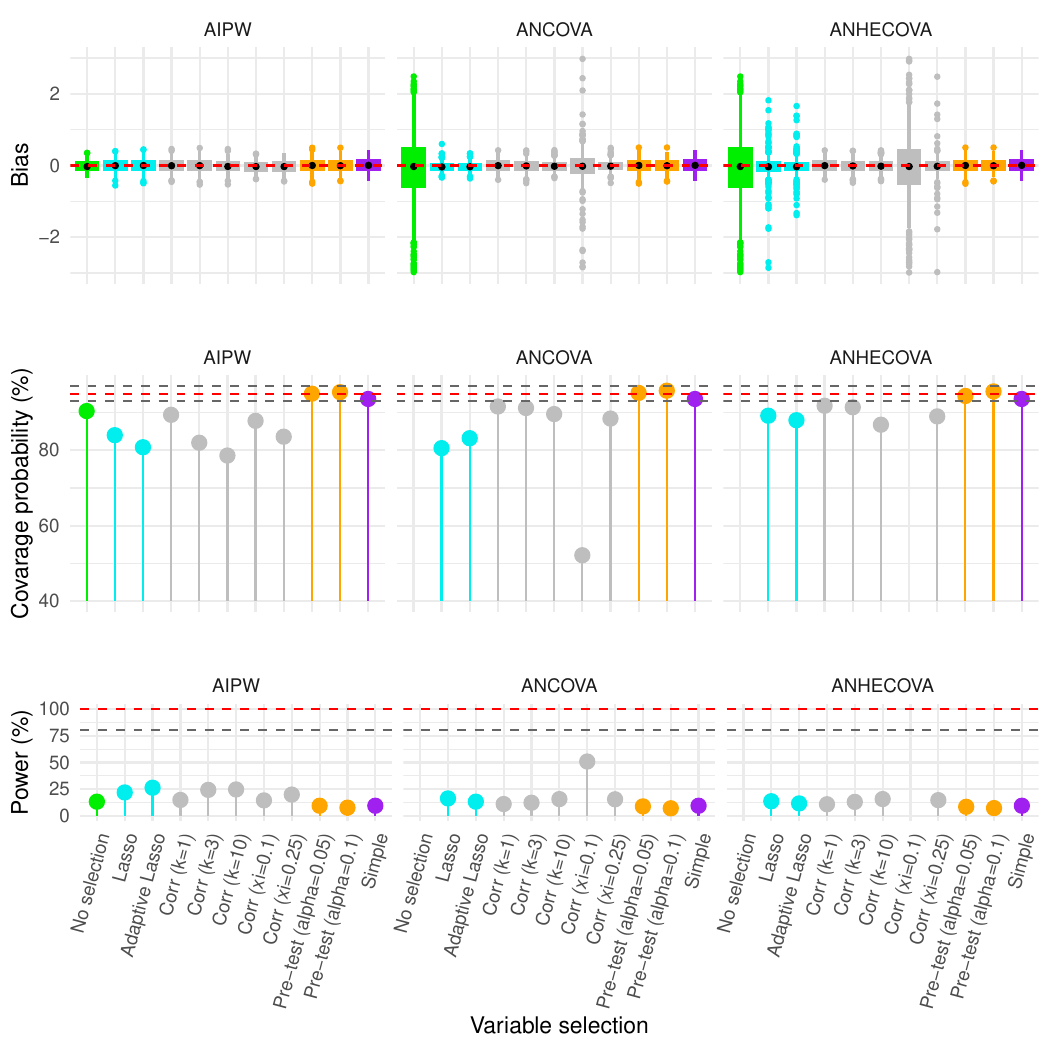}
    \caption{Simulation results under binary outcome, nonlinear $\delta(\mb X)$ and $N=40$. In the CP\% plots, the \textcolor{red}{red dashed} line indicates 95\% coverage level, and the two \textcolor{gray}{gray dashed} lines indicate 93\% and 97\% coverage levels. In the power plots, the \textcolor{red}{red dashed} line indicates 100\% power, and the \textcolor{gray}{gray dashed} line indicates 80\% power. }
    \label{fig:nonlin40_bin}
\end{figure}

\begin{figure}
    \centering
    \includegraphics[width=0.8\linewidth]{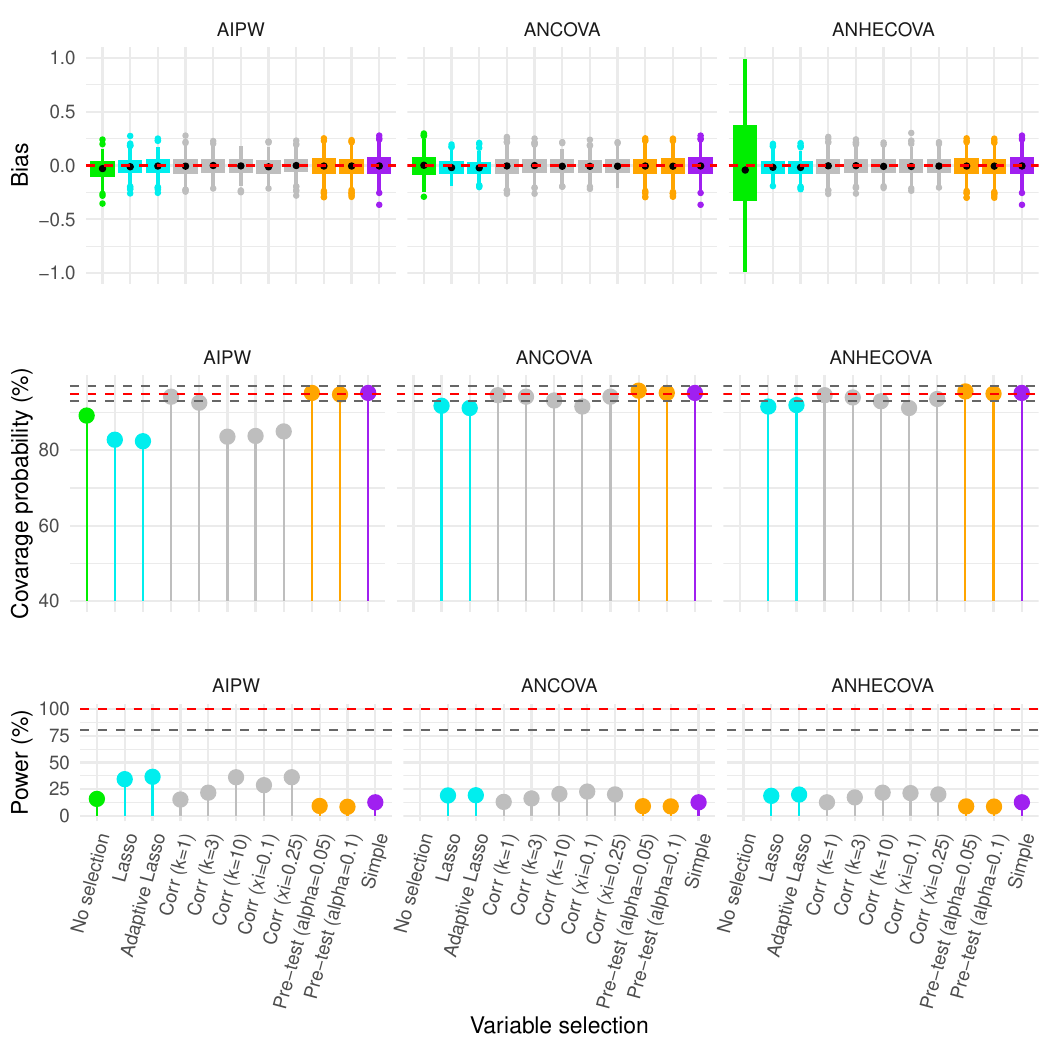}
    \caption{Simulation results under binary outcome, nonlinear $\delta(\mb X)$ and $N=100$. In the CP\% plots, the \textcolor{red}{red dashed} line indicates 95\% coverage level, and the two \textcolor{gray}{gray dashed} lines indicate 93\% and 97\% coverage levels. In the power plots, the \textcolor{red}{red dashed} line indicates 100\% power, and the \textcolor{gray}{gray dashed} line indicates 80\% power. }
    \label{fig:nonlin100_bin}
\end{figure}

\begin{figure}
    \centering
    \includegraphics[width=0.8\linewidth]{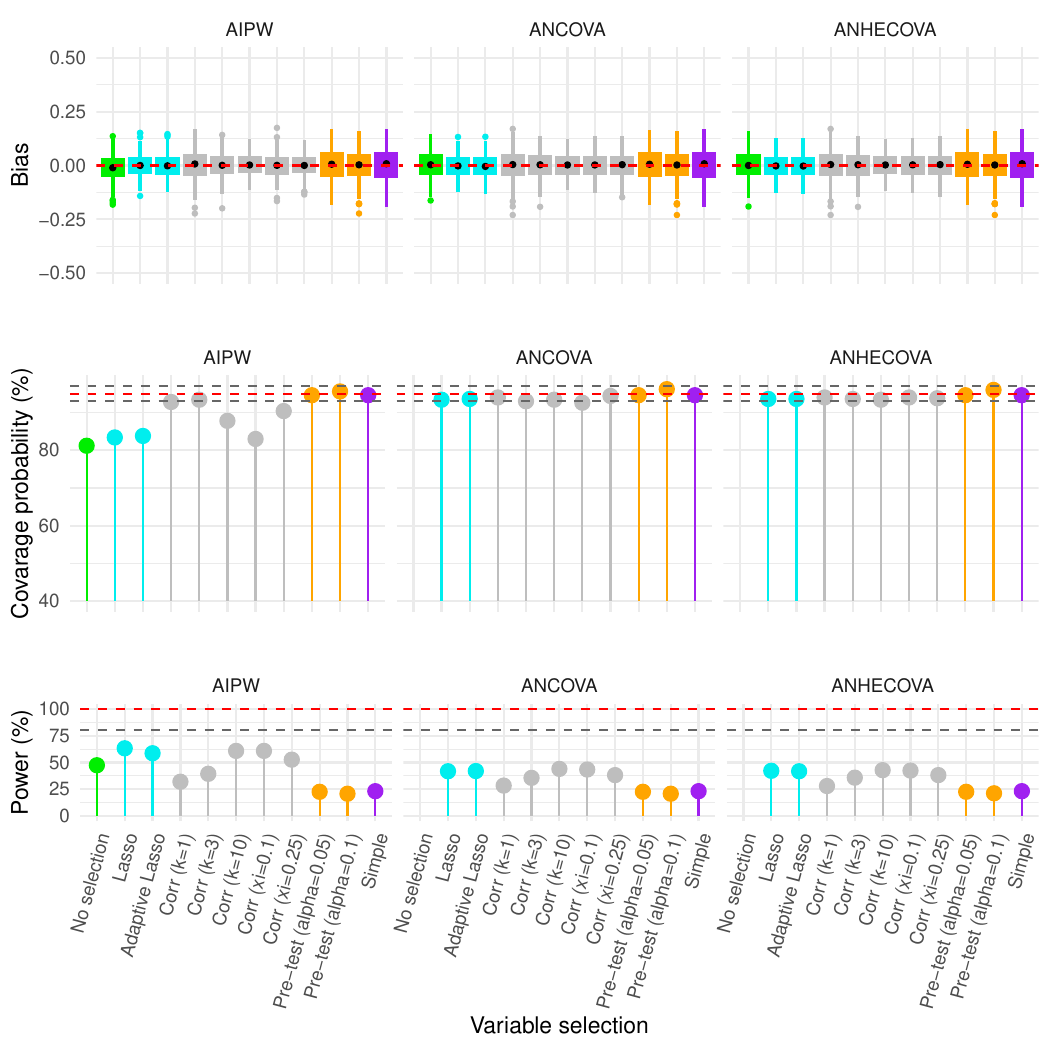}
    \caption{Simulation results under binary outcome, nonlinear $\delta(\mb X)$ and $N=200$. In the CP\% plots, the \textcolor{red}{red dashed} line indicates 95\% coverage level, and the two \textcolor{gray}{gray dashed} lines indicate 93\% and 97\% coverage levels. In the power plots, the \textcolor{red}{red dashed} line indicates 100\% power, and the \textcolor{gray}{gray dashed} line indicates 80\% power. }
    \label{fig:nonlin200_bin}
\end{figure}

\begin{figure}
    \centering
    \includegraphics[width=0.8\linewidth]{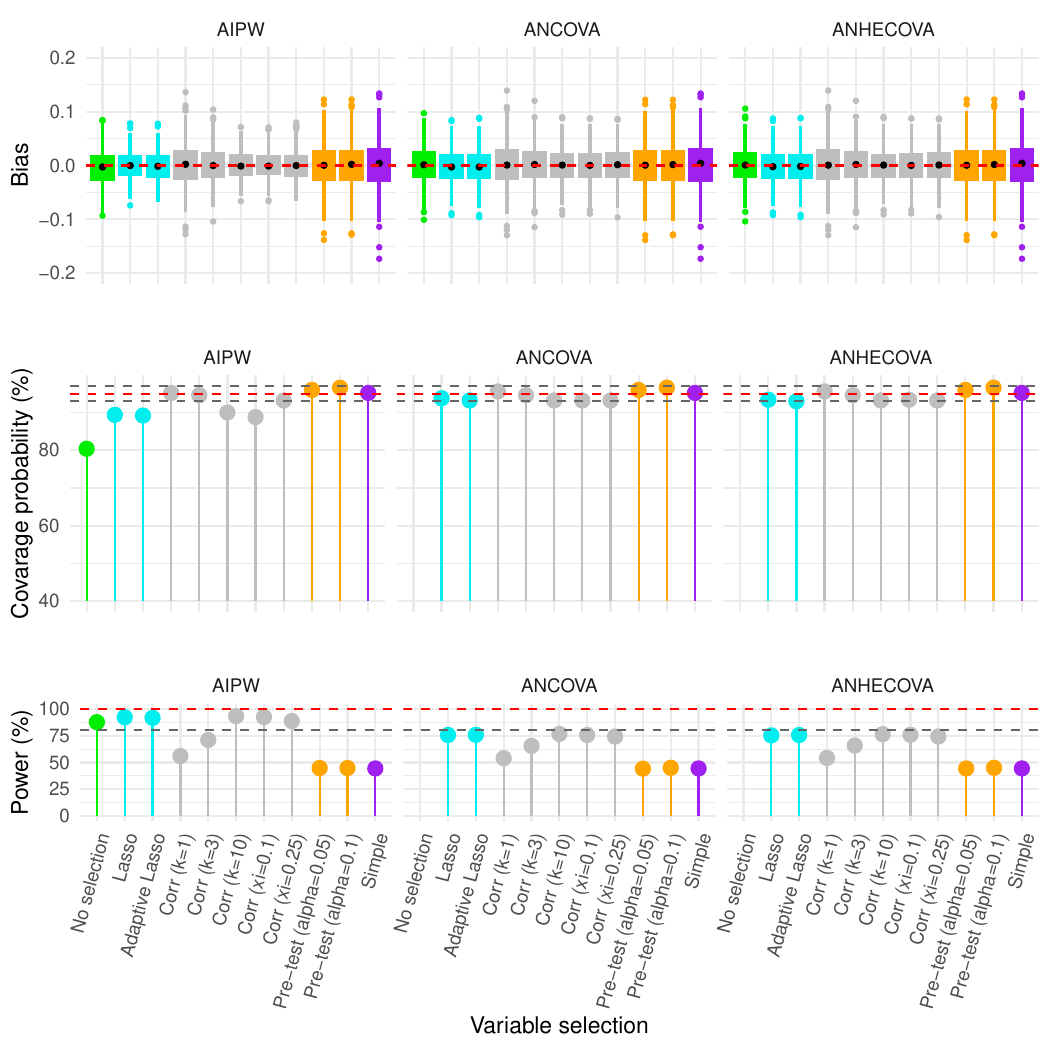}
    \caption{Simulation results under binary outcome, nonlinear $\delta(\mb X)$ and $N=500$. In the CP\% plots, the \textcolor{red}{red dashed} line indicates 95\% coverage level, and the two \textcolor{gray}{gray dashed} lines indicate 93\% and 97\% coverage levels. In the power plots, the \textcolor{red}{red dashed} line indicates 100\% power, and the \textcolor{gray}{gray dashed} line indicates 80\% power.  }
    \label{fig:nonlin500_bin}
\end{figure}

In Figures \ref{fig:lin40_cont}--\ref{fig:nonlin500_bin}, we present full results (Bias, CP\% and Power\%) of simulation in Section 4 of the main text.  

\subsection{Additional simulation results of estimate distributions}

\begin{figure}[H]
    \centering
    \includegraphics[width=0.8\linewidth]{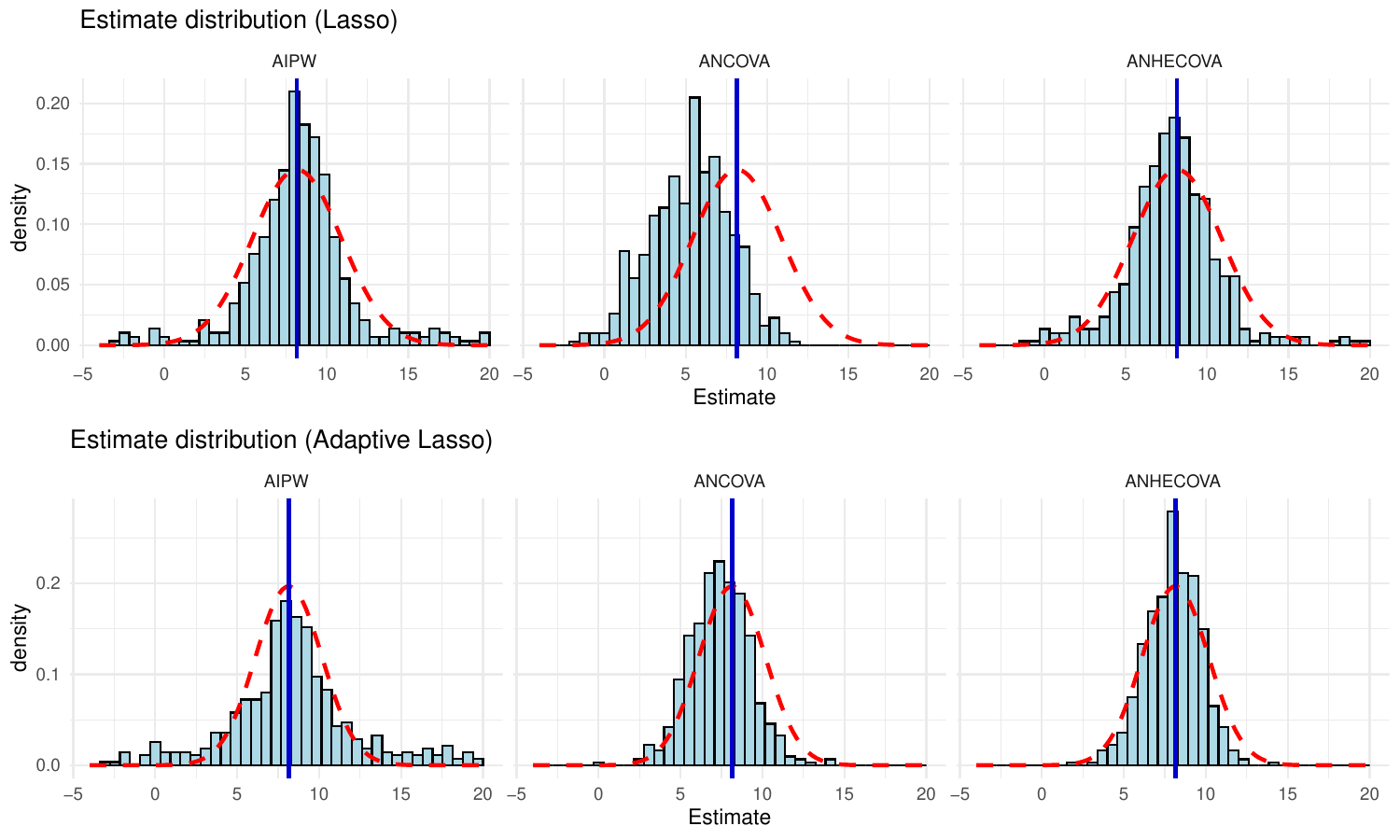}
    \caption{Simulation results for estimation distributions by Lasso and adaptive Lasso variable selections, under continuous outcome, linear $\delta(\mb X)$ and $N=40$. The \textcolor{red}{red dashed curves} are the theoretical normal densities curves, and the \textcolor{blue}{blue lines} indicate the true ATE. }
    \label{fig:est-lin40-cont}
\end{figure}

\begin{figure}[H]
    \centering
    \includegraphics[width=0.8\linewidth]{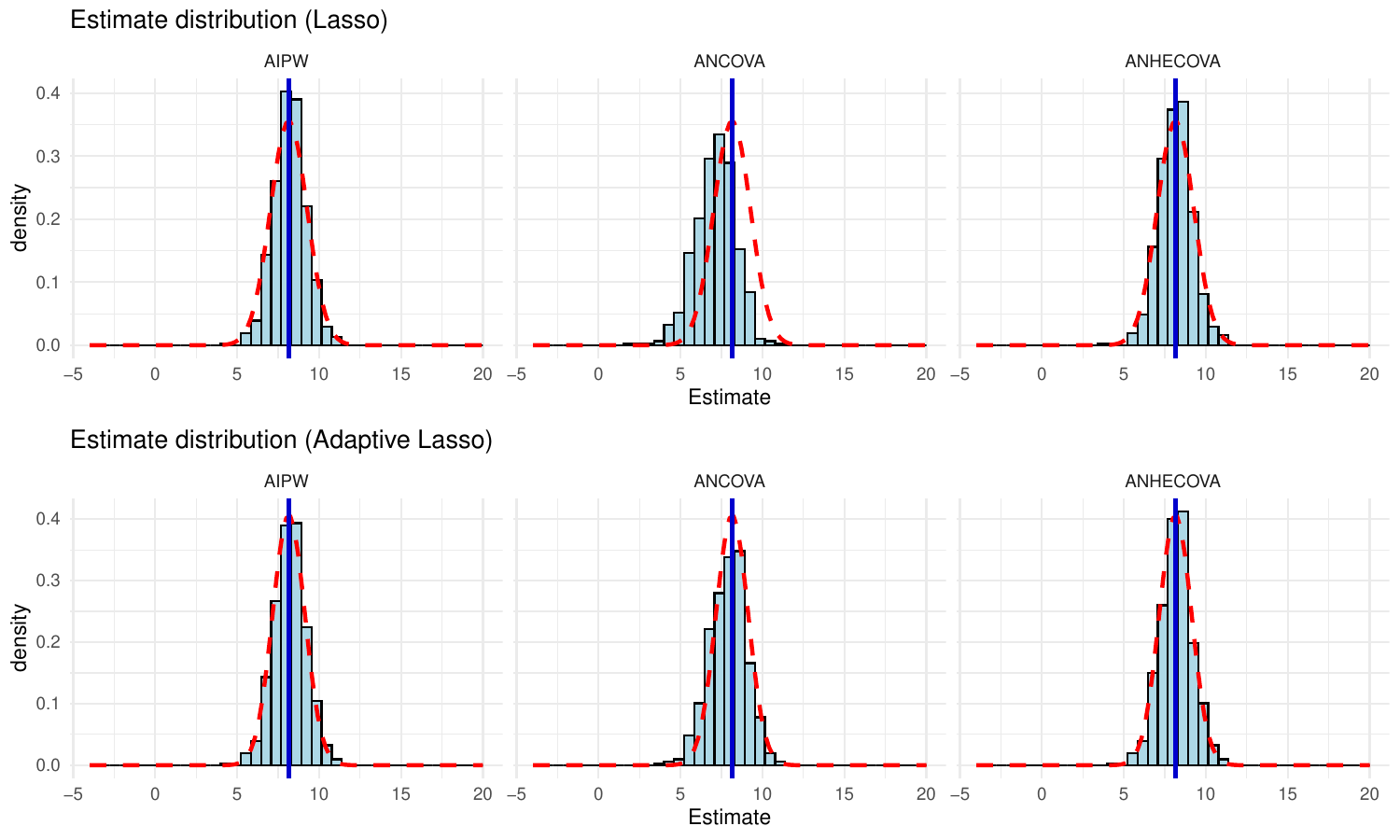}
    \caption{Simulation results for estimation distributions by Lasso and adaptive Lasso variable selections, under continuous outcome, linear $\delta(\mb X)$ and $N=100$. The \textcolor{red}{red dashed curves} are the theoretical normal densities curves, and the \textcolor{blue}{blue lines} indicate the true ATE.  }
    \label{fig:est-lin100-cont}
\end{figure}

\begin{figure}[H]
    \centering
    \includegraphics[width=0.8\linewidth]{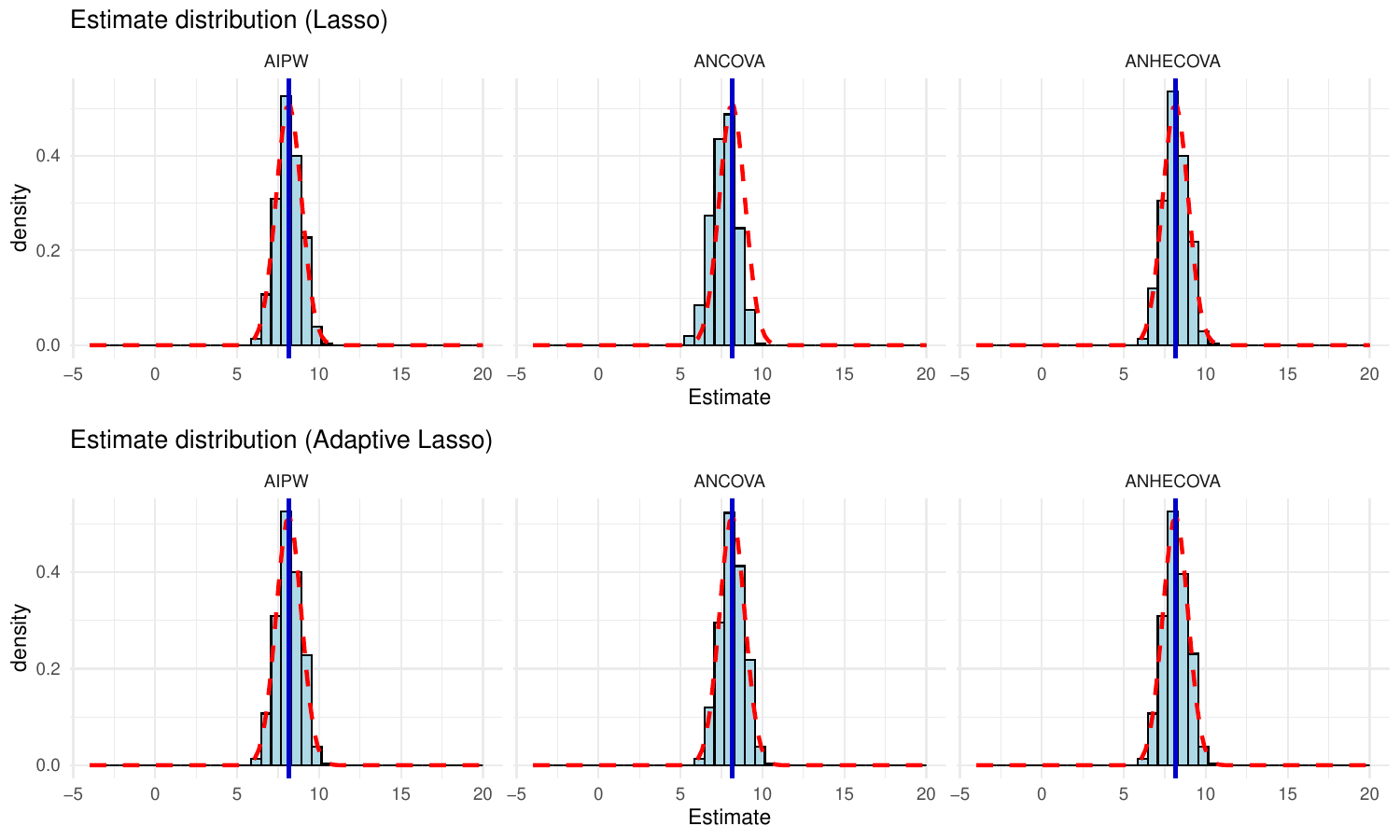}
    \caption{Simulation results for estimation distributions by Lasso and adaptive Lasso variable selections, under continuous outcome, linear $\delta(\mb X)$ and $N=200$. The \textcolor{red}{red dashed curves} are the theoretical normal densities curves, and the \textcolor{blue}{blue lines} indicate the true ATE.  }
    \label{fig:est-lin200-cont}
\end{figure}

\begin{figure}[H]
    \centering
    \includegraphics[width=0.8\linewidth]{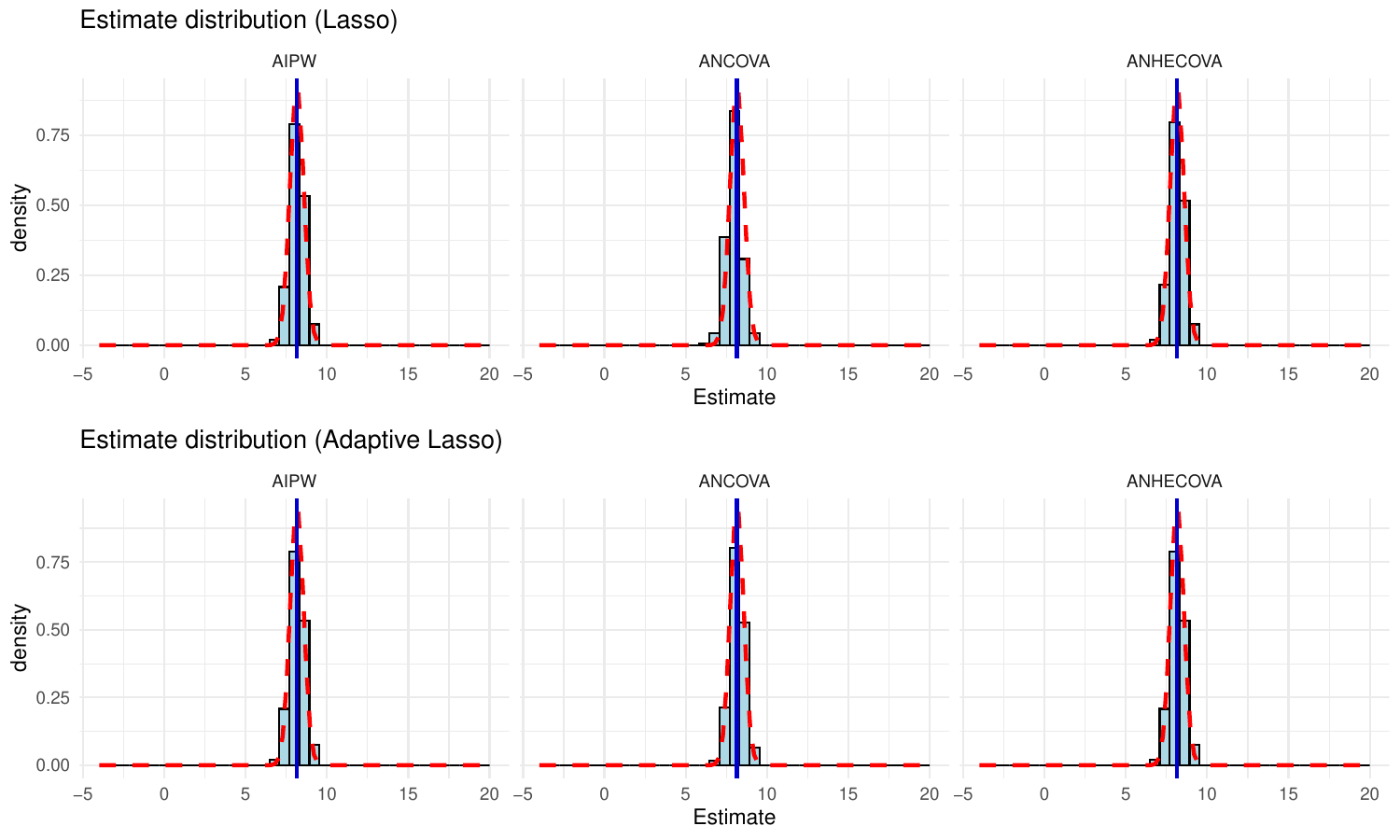}
    \caption{Simulation results for estimation distributions by Lasso and adaptive Lasso variable selections, under continuous outcome, linear $\delta(\mb X)$ and $N=500$. The \textcolor{red}{red dashed curves} are the theoretical normal densities curves, and the \textcolor{blue}{blue lines} indicate the true ATE.  }
    \label{fig:est-lin500-cont}
\end{figure}

\begin{figure}[H]
    \centering
    \includegraphics[width=0.8\linewidth]{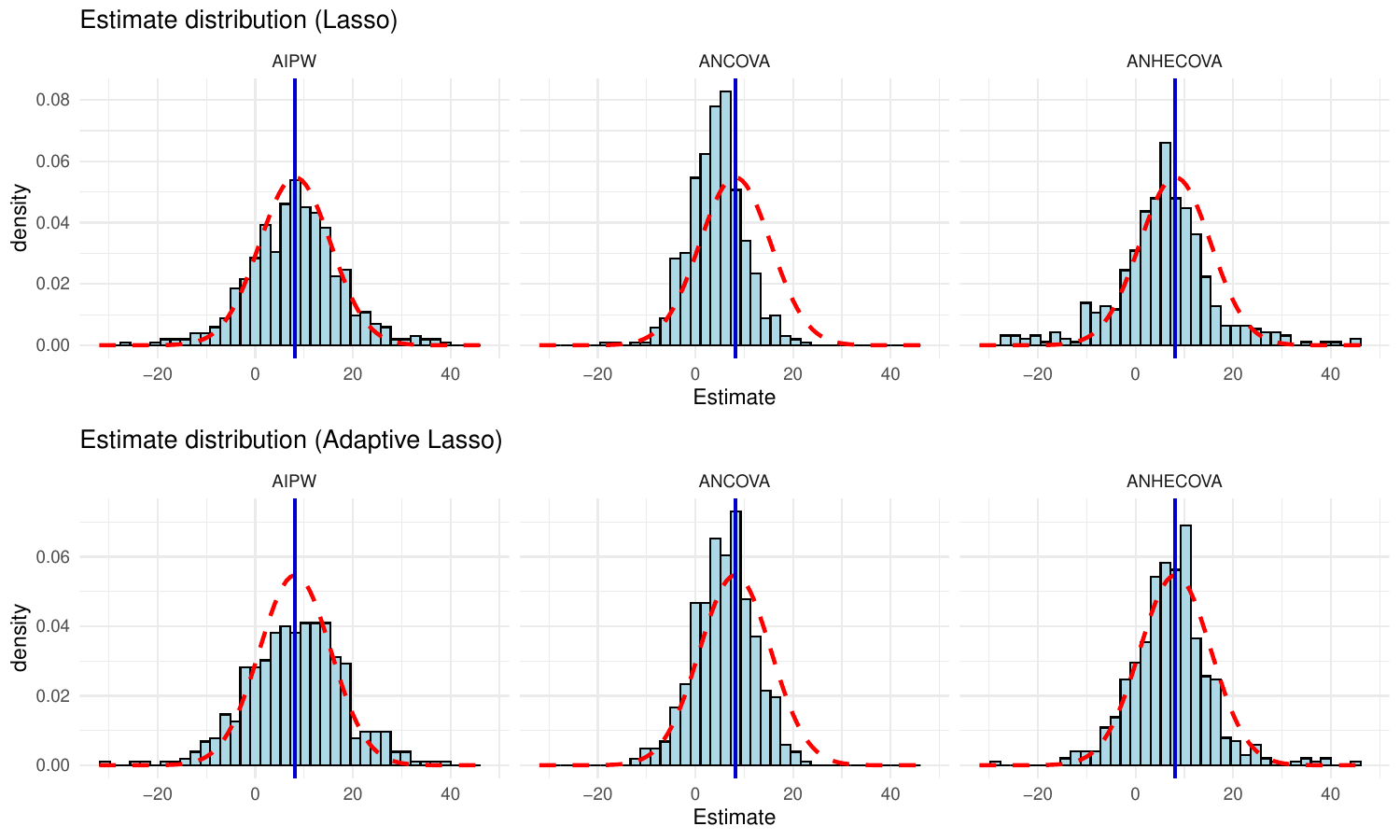}
    \caption{Simulation results for estimation distributions by Lasso and adaptive Lasso variable selections, under continuous outcome, nonlinear $\delta(\mb X)$ and $N=40$. The \textcolor{red}{red dashed curves} are the theoretical normal densities curves, and the \textcolor{blue}{blue lines} indicate the true ATE. }
    \label{fig:est-nonlin40-cont}
\end{figure}

\begin{figure}[H]
    \centering
    \includegraphics[width=0.8\linewidth]{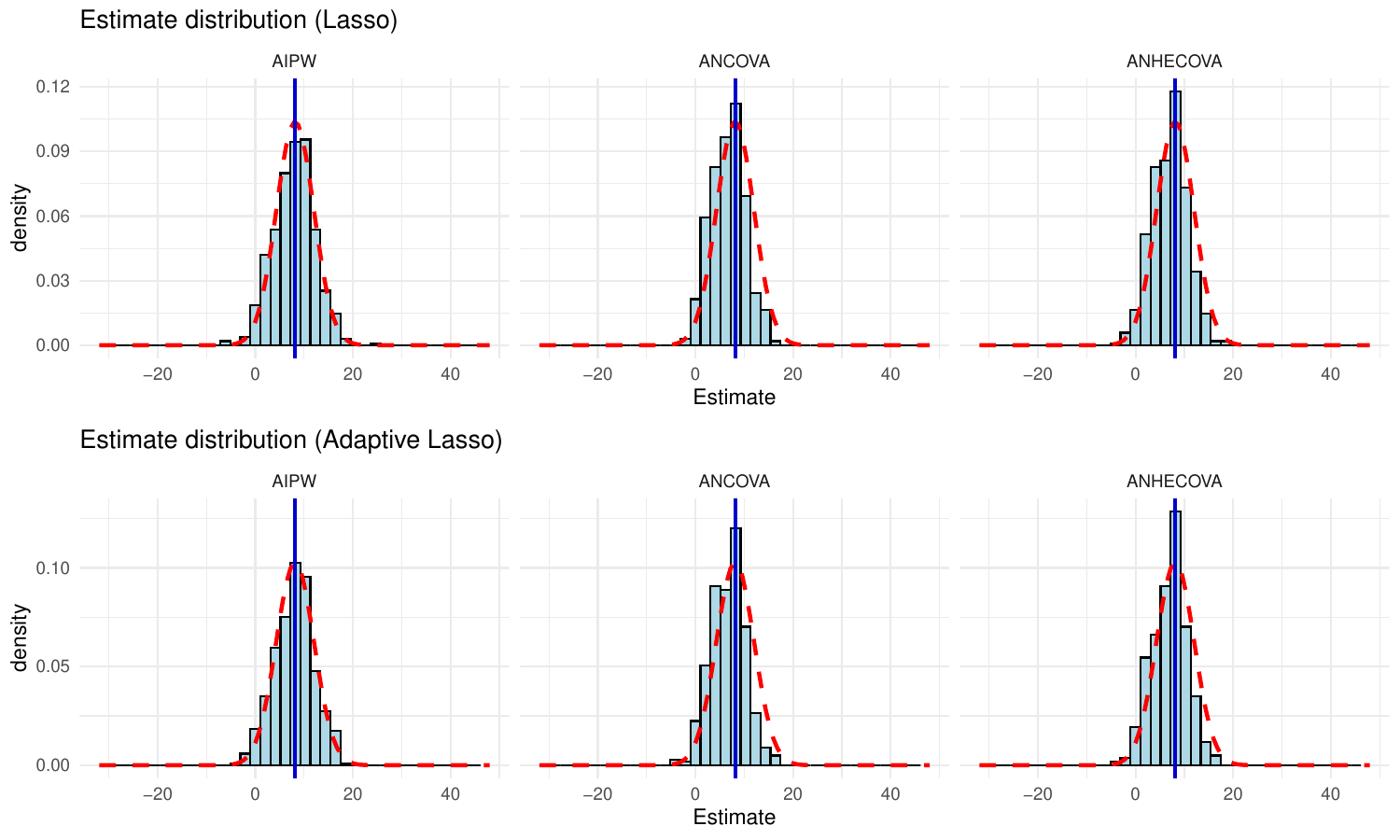}
    \caption{Simulation results for estimation distributions by Lasso and adaptive Lasso variable selections, under continuous outcome, nonlinear $\delta(\mb X)$ and $N=100$. The \textcolor{red}{red dashed curves} are the theoretical normal densities curves, and the \textcolor{blue}{blue lines} indicate the true ATE. }
    \label{fig:est-nonlin100-cont}
\end{figure}

\begin{figure}[H]
    \centering
    \includegraphics[width=0.8\linewidth]{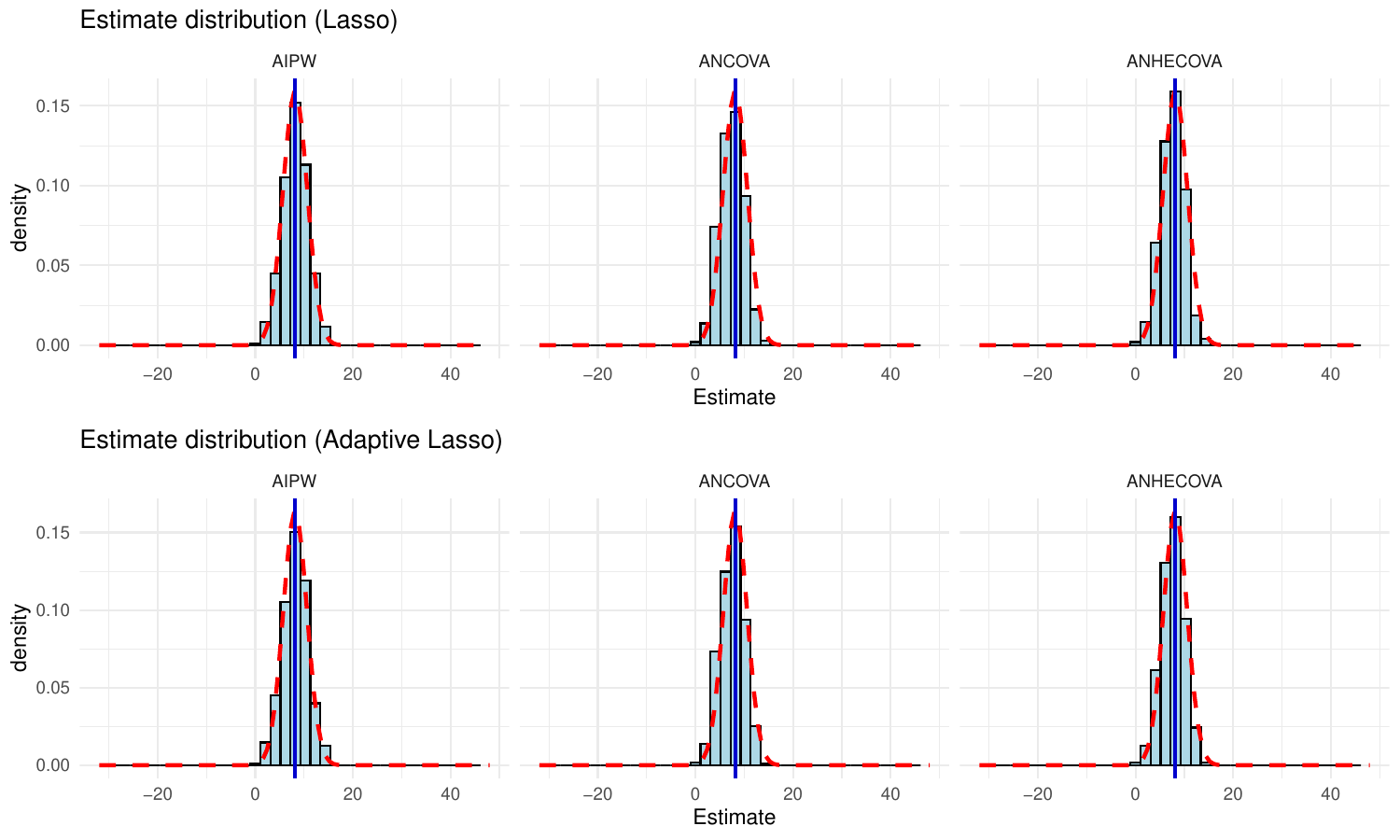}
    \caption{Simulation results for estimation distributions by Lasso and adaptive Lasso variable selections, under continuous outcome, nonlinear $\delta(\mb X)$ and $N=200$. The \textcolor{red}{red dashed curves} are the theoretical normal densities curves, and the \textcolor{blue}{blue lines} indicate the true ATE. }
    \label{fig:est-nonlin200-cont}
\end{figure}

\begin{figure}[H]
    \centering
    \includegraphics[width=0.8\linewidth]{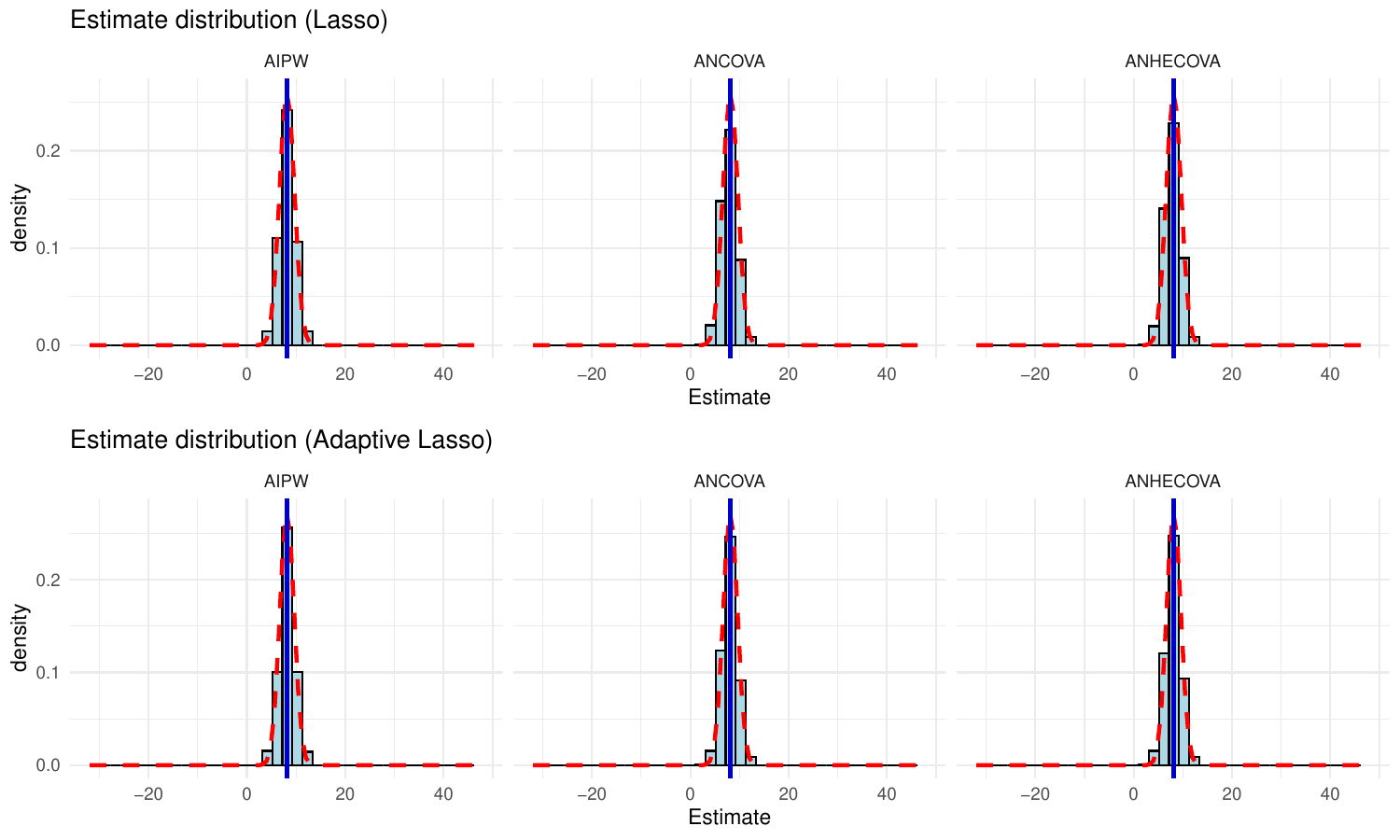}
    \caption{Simulation results for estimation distributions by Lasso and adaptive Lasso variable selections, under continuous outcome, nonlinear $\delta(\mb X)$ and $N=500$. The \textcolor{red}{red dashed curves} are the theoretical normal densities curves, and the \textcolor{blue}{blue lines} indicate the true ATE. }
    \label{fig:est-nonlin500-cont}
\end{figure}

\begin{figure}[H]
    \centering
    \includegraphics[width=0.8\linewidth]{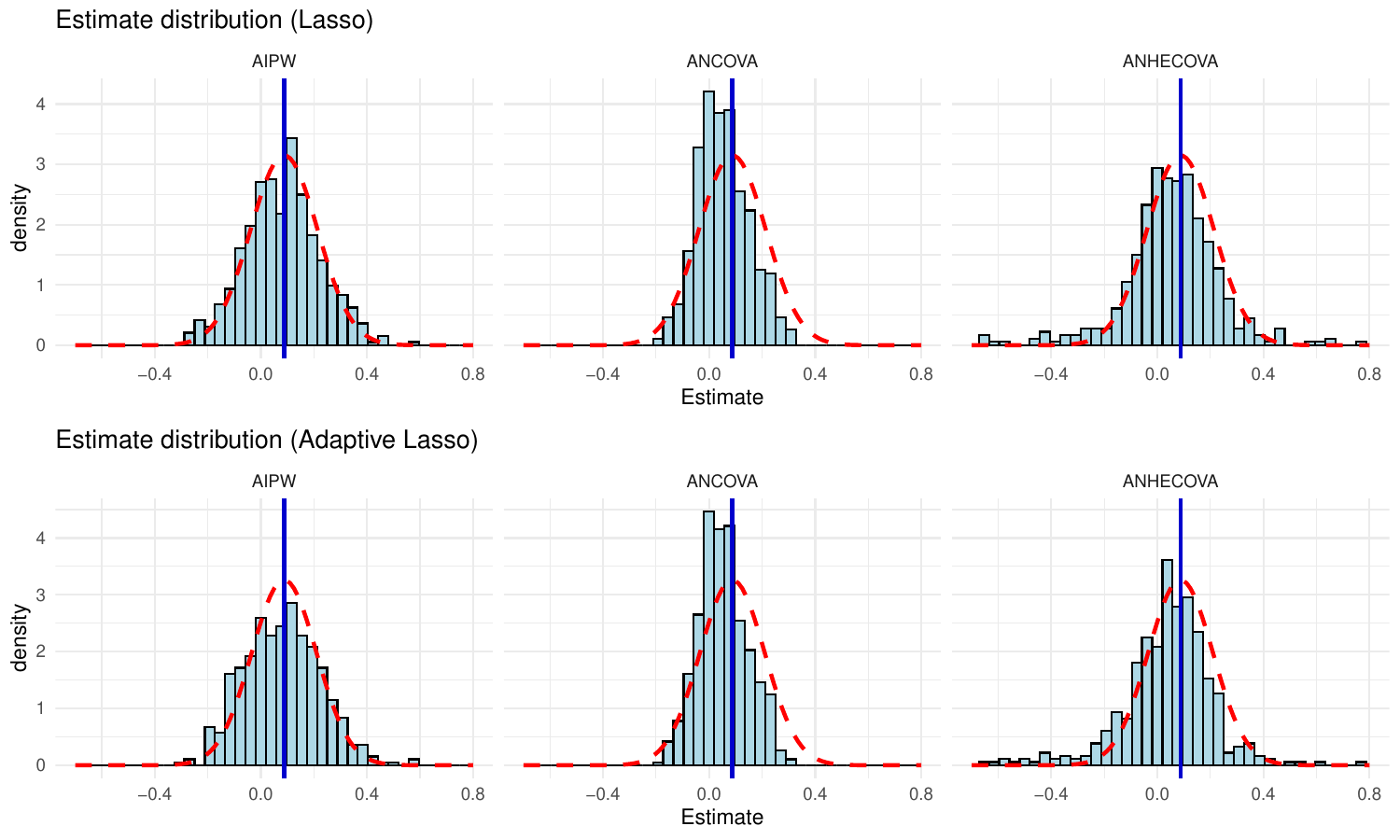}
    \caption{Simulation results for estimation distributions by Lasso and adaptive Lasso variable selections, under binary outcome, linear $\delta(\mb X)$ and $N=40$. The \textcolor{red}{red dashed curves} are the theoretical normal densities curves, and the \textcolor{blue}{blue lines} indicate the true ATE. }
    \label{fig:est-lin40-bin}
\end{figure}

\begin{figure}[H]
    \centering
    \includegraphics[width=0.8\linewidth]{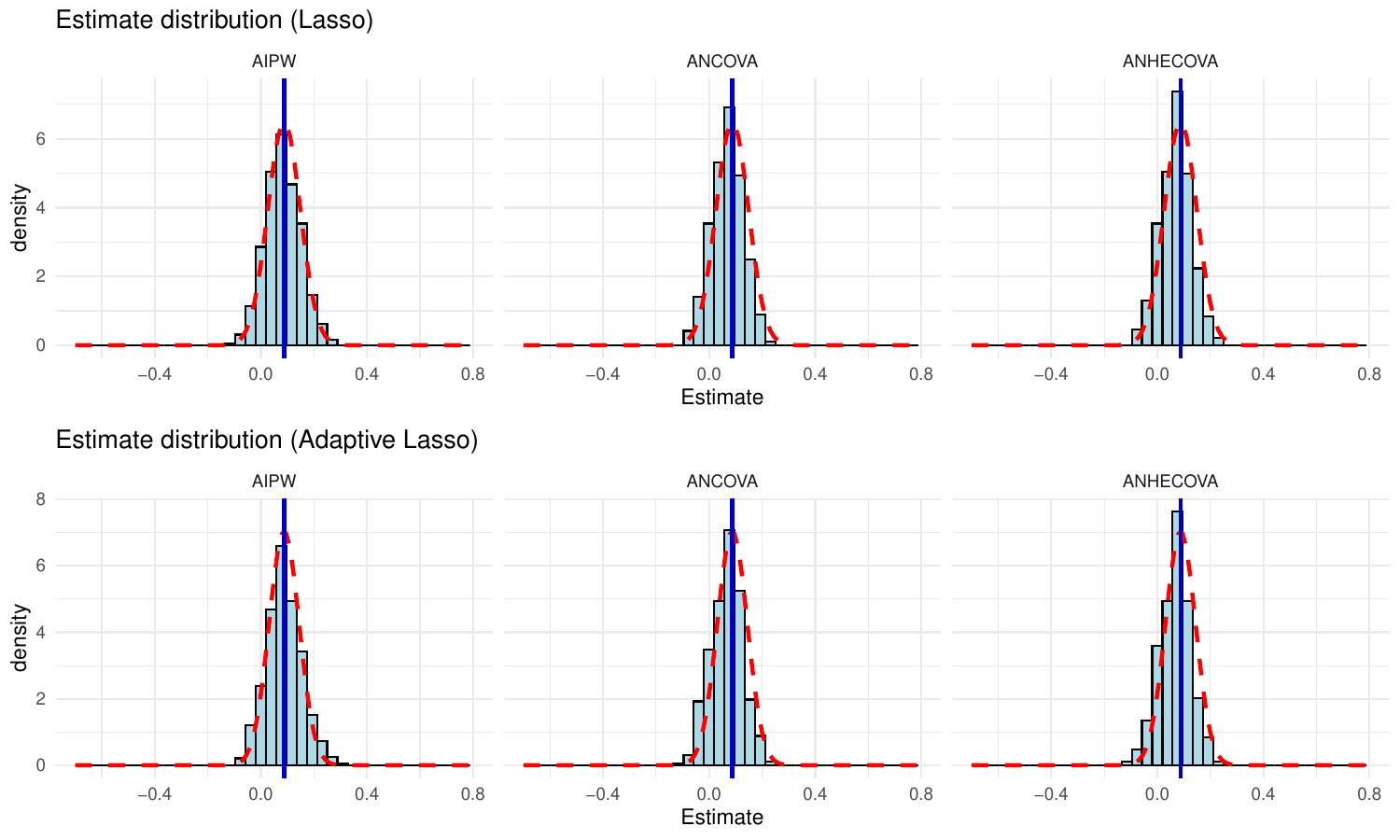}
    \caption{Simulation results for estimation distributions by Lasso and adaptive Lasso variable selections, under binary outcome, linear $\delta(\mb X)$ and $N=100$. The \textcolor{red}{red dashed curves} are the theoretical normal densities curves, and the \textcolor{blue}{blue lines} indicate the true ATE. }
    \label{fig:est-lin100-bin}
\end{figure}

\begin{figure}[H]
    \centering
    \includegraphics[width=0.8\linewidth]{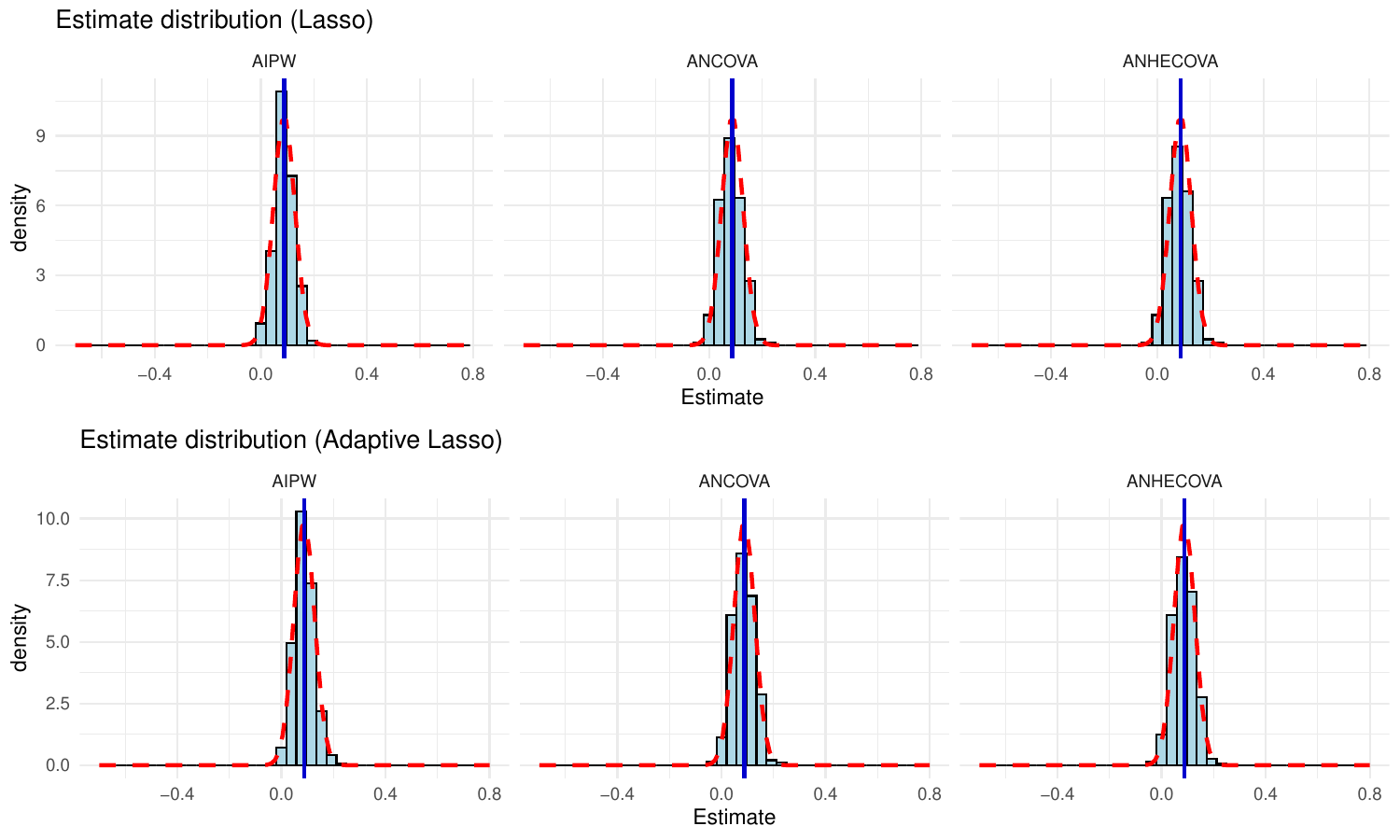}
    \caption{Simulation results for estimation distributions by Lasso and adaptive Lasso variable selections, under binary outcome, linear $\delta(\mb X)$ and $N=200$. The \textcolor{red}{red dashed curves} are the theoretical normal densities curves, and the \textcolor{blue}{blue lines} indicate the true ATE. }
    \label{fig:est-lin200-bin}
\end{figure}

\begin{figure}[H]
    \centering
    \includegraphics[width=0.8\linewidth]{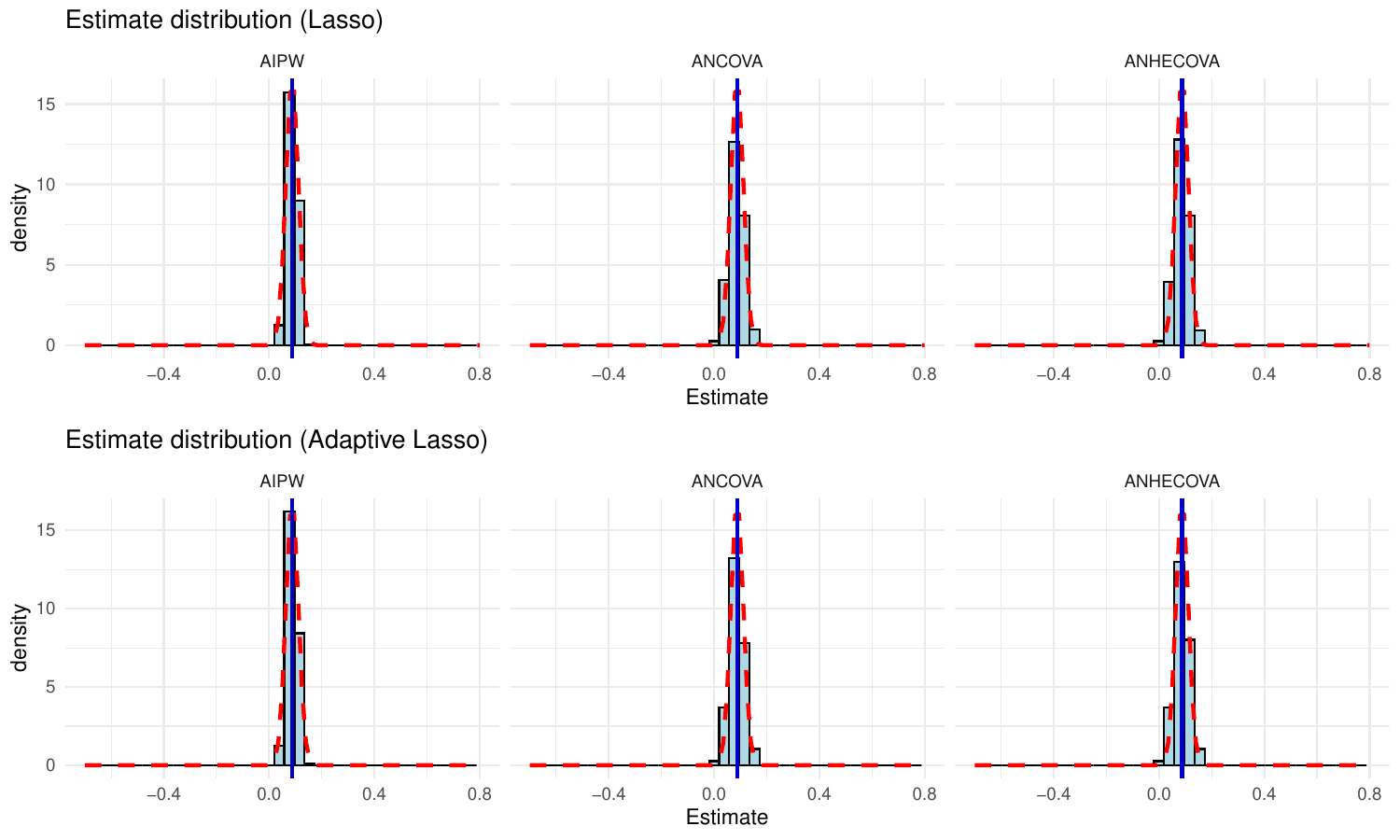}
    \caption{Simulation results for estimation distributions by Lasso and adaptive Lasso variable selections, under binary outcome, linear $\delta(\mb X)$ and $N=500$. The \textcolor{red}{red dashed curves} are the theoretical normal densities curves, and the \textcolor{blue}{blue lines} indicate the true ATE. }
    \label{fig:est-lin500-bin}
\end{figure}

\begin{figure}[H]
    \centering
    \includegraphics[width=0.8\linewidth]{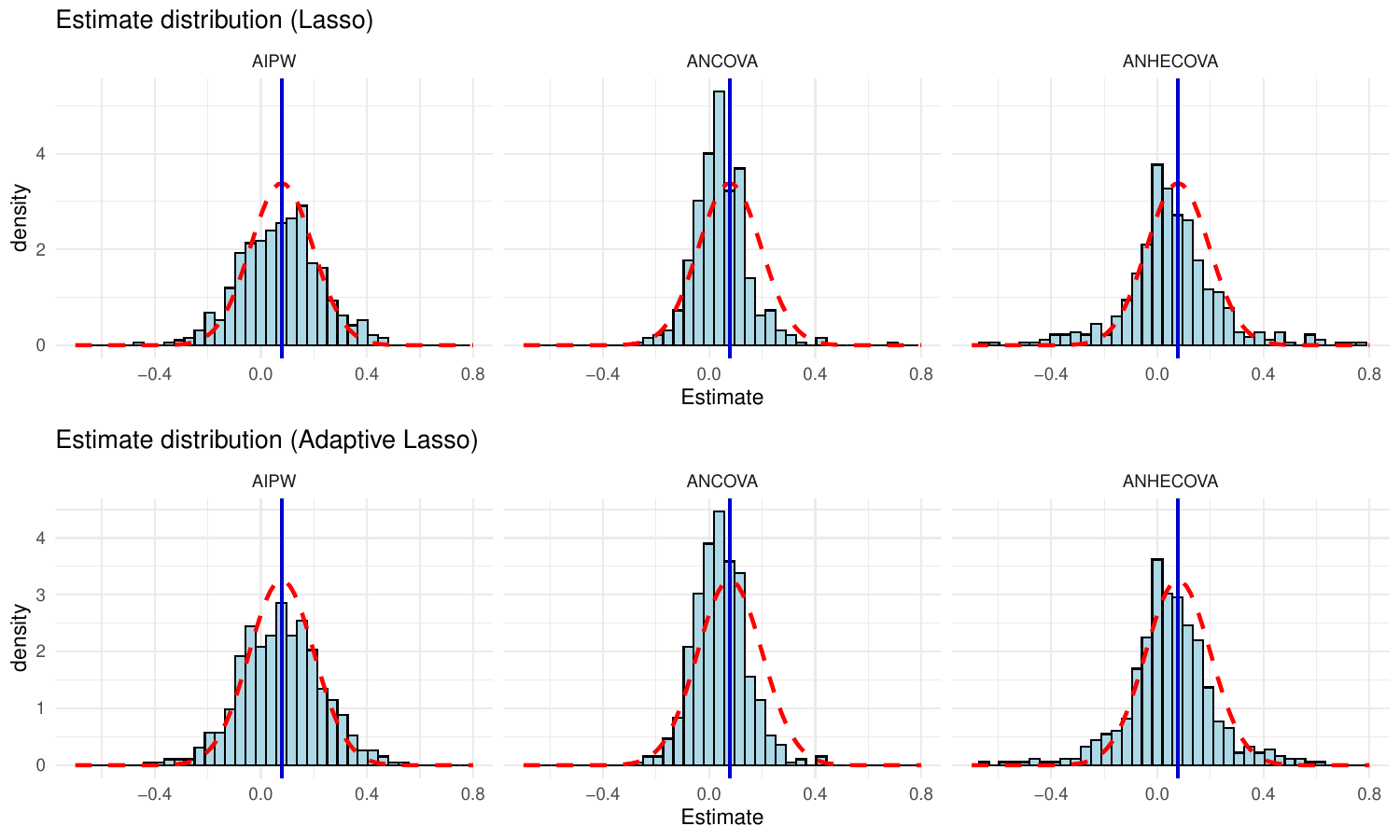}
    \caption{Simulation results for estimation distributions by Lasso and adaptive Lasso variable selections, under binary outcome, nonlinear $\delta(\mb X)$ and $N=40$. The \textcolor{red}{red dashed curves} are the theoretical normal densities curves, and the \textcolor{blue}{blue lines} indicate the true ATE. }
    \label{fig:est-nonlin40-bin}
\end{figure}

\begin{figure}[H]
    \centering
    \includegraphics[width=0.8\linewidth]{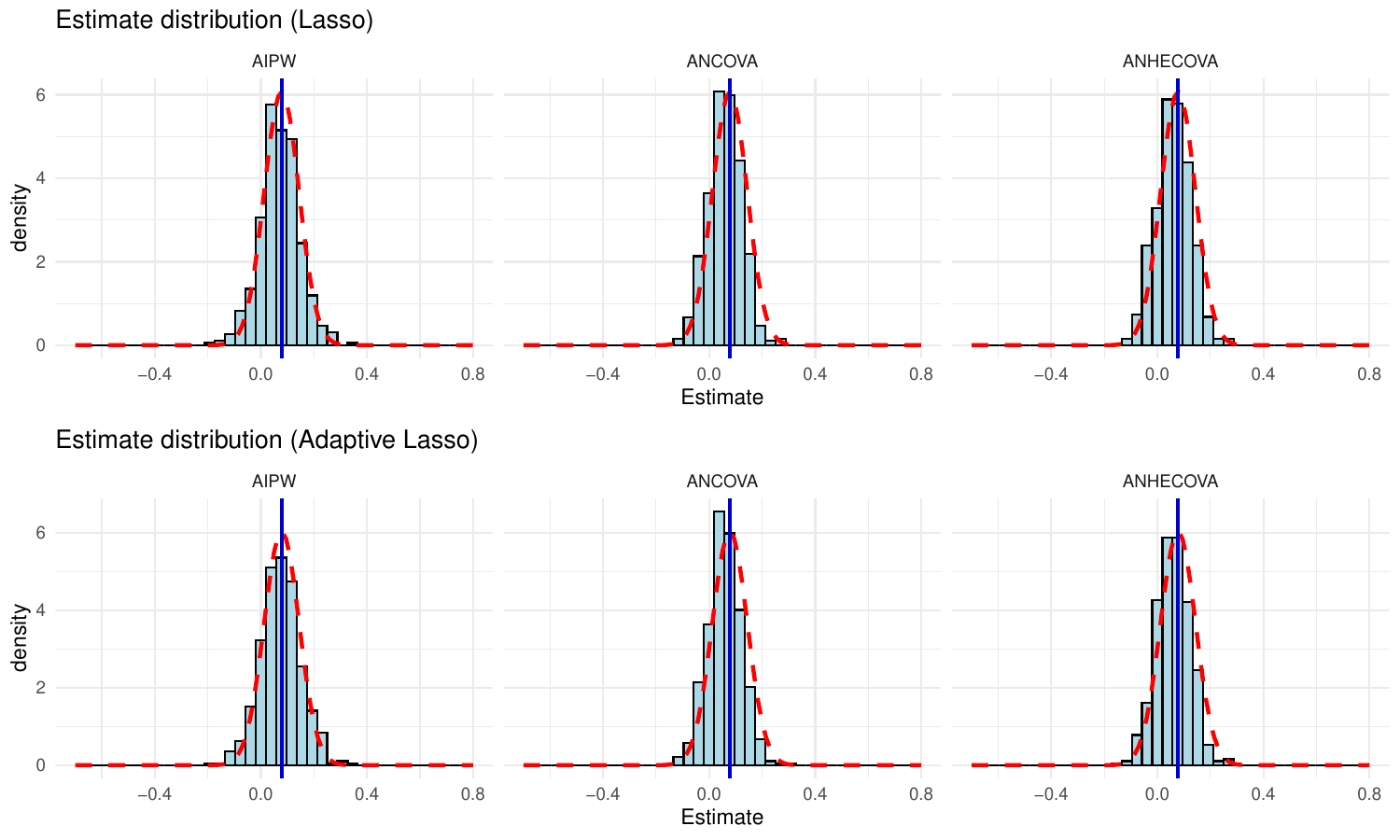}
    \caption{Simulation results for estimation distributions by Lasso and adaptive Lasso variable selections, under binary outcome, nonlinear $\delta(\mb X)$ and $N=100$. The \textcolor{red}{red dashed curves} are the theoretical normal densities curves, and the \textcolor{blue}{blue lines} indicate the true ATE.  }
    \label{fig:est-nonlin100-bin}
\end{figure}

\begin{figure}[H]
    \centering
    \includegraphics[width=0.8\linewidth]{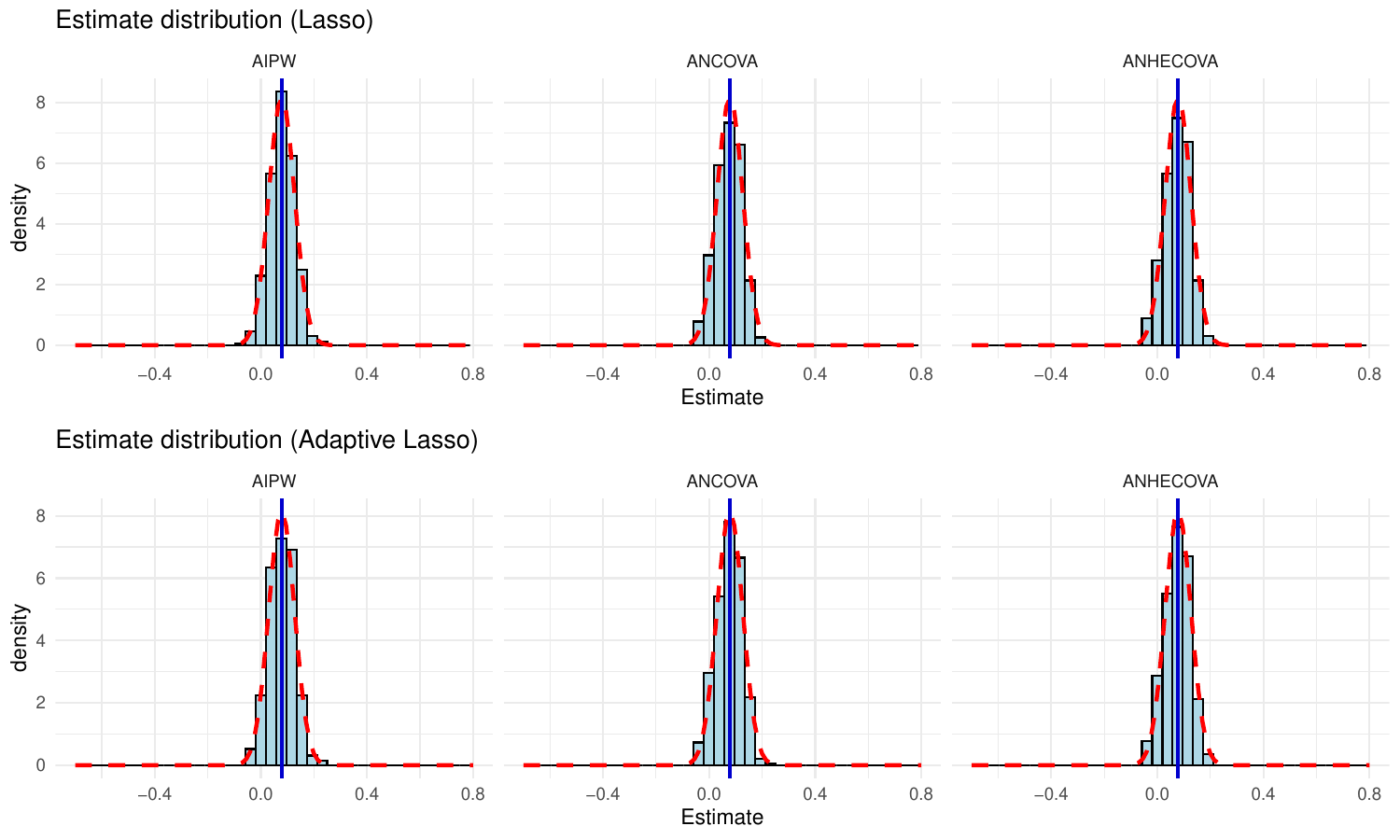}
    \caption{Simulation results for estimation distributions by Lasso and adaptive Lasso variable selections, under binary outcome, nonlinear $\delta(\mb X)$ and $N=200$. The \textcolor{red}{red dashed curves} are the theoretical normal densities curves, and the \textcolor{blue}{blue lines} indicate the true ATE. }
    \label{fig:est-nonlin200-bin}
\end{figure}

\begin{figure}[H]
    \centering
    \includegraphics[width=0.8\linewidth]{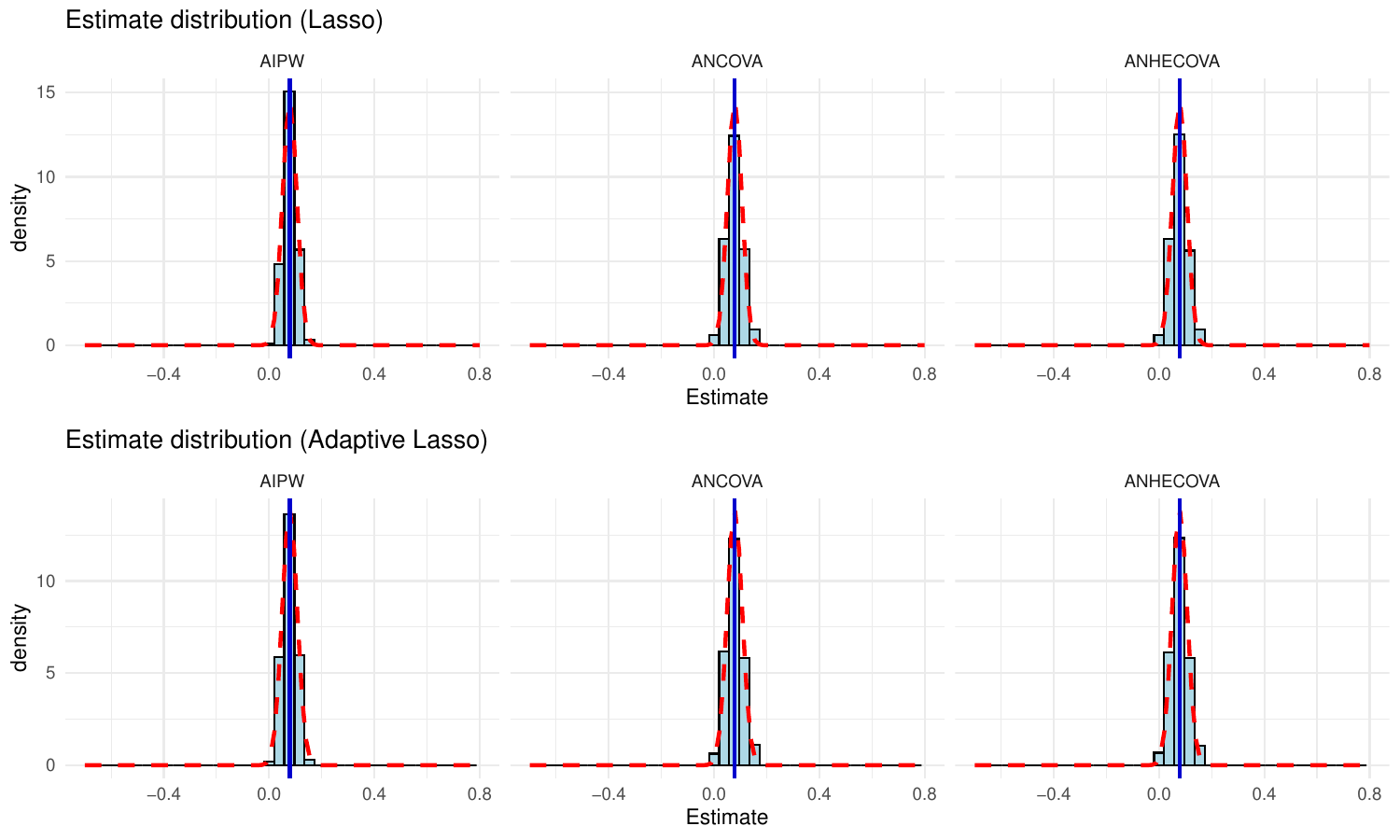}
    \caption{Simulation results for estimation distributions by Lasso and adaptive Lasso variable selections, under binary outcome, nonlinear $\delta(\mb X)$ and $N=500$. The \textcolor{red}{red dashed curves} are the theoretical normal densities curves, and the \textcolor{blue}{blue lines} indicate the true ATE. }
    \label{fig:est-nonlin500-bin}
\end{figure}

\end{document}